\documentclass[12pt,preprint]{aastex}
\usepackage{subfigure}
\shorttitle{CAPMAP and PIQUE CMB Polarization Instruments}
\shortauthors{Barkats et~al.}

\begin{document}
\title{CMB Polarimetry using Correlation Receivers with the PIQUE and CAPMAP Experiments}
\author{D.~Barkats\altaffilmark{1,5}, C.~Bischoff\altaffilmark{2}, P.~Farese\altaffilmark{1,7}, T.~Gaier\altaffilmark{3}, J.~O.~Gundersen\altaffilmark{4}, M.~M.~Hedman\altaffilmark{2,6,8}, L.~Hyatt\altaffilmark{1}, J.~J.~McMahon\altaffilmark{1}, D.~Samtleben\altaffilmark{2,8}, S.~T.~Staggs\altaffilmark{1}, E.~Stefanescu\altaffilmark{4}, K.~Vanderlinde\altaffilmark{2}, B.~Winstein\altaffilmark{2}}
\altaffiltext{1}{Department of Physics, Princeton University, Princeton, NJ, 08544}
\altaffiltext{2}{Kavli Institute of Cosmological Physics and Enrico Fermi Institute, University of Chicago, Chicago, IL, 60637}
\altaffiltext{3}{Jet Propulsion Laboratory, California Institute of Technology, Oak Grove Drive, Pasadena, CA, 91109}
\altaffiltext{4}{Department of Physics, University of Miami, Coral Gables, FL, 33146}
\altaffiltext{5}{now at: Department of Physics, California Institute of
Technology, Pasadena, CA, 91125}
\altaffiltext{6}{now at: Department of Astronomy, Cornell University, Ithaca,
NY, 14853}
\altaffiltext{7}{Dicke Fellow}
\altaffiltext{8}{Kavli Fellow}
\begin{abstract}
The Princeton IQU Experiment (PIQUE) and the Cosmic Anisotropy
Polarization MAPper (CAPMAP) are experiments designed to measure the
polarization of the Cosmic Microwave Background (CMB) on sub-degree scales
in an area within $1^{\circ}$ of the North Celestial Pole using heterodyne
correlation polarimeters and off-axis telescopes located in central New
Jersey. PIQUE produced the tightest limit on the CMB polarization prior
to its detection by DASI, while CAPMAP has recently detected polarization at
$\ell \sim 1000$. The experimental methods and instrumentation for these two
projects are described in detail with emphasis on the particular
challenges involved in measuring the tiny polarized component of the CMB.
\end{abstract}

\section{INTRODUCTION}
The polarized component of the Cosmic Microwave Background (CMB) provides
abundant information about the structure and dynamics of the early
universe. Measurements of the E-mode polarization
complement temperature anisotropy observations,
and the consistency between these data sets should provide strong
confirmation of the standard hot Big Bang cosmology. Measurements of the
B-mode polarization can both reveal gravitational lensing of the CMB \citep{Zelda:1998,Seljak:2004,Smith:2004}
and probe the gravity wave content of the universe \citep{Seljak:1997gy,Seljak:1998nu,Kamionkowski:1998av}.
Recently, a polarized component of the CMB was detected by the
DASI \citep{Kovac:2002,Leitch:2002} and WMAP \citep{Bennet:2003a} experiments.
These early data were extended last year by new results from CAPMAP \citep{Barkats:2004he}, DASI \citep{Leitch:2004gd}, and CBI \citep{Readhead:2004xg}. Now about a dozen experiments are underway to characterize this elusive signal.
These various projects employ a variety of detection techniques to
contend with the small size of the
polarized signal (a few $\mu$K rms), which demands not only high
sensitivity but also stringent control of systematic effects.

This paper describes the experimental methods and instrumentation for the Princeton IQU Experiment
(PIQUE) and the Cosmic Anisotropy Polarization MAPper (CAPMAP),
both of which have used correlation polarimetry at 90 and 40 GHz and
off-axis telescopes to search for the E-mode polarization of the CMB. 
The PIQUE and CAPMAP instruments have been described briefly in their respective results papers \citep{Hedman:2001,Hedman:2002ck,Barkats:2004he}. Here we give a detailed description of their characterization and optimization, with emphasis on both unexpected challenges met and new techniques developed which will be relevant for future large-scale experiments. Further discussion on specific topics may be found in \citet{Hedman:02T} and \citet{Barkats:04T}. After an
overview of the projects (\S~\ref{overview}) and the Stokes
parameters in the context of CMB observations (\S~\ref{stokes}), we
describe the
polarimeters (\S~\ref{polarimeters}) and the optics for the two
experiments (\S~\ref{optics}), as well as the atmospheric
characteristics of the observing
sites (\S~\ref{atmos}). We conclude with
detailed discussions on the operation, performance, and calibration of the
instruments (\S~\ref{operperf}), with a special focus on
polarization-specific systematic effects (\S~\ref{polfid}), including those that---although small enough to be neglected in these experiments--will prove crucial for future campaigns.

Table~\ref{acronyms} summarizes the acronyms
used throughout this paper.

\section{OVERVIEW}
\label{overview}
A summary of the main characteristics of the PIQUE and CAPMAP systems is provided in Table~\ref{tbl:senfac}.

PIQUE consisted of two independent
heterodyne correlation polarimeters using cryogenic high electron mobility transistor (HEMT) amplifiers
coupled through corrugated feed horns to a 1.2-meter off-axis
parabolic primary mirror. One polarimeter
operated at W-band (84--100 GHz) with a nearly Gaussian, $15\arcmin$ full width
at half maximum (FWHM) beam. The other
operated at Q-band (35--45 GHz) with a $30\arcmin$ FWHM beam. Construction
of the PIQUE instrument began in 1997, and observations were made from the roof of Jadwin Hall in Princeton, New Jersey on two successive
winters in 2000 and 2001. Observations at W-band near the
North Celestial Pole (NCP), published
in \citet{Hedman:2001,Hedman:2002ck}, provided the tightest limits on the polarized component
of the CMB prior to its detection by the DASI experiment \citep{Kovac:2002}.

CAPMAP extends the fundamental technologies developed
for PIQUE to smaller angular scales and higher sensitivity.
The full CAPMAP instrument is an array of 16
polarimeters installed in the focal plane of the 7-meter
antenna \citep{Chu:1978} at Lucent Technologies in Holmdel, NJ. Twelve of the polarimeters operate
at W-band and four at Q-band. This combination provides nearly equal
sensitivity at two frequencies and will allow galactic foregrounds to be
identified and removed if needed. The beam FWHMs are $4\arcmin$ and $6\arcmin$ at
W-band and Q-band, maximizing CAPMAP's sensitivity to angular
scales where the E-mode polarization
is expected to peak ($\ell \sim 1000$).
The CAPMAP instrument was designed to be fielded in a staged deployment.
In the winter of 2002--2003, four W-band receivers (denoted $A$--$D$) in a
single dewar made observations. The results from the first season of CAPMAP are published in \citet{Barkats:2004he}.
Twelve polarimeters in three dewars recorded data in the winter of
2003--2004, and analysis of these data is underway. The full CAPMAP system,
with 16 polarimeters, has now been deployed for the winter of 2004--2005.

\section{STOKES PARAMETERS}
\label{stokes}
\begin{figure}
\plotone{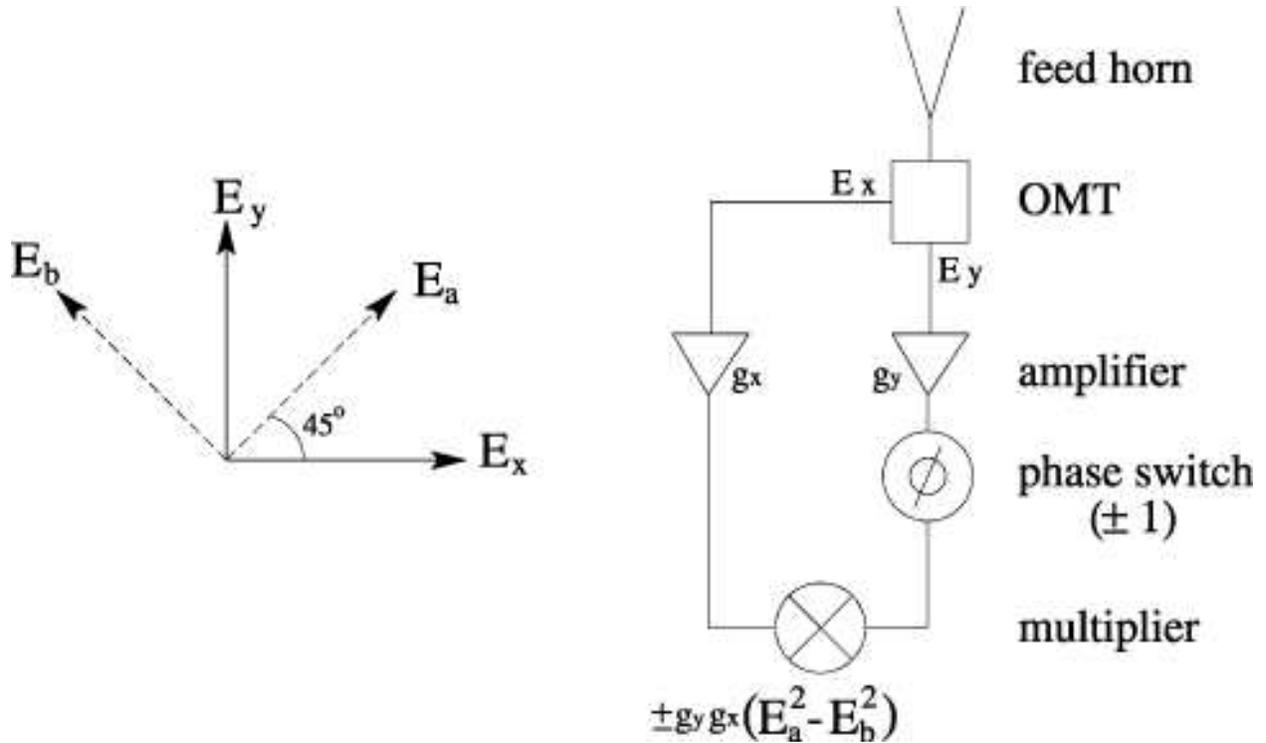}
\caption{\label{fig:correlation}Principles of correlation polarimetry.
The polarimeter couples the two orthogonal components of
the incident electric field ($E_x$ and $E_y$) into separate arms. After amplification, both components
are fed into a multiplier, which produces an output voltage proportional to the product of the two
input fields. This product is proportional to the Stokes parameter $U=\left\langle 2E_x E_y \right\rangle=\left\langle E_a^2\right\rangle-\left\langle E_b^2\right\rangle$, where $E_a$ and $E_b$ are another set of orthogonal components of the radiation illustrated above.
The signal from the correlation polarimeter can be modulated rapidly using a $0^{\circ}$--$180^{\circ}$ phase switch in one of the arms, which takes $U$ to $-U$. Fluctuations in the gains $g_x$ and $g_y$ alter only the small polarized component rather than the total intensity.}
\end{figure}

The Stokes parameters $I$, $Q$, $U$, and $V$ completely determine the
polarization state of an electromagnetic wave, and they form the
basic formalism of astronomical polarimetry \citep{Hamaker:1996,Kraus:1986,jackson:book}. The CMB is not
expected to have a circularly polarized component, so in what follows we consider the case $V=0$.
The remaining parameters can be quantified using a particular
coordinate system. Define two orthogonal axes $x$ and $y$ in the plane
perpendicular to the incident wave vector.
Next, define the orthogonal axes $a$ and $b$ which are rotated
$45^{\circ}$ with respect to $x$ and $y$ as illustrated in
Figure~\ref{fig:correlation}. The components of the electric field
polarized along each of these axes are $E_x$, $E_y$, $E_a$, and $E_b$ such that $E_x =\frac{1}{\sqrt{2}}\left(E_a-E_b\right)$ and $E_y=
\frac{1}{\sqrt{2}}\left(E_a + E_b\right)$. The Stokes parameters
can then be written as
\begin{equation}\begin{array}{ccccc}\label{iqu}
I &=& \left\langle E_x^2\right\rangle + \left\langle E_y^2\right\rangle &=& \left\langle E_a^2\right\rangle + \left\langle E_b^2\right\rangle,\\
Q &=& \left\langle E_y^2\right\rangle - \left\langle E_x^2\right\rangle &=& 2 \left\langle E_a E_b \right\rangle,\\
U &=& 2 \left\langle E_x E_y\right\rangle &=& \left\langle E_a^2\right\rangle-\left\langle E_b^2\right\rangle.
\end{array}
\end{equation}
The angle brackets denote a time average over the sampling period of the detector. The interpretation of
the Stokes parameters in this basis is clear. The
parameter $I$ is the total intensity of the radiation (up to a
constant factor), while $Q$ and $U$ quantify the linear polarization.
The polarization fraction $P = \sqrt{Q^2+U^2}/I$
and the polarization orientation $\theta=\frac{1}{2}\arctan(U/Q)$ can be
derived from these parameters.
The Stokes parameters provide a natural formalism for CMB polarimetry
because the output of a polarimeter is generally proportional to some
linear combination of $Q$ and $U$ (see Figure~\ref{fig:correlation}). However, this formalism
also requires (1) that the parameters be defined self-consistently and (2)
that a coordinate system be specified appropriately.

As defined in Equation~\ref{iqu}, $I$, $Q$, and $U$ are self-consistent under
transformations of the coordinate system and other operations on the state
of the radiation field, which can be implemented using the Jones matrix or Mueller matrix
formalism \citep{Heiles:2001,Tinbergen:96,O'dell:02}. A similarly self-consistent set of
parameters can be generated by multiplying all of these terms by the same
constant factor. Thus we can generate Stokes parameters with units
of temperature as follows: if $T_{x}$ is the physical
temperature of a blackbody which emits the observed
value of $\left\langle E_x^2 \right\rangle$ (and similarly for $T_y$), then the Stokes parameters
can be expressed as $I=\left(T_x+T_y\right)/2$ and $Q=\left(T_x-T_y\right)/2$. With this convention, 
$I$ is the temperature of the object which produces the observed
unpolarized intensity, and $Q$ is the consistent measure of the linear
polarization.

Two natural coordinate systems are typically used to define $Q$
and $U$ in astronomical polarimetry. One system, established by the \citet{IAU},
defines a coordinate system at every point in the sky such that a signal
polarized parallel to a great arc connecting the celestial poles has
positive $Q$ and zero $U$. This system is useful for combining and
comparing measurements from different instruments, but for individual
polarimeters there is also a coordinate system defined by the instrument
itself. The latter is useful when discussing the performance
and characteristics of the experiment and is used exclusively in
this paper.

\section{POLARIMETERS}
\label{polarimeters}
The CAPMAP and PIQUE projects use phase-switched correlation polarimeters. Here we give an overview of correlation
polarimetry followed by details of its implementation in PIQUE and CAPMAP.
\subsection{Principles of Correlation Polarimetry}
\label{corpolprin}
\begin{figure}
\plotone{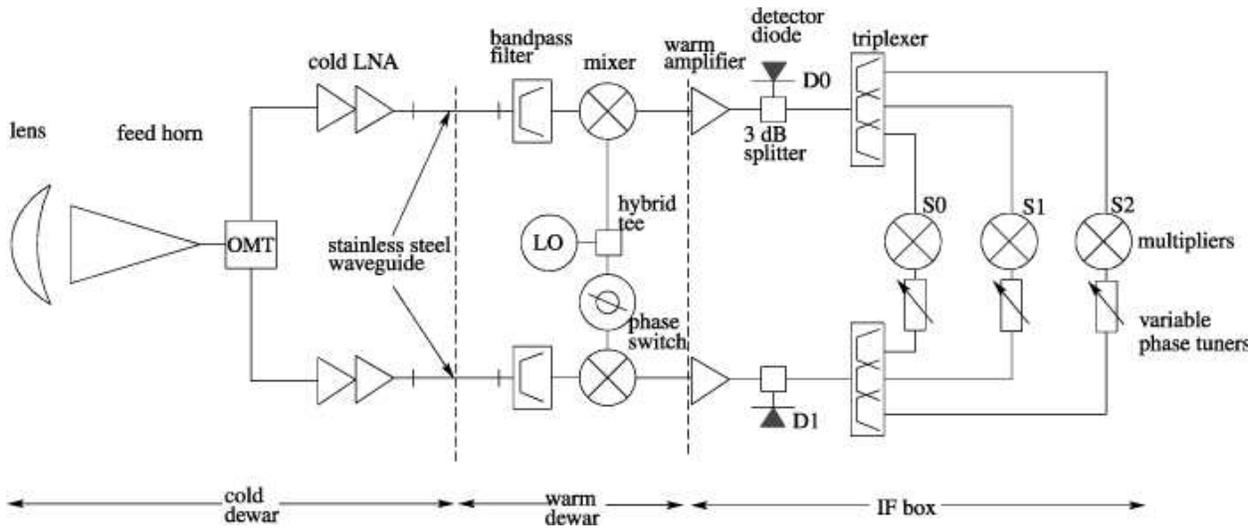}
\caption{\label{blockrad}Block diagram of the PIQUE and CAPMAP polarimeters. In CAPMAP the components inside the dewar are either cryogenically cooled (cold dewar), or kept at room temperature (warm dewar). In PIQUE, the LO, phase switch, and hybrid tee were located outside the dewar while the filters and mixers were cooled to 50~K--80~K (see Table~\ref{temps}). After the mixers, the signals exit the dewar via coaxial cable and are further processed in the IF box. The IF box is an RF-tight enclosure, mechanically attached to the dewar. The three polarimetry channels are S0, S1,
and S2, while the two total power (intensity) channels are D0 and D1. Refer to \S~\ref{radconstruction} for a detailed description of the polarimeter.}
\end{figure}

Both PIQUE and CAPMAP use phase-switched correlation polarimeters to extract small polarized
signals from largely unpolarized radiation. Figure~\ref{fig:correlation} sketches the key details of such a polarimeter.
It has two identical ``arms'', each of which carries radiation coherently from an
ortho-mode transducer (OMT) to a multiplier.
The OMT defines a set of coordinate axes $x$ and $y$. The $x$ component of the
incident electric field, $E_x$, is coupled through one output of the
OMT into one arm of the polarimeter, while $E_y$ is coupled into the
other arm. The two arms (see also Figure~\ref{blockrad}) contain devices which amplify and filter these
components of the radiation before they reach the multiplier. The output
voltage of the multiplier is proportional to the
product $\left\langle E_x E_y \cos\theta \right\rangle$, where the angle brackets denote a low-pass-filtered time average,
and $\theta$ is the phase difference between $E_x$ and $E_y$. In the case of $V=0$, this phase difference arises entirely from mismatches between the receiver arms. The output is proportional to the $U$ Stokes
parameter in the instrumental coordinate system defined by the OMT. In contrast to the correlation polarimeters used in PIQUE and CAPMAP,
a number of CMB polarization experiments use differencing
polarimeters, which measure $\left\langle E_x^2 \right\rangle$ and $\left\langle E_y^2 \right\rangle$ separately and
then difference to obtain an estimate of $Q$ \citep{Johnson:03,Montroy:03,Keating:2003,Church:03}.
Differencing polarimeters using bolometers from balloons and satellites can achieve better sensitivities than correlation
polarimeters, which rely on coherent amplifiers. However,
correlation polarimetry has advantages in the control of
systematic effects.
For both types of polarimeter there are coupling coefficients $c_{x,y}$ between
the components of the electric field and the detector elements, so that the
outputs of the polarimeters are:
\begin{equation}
2 c_x c_y \left\langle E_x E_y \right\rangle
\end{equation}
for a correlation system and
\begin{equation}
c_x^2 \left\langle E_x^2 \right\rangle - c_y^2 \left\langle E_y^2 \right\rangle
\end{equation}
for a differencing polarimeter.
The coupling coefficients will in general drift in time.
If the radiation is unpolarized ($\langle E_x^2 \rangle - \langle E_y^2 \rangle = \langle E_x E_y \rangle =
0$), these drifts do not generate a spurious nonzero output in a correlation
polarimeter (to first order).

In practice, either type of polarimeter has a time-variable
offset. The stability requirements on these offsets can be greatly
reduced by modulating the input signal. If the signal is modulated
at a frequency $f$, then only drifts in the offset on time scales shorter than
$1/f$ are relevant. The phase switch in a correlation polarimeter multiplies the electric field in
one arm by $\pm 1$ by adding alternately $0^{\circ}$ or $180^{\circ}$ of extra phase. This switching can be done at frequencies of a few kHz.
The demodulated output of the correlation polarimeter is therefore immune to nearly all $1/f$ noise caused by gain drifts in the receiver.

\subsection{Construction}
\label{radconstruction}

Most of the heterodyne correlation polarimeters used in PIQUE and CAPMAP are sensitive
to W-band radiation, where the
contribution from astrophysical foregrounds is expected to be minimal \citep{Tegmark:2000}. Q-band polarimeters
provide data at a second frequency for foreground identification.
The basic architecture
of all of these instruments is essentially the same, so the following
description focuses on the W-band polarimeters; the Q-band instruments are discussed only
insofar as they differ from the W-band systems.
Figures \ref{blockrad} and \ref{actrad} illustrate the architecture of
the PIQUE and CAPMAP polarimeters. The incident radiation is
first coupled into the radio frequency (RF) section, where the first stage of amplification
and bandpass filtering is performed. A mixer and local oscillator (LO)
system introduces the 4 kHz phase modulation and down-converts the radiation in both arms
to the intermediate frequency (IF) band at 2--18~GHz. This radiation enters the IF section where the signal is
further amplified and divided by filter banks into three (for W-band) or
two (for Q-band) frequency sub-bands. For each sub-band, an analog multiplier receives the
radiation from both arms and produces an output voltage.
The back-end electronics extract the component of this voltage
modulated at 4 kHz, which is proportional to the $U$ Stokes parameter of
the incident radiation. Finally, a data acquisition program
stores the amplitude of the modulated signal to disk. The signal levels
carried through each device are summarized in Table~\ref{gainstab}. A list of the actual components used
in the polarimeters is provided in Table~\ref{compolist}.

\subsubsection{RF Section}
\label{rfdesc}

Radiation enters the polarimeter through a feed horn (see
\S~\ref{optics}), and the OMT couples two orthogonal polarized
components of the radiation into the two arms of the polarimeter via
waveguide. The isolation between the outputs of the OMT is an
important factor in determining the polarization fidelity of the polarimeter
and is discussed at length in \S~\ref{poltrans}.

Low-noise amplifiers (LNAs) are attached
directly to both outputs of the OMT. Both PIQUE and CAPMAP use cryogenic
amplifiers based on indium-phosphide (InP)
HEMTs to achieve
high gain with low noise. For PIQUE, the LNAs were designed by M.
Pospieszalski for the WMAP satellite \citep{Jarosik:2003}. While these components
were acceptable LNAs, they were not available in large numbers because each amplifier was
laboriously assembled from multiple individual HEMT devices by hand. CAPMAP's
LNAs use monolithic microwave integrated circuits (MMICs) made from InP HEMTs \citep{weinreb:1999}
designed at JPL and Northrop Grumman Corporation. MMIC
technology circumvents the time-consuming and expensive assembly procedures
and enables the large number of LNAs
required by the CAPMAP experiment to be produced quickly, reproducibly, and at
relatively low cost.

The OMTs, LNAs, and horns are cooled to cryogenic temperatures (25--40~K) with a
closed cycle helium refrigerator (see \S~\ref{thermal}). Stainless steel waveguide
provides the thermal break between these components and the remainder of the
polarimeter, which operates at warmer temperatures. The rest of the
RF section consists of various waveguide sections and a band-defining filter, followed by the mixer. In PIQUE, these components
were cooled by the first stage of the refrigerator. In CAPMAP, these
components were kept inside the dewar but operated at room temperature
(see Table~\ref{temps}).

\subsubsection{Mixer-LO System}
\label{lodesc}

The mixers require a narrow-band, high-power tone to down-convert the RF
signal to the IF band. This signal is provided by the LO,
which is coupled to both mixers via a waveguide hybrid tee and two waveguide paths.
A phase switch in one path modulates the electric field at one mixer by
$\pm 1$. The modulation frequency is 4 kHz, well above the $1/f$ knee of
the polarimeter noise power spectrum, so that the raw polarimetry data are
stable on time scales of many minutes (see \S~\ref{polperf}).

Both PIQUE and CAPMAP use baseband rather than harmonic mixers. The LO
tones are therefore at a frequency comparable to the RF signals: 30.5
GHz for Q-band, 82 GHz for W-band. This imposes
strong constraints on the LO system design, because the high-frequency
signals can only be transmitted efficiently over rigid waveguide, and the two LO paths must have the same length
to within a
millimeter to insure optimal performance (see \S~\ref{radopt}).
Furthermore, at 90 GHz the passive loss through
waveguide is relatively high ($\sim 1.5$~dB/ft for warm Cu waveguide), and the RF power
production efficiency is relatively low, making it more difficult to
deliver the required power to the mixers.

\begin{figure}
\begin{center}
\includegraphics[width=.65\textwidth]{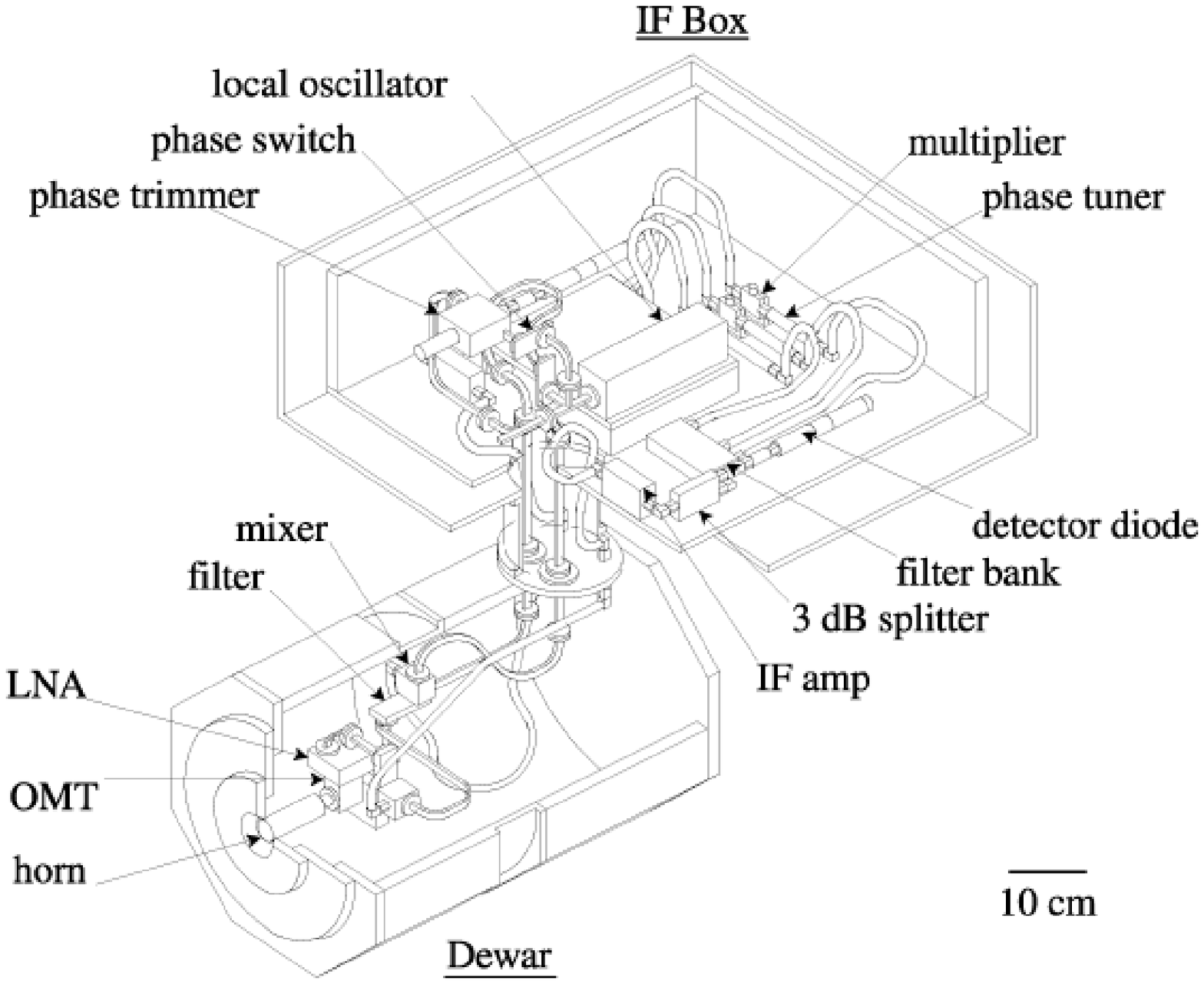}
\\
\includegraphics[width=.75\textwidth]{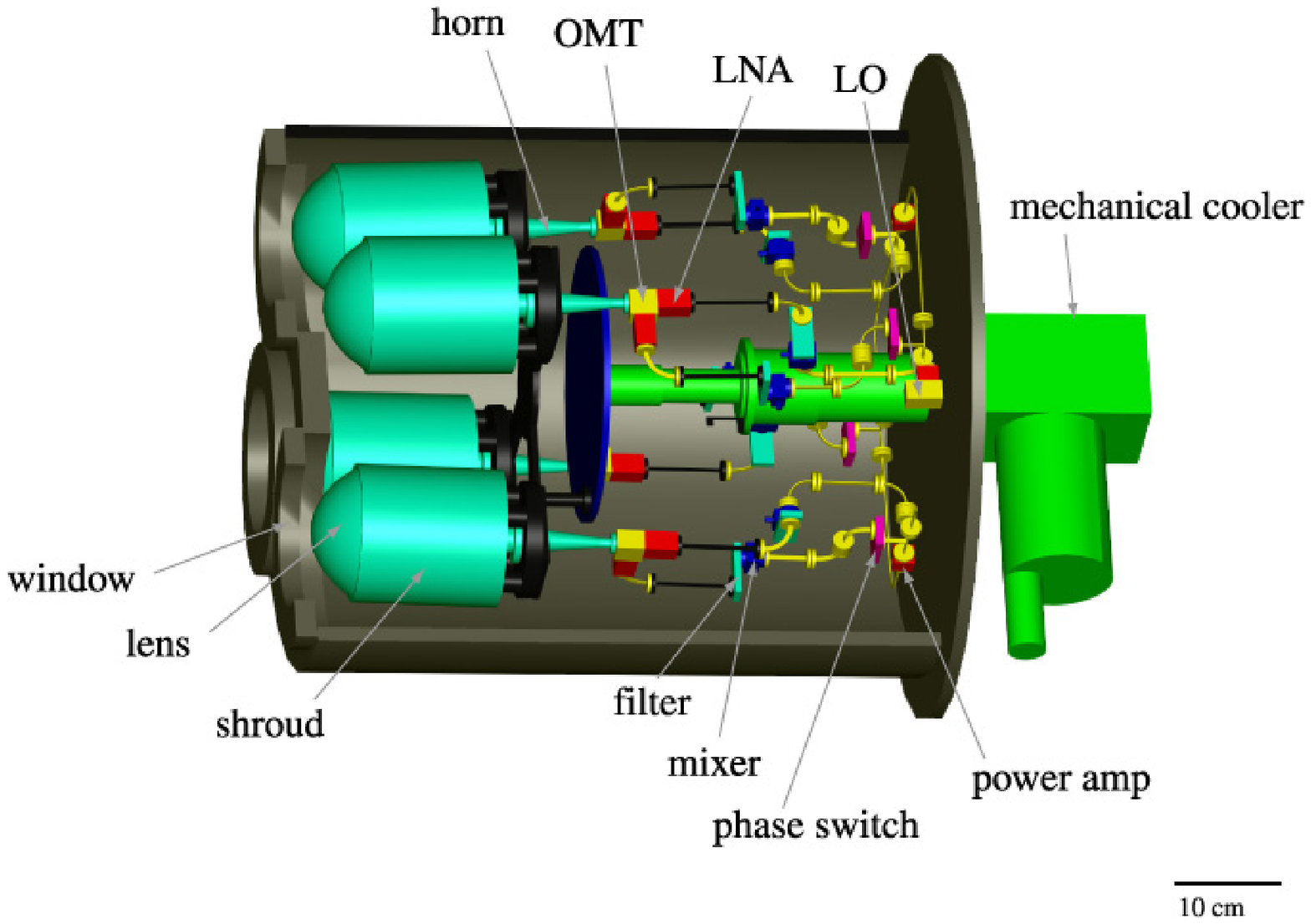}
\end{center}
\caption{\label{actrad}Cutaway drawings of the PIQUE \textbf{(top)} and CAPMAP \textbf{(bottom)} polarimeters. In CAPMAP, each receiver looks through
 its own 6~in (15.24 cm) diameter window, made of a 0.75~mil
 (19~$\micron$) layer of polypropylene backed with a 125~mil (3.127~mm) layer of Goretex
 (RA7957 from W.~L.~Gore). The single PIQUE 1~in (2.54~cm) diameter window consists of
 2~mil (50~$\micron$) polypropylene and 20~mil (0.5~mm) Goretex. Above
 the horn of each CAPMAP receiver are the lens and lens shroud, described
 in more detail in \S~\ref{capopt}. Note that neither the thermal
 straps nor the wiring to the active
 devices is shown. The phase tuners in the IF section allow the path
 length in one arm to be adjusted to match that of the other
 arm. The box containing the IF section of the polarimeters is not 
shown in the CAPMAP system; it contains the same
components as the PIQUE system in a more compact configuration.}
\end{figure}

In PIQUE, each polarimeter had a powerful 100~mW LO located outside the dewar.
A portion of the LO section was therefore
accessible for phase tuning when the dewar was closed. The drawback of this LO system was that it consisted of
convoluted, several-foot-long paths of lossy waveguide (see
Figure~\ref{actrad}).

In the CAPMAP array, the entire mixer-LO system is
contained in the dewars (see Figure~\ref{actrad}) and uses smaller, lower power LOs along with external
MMIC power amplifiers \citep{Wang_huei:2001}. Initially, each polarimeter had an independent
LO for greater modularity. In this configuration, the outputs of
the polarimeters had significant offsets that were extremely sensitive
to mechanical vibrations. These
offsets occurred because the proximity between the LO tone and the RF band made it possible for LO signals
leaking out of the waveguide to couple optically into the RF sections of the polarimeters. Since the LOs
were not locked to the same frequency, the tone leaking from one
LO into other polarimeters was down-converted to a few hundred MHz at the output of the mixers, saturating the IF amplifiers.

This problem was greatly
reduced by running all the W-band polarimeters in a dewar from a single LO, using power
amplifiers to insure the delivery of sufficient power to each polarimeter (see
Figure~\ref{actrad}). However, the
leaking\footnote{We have found that a correctly closed waveguide joint can leak at the $-30$~dB to $-40$~dB power level.}
LO tone could still degrade the
performance of the polarimeter by saturating the LNAs, so all waveguide
joints were covered in microwave absorber\footnote{Emerson \& Cuming (Eccosorb GDS/SS-6M): \texttt{http://www.eccosorb.com}}
and aluminum tape, and a portion of the inside of the dewar was lined with absorber to control further leakage. These procedures have been followed in
subsequent seasons of CAPMAP, where both Q-band and W-band polarimeters
exist in a single dewar. (The third harmonic of the Q-band oscillator is in
the middle of the W-band polarimeters' RF band, so the leakage between
these radiometers must also be controlled.)

\subsubsection{IF Section}
\label{ifdesc}

The IF sections of PIQUE and CAPMAP are virtually identical (except
for the models of the IF amplifiers and phase tuners). After
down-conversion, the 2--18 GHz signals are
carried on semi-rigid SMA coaxial
cables out of the dewar into an ``IF box''. Here IF
amplifiers provide additional gain before 3~dB splitters
direct half the power in each arm to
detector diodes, which measure the total power in the arm (defined as the D0 and D1 channels regardless of their orientation on the sky).
The other half of the power enters a
filter bank, which splits the radiation into three roughly equal IF
bands corresponding to RF frequency bands of 84--89, 89--94.7, and 94.7--100 GHz
for the W-band polarimeters. In the Q-band polarimeters they correspond
to 35--40 and 40--45 GHz. The radiation from both arms of each of these sub-bands is
fed into an analog multiplier. In front of each
multiplier, a phase tuner allows the length of
one arm of the polarimeter to be changed in order to match the phase length in each arm (see \S~\ref{radopt}). Each
multiplier outputs a voltage proportional to the product of the
two input fields (see \S~\ref{corpolprin}). The
outputs of the multipliers for the three sub-bands are denoted S0, S1,
and S2 in order of increasing frequency; S2 does not exist in the Q-band
polarimeters.

\subsubsection{Back-End Electronics/DAQ}
\label{daqdesc}
\begin{figure}
\plotone{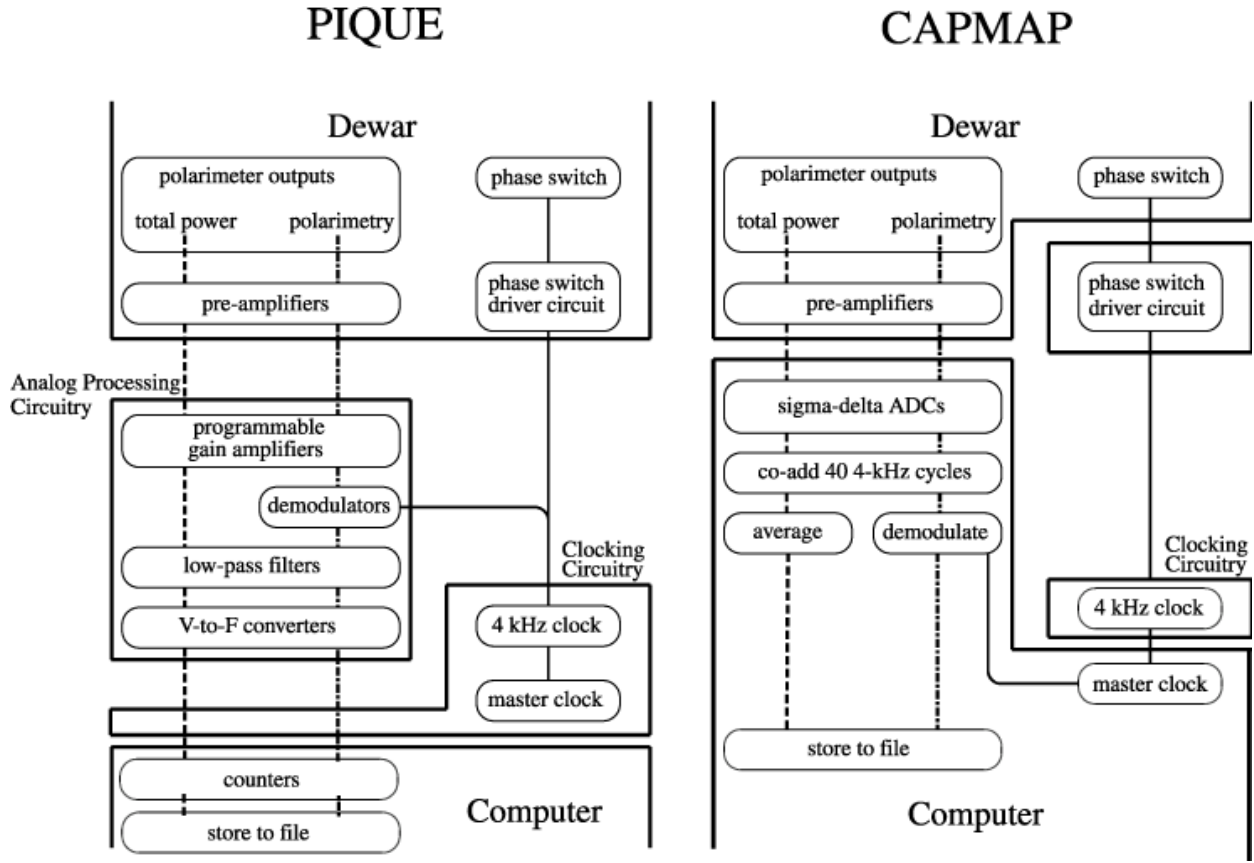}
\caption{\label{fig:beflow}Block diagrams of the data acquisition systems of the PIQUE and
CAPMAP experiments. These electronics provide the 4-kHz signal to
the phase switch and synchronously demodulate the data from the
polarimetry channels. In PIQUE these tasks were performed by custom-made,
largely analog circuitry, while CAPMAP used commercial ADCs, with
demodulation in software.}
\end{figure}

Both PIQUE and CAPMAP use pre-amplifiers to buffer and amplify by 100 the
output voltages from the multipliers and detector diodes before they are
transmitted to the processing electronics and the data acquisition system.
These electronics are responsible for providing the clocking signal to
the phase switch and for synchronously demodulating the polarimetry data
for storage to disk. PIQUE and CAPMAP use different techniques to
perform these two functions, as illustrated in Figure~\ref{fig:beflow}.

In PIQUE, a series of custom-made analog circuits
demodulate, filter, and digitize the polarimeter output voltages
before they
are transmitted and stored on a computer. First, a programmable-gain
amplifier circuit buffers and amplifies the signals from the pre-amps. Then
the polarimetry channel data is demodulated using an analog circuit
based on a modulator/demodulator chip\footnote{Analog Devices (AD630): \texttt{http://www.analog.com}}. Finally, both the
polarimetry and the total power
channel signals are filtered before reaching a voltage-to-frequency
ADC\footnote{Analog Devices (AD652AQ)}. The digital outputs of this circuit are then transmitted and stored
to file on a PC running MS-DOS. The master clock for this system is a
4.096-MHz oscillator located outside the computer. This clock is
divided by 1024 to 4~kHz, yielding the reference
clock for the phase switch driving circuit and the demodulation circuit.
This clock signal is also divided down to the 31.25-Hz sampling rate to provide
interrupts to the data acquisition program.

In CAPMAP the output voltages from the pre-amps are
transmitted directly to a PC running Linux, where they are
digitized at 100 kHz by a commercial 24-bit sigma-delta ADC with 32 differential channels\footnote{Interactive Circuits and Systems Ltd (ICS-610): \texttt{http://www.ics-ltd.com}}. This sampling
rate yields 24 samples per 4-kHz cycle and can adequately record the 4-kHz square waves.
These data are then digitally demodulated and down-sampled to 100 Hz in
software before being recorded to file. In the first season, a second set of ``raw'', undemodulated files was recorded containing averages of 40 4-kHz cycles for every channel. This digital demodulation
enables a variety of ``quadrature'' demodulations to be implemented for
systematic checks and debugging purposes. The master clock is a
12.8-MHz oscillator on the ADC, which is down-converted to 4 kHz in an
external ``clock box'' before being sent to the phase switch driver
circuitry. During the first CAPMAP observing season, although the 4-kHz clock was synchronized with the data acquisition and demodulation
programs, the phase between these two clocks varied each time the acquisition program was restarted between 
observing sessions, so the data required additional off-line processing involving the raw files. The
software and clock-box were upgraded for the subsequent observing season to
eliminate this problem.

\subsubsection{Radiometer Environment}
\label{thermal}
\paragraph{Cryogenics}
A closed-cycle refrigerator\footnote{Helix Technology 
(CTI-1020): \texttt{http://www.helixtechnology.com}} cools
the LNAs, horns, and OMTs to cryogenic temperatures. For PIQUE the LNAs were
attached to
the second stage of the refrigerator, while the filters and mixers were attached
to the first stage\footnote{The first and second stages of the refrigerator 
correspond to approximately
70~K and 20~K respectively.}. During the first season of PIQUE, IF
amplifiers were
located inside the dewar, attached to the first stage of the
refrigerator. This stage of amplification was unnecessary and was removed
for the second season, which reduced the loading on the
cold head and lowered the operating temperature of the other cryogenic
components. Table~\ref{temps} shows typical operating temperatures during
the two observing seasons.

Performance of the polarimeters does not improve when the mixers and
filters are
cooled, so for CAPMAP it was decided to operate these components at room
temperature inside the dewar. The LNAs operated at higher temperatures
(see Table~\ref{temps}) than in PIQUE due to the increased thermal load
on the refrigerator with four polarimeters and the larger windows
which were required for the lenses (see \S~\ref{optics}).
Modifications to improve the thermal isolation between
the cold stages have been completed and have reduced the LNA
temperatures to between 20~K and 30~K for the CAPMAP 2004 season.

For PIQUE and the first season of CAPMAP, the dewars lacked
thermal regulation, so the various components came
to temperatures determined by the cooling power of the cold head. During
the course of three months of observing, the temperature of the LNAs
drifted by $\sim 3$~K. These
drifts did not affect the gain of the polarimeters by more than a few
percent (see \S~\ref{polresp}) and did not significantly affect the
data quality. A system for thermal regulation was built into the CAPMAP
dewars but was not activated until the 2004 season.

In both PIQUE and CAPMAP, the feeds look out of the dewar through polypropylene windows (see Figure~\ref{actrad}). To prevent 
moisture in the air from condensing on these windows, another 
polypropylene sheet 
is used to create an isolated air space, which is kept dry using 
Drierite\footnote{W. A. Hammond Drierite Company: \texttt{http://www.drierite.com}} pellets. In PIQUE the pellets were 
placed in the dry space itself, while CAPMAP used a recirculating air 
system including a cylinder packed with Drierite.

\paragraph{Operational Configuration and Thermal Environment}
For PIQUE the IF box containing the LO and the IF components was
exposed to the outdoor environment. Most of the back-end electronics and 
associated power supplies were also located outside on the telescope base, 
while the DAQ computer and the compressor running the refrigerator were kept 
in boxes within $\sim 5$~m of the telescope. The system was monitored from a 
terminal in a nearby room. The computer and compressor were kept warm by their 
own waste heat, and their temperatures could be controlled by opening 
and closing hatches in the appropriate boxes. Analog heater circuits 
regulated the temperatures of the IF box, the LO, and the back-end electronics.
This system worked well during the early parts of the
season, when the ambient temperature was below roughly 10~$^{\circ}$C.
However, as the outside temperature rose, the exterior of the IF box had to be
cooled for the regulated heaters to operate.

The CAPMAP dewars are located in an enclosed, temperature-controlled
room on the Crawford Hill telescope.
A 2~m~$\times$~1.5~m styrofoam window in one of the walls enables the 
polarimeters to see outside to the secondary mirror of the telescope. 
All the back-end electronics, the DAQ computer, and the associated power supplies are located
in two racks a few meters behind the dewars. The computer control is
performed remotely via ethernet. The compressors running the
mechanical fridges are kept in a separate room to prevent them from
overheating the receiver room and are connected to the fridges via
10~m helium flex lines. When the temperature regulation is functioning correctly, the room temperatures can be
stabilized to better than 1~K. 
During the first observing season,
digital temperature controllers\footnote{OMRON (PN E5GN): \texttt{http://oeiweb.omron.com}} attached to thermoelectric
coolers provided additional regulation of the
LO and the IF components. The controllers ran in a PID configuration by
controlling the duty cycle of the
output. For the second observing season the temperature controllers were
upgraded to units with true PID circuits\footnote{WEST (PN 6100+): \texttt{http://www.westinstruments.com}}.

When the temperature in the cabin was moderately stable, the regulators
could hold the temperature of the IF box and the LOs to within 0.1~K.
Unfortunately, the thermal regulation in the room failed partway through
the first observing season, which caused the IF box regulation to fail and
the temperature to fluctuate by up to 5~K. The temperature change of
the IF electronics affected the polarimetry channel gains by up to
3\%/K but was stable to 1~K for 70\% of the data; the room temperature regulation was restored during subsequent observing seasons.

\subsection{Performance}
\label{radperformance}

We have developed a series of tests and measurements that optimize
and quantify the performance of our polarimeters. As shown in \S~\ref{operperf}, these tests yield data consistent
with the performance of the polarimeters operating on the telescope.

This section focuses primarily on the W-band polarimeters
from PIQUE and the first season of CAPMAP, as the performance of these
instruments has been studied more thoroughly. Additional data from Q-band
polarimeters is included for comparison.

\subsubsection{Optimization of Polarimeters}
\label{radopt}
The smallest signal which can be measured by a correlation polarimeter is
determined by a generalized version of the radiometer equation given in \citet{Kraus:1986}:
\begin{equation}
\Delta X_{min} = K\frac{T_{sys}}{\sqrt{\tau\Delta\nu}} \equiv \frac{S}{\sqrt{\tau}}.
\label{Tsys}
\end{equation}
Here $\Delta X_{min}$ is the smallest detectable value of the
observed Stokes parameter $X=T,Q,U,$ etc, measured in units of temperature
following the conventions in \S~\ref{stokes}. The other parameters are the
same as in the standard radiometer equation: $K$ is a dimensionless constant of order unity
which depends on the type of radiometer, $T_{sys}$ is the effective
system temperature of the instrument, $\Delta \nu$ is the effective
bandwidth, $\tau$ is the post-detection integration time, and $S$ is
the sensitivity, defined as the smallest signal that is detectable at $1\sigma$ in a unit of integration time.

This equation is a fundamental consequence of Gaussian statistics and applies to a large class of radiometers which measure
signals dominated by white noise. The only differences
between various radiometer types are the precise
interpretation of the system temperature and the value of $K$, which can
be computed based on the statistics of the input radiation.
As discussed in standard references such as \citet{Kraus:1986}, $K=1$ for a simple total power receiver with a square law
detector, while a simple correlation receiver with a matched load has
$K=\sqrt{2}$ and $T_{sys}=\sqrt{T_1 T_2}$, where $T_1$ and $T_2$ are the
system temperatures of the two receiver arms. Both of
these devices measure the Stokes parameter $X=T$ and are sensitive to only
a single polarized component of the incident radiation. By contrast,
correlation polarimeters measure the Stokes parameter $X=U$ and receive
both polarized components simultaneously, which improves the
signal-to-noise ratio by a factor of $\sqrt{2}$ over a
simple correlation receiver. Therefore $K=\sqrt{2}/\sqrt{2}=1$ for the
PIQUE and CAPMAP polarimeters, and the radiometer equation is:
\begin{equation}
\Delta U_{min} = \frac{T_{sys}}{\sqrt{\tau\Delta \nu}} \equiv \frac{S}{\sqrt{\tau}}.
\label{Usys}
\end{equation}
Equation~\ref{Usys} shows that in order to achieve optimal sensitivity, the 
system temperature of each polarimeter arm must be as small as possible, and the 
bandwidth must be as large as possible. Both of these parameters have therefore been measured and tuned using the following procedures.
\paragraph{System Temperature}
The system temperature is a measure of
the total power carried in the arms of the operational polarimeter.
It includes contributions from the instrument, the atmosphere, and the CMB
itself. The contribution from the polarimeter is called the receiver
noise temperature $T_{rec}$ and is due to the noise power emitted by the
various components in the instrument. Because the power generated by the
first-stage LNAs is amplified by every subsequent stage with gain, the input LNAs
account for 95\% of the total receiver noise. Note that
the noise temperature of total power channels is directly measured, while the
noise temperature of polarimetry channels can be calculated by taking
the geometric mean of the noise temperatures of the two total power channels.

The LNAs used for the first stage amplifiers in PIQUE and
CAPMAP are designed to have low noise temperatures. The noise
temperatures of both the LNAs in isolation and the assembled
polarimeters are measured using
Y-factor measurements, in which the device views two
different sources of radiation at two different temperatures $T_{hi}$ and
$T_{low}$. The output of the device in these two cases is then
\begin{eqnarray}
O_{hi}&=&R\left(T_{hi}+T_{rec}\right),\\
O_{low}&=&R\left(T_{low}+T_{rec}\right),
\end{eqnarray}
from which we obtain the responsivity in V$\cdot$K$^{-1}$,
\begin{equation}
R = \frac{O_{hi}-O_{low}}{T_{hi}-T_{low}},
\end{equation}
and the noise temperature in K,
\begin{equation}
T_{rec}=\frac{T_{hi}-Y\;T_{low}}{Y-1},
\end{equation}
where $Y=O_{hi}/O_{low}$.

Y-factor tests were performed on the CAPMAP MMICs in a test
chamber at JPL, where a variety of sources could be coupled into the amplifiers. The output power from each device was measured in 200-MHz-wide bands
throughout the relevant frequency range (80--100 GHz for W-band) using a superheterodyne down-conversion system
to give the responsivity and system temperature of the LNA as a function of frequency.
Two different Y-factor tests were performed in the test chamber.
First, a signal from a room-temperature noise diode was coupled into the
MMIC through a 23~dB coupler. When this diode is turned on and off, the
effective temperature of the source changes, providing a relative
measure of the noise temperature. The noise diode can be turned on and
off rapidly, so this test can be repeated many times with many different
bias settings in order to find efficiently the settings which minimize the
noise temperature. After the optimal
settings were found, the absolute noise temperature was measured by
coupling a regulated load to the input of the MMIC. The temperature of
this load was varied between 20~K and 50~K and the frequency-dependent
output was recorded to obtain the noise temperature. These numbers are
given in the first column of Table~\ref{noisetemp}.

The noise temperature of the assembled polarimeters for PIQUE and the
first season of CAPMAP were measured in a similar way, with a circular
waveguide load attached directly
to the OMT in place of the feed horn. The detector diodes monitor the total
power output of the polarimeter with the load at different temperatures;
the derived values of $T_{rec}$ are given in the second column of
Table~\ref{noisetemp}. The two sets of numbers are within a
few Kelvin of each other, demonstrating that the noise temperature of
the polarimeter is dominated by the LNAs.

The above method of measuring the noise temperature is invasive and time
consuming. A simpler Y-factor test is performed on the completely
assembled polarimeter using two external loads---one at room temperature
and one immersed in liquid nitrogen---which are held in front of the horn
or lens. This test is easier to perform, but the room
temperature load can saturate the detector diodes in the polarimeter,
leading to an overestimate of the noise temperature. After the first season of 
CAPMAP, additional attenuation in the total power channels largely eliminated 
this complication.

For PIQUE, the receiver noise temperature at W-band was roughly 75~K, while for the first season of CAPMAP it was closer to
50~K. The noise figures in CAPMAP are typical of the LNAs
available today \citep{gaier:03}.

\paragraph{Bandwidth}

The bandwidth of a radiometer is normally computed from its
frequency-dependent gain $g(\nu)$ \citep{Kraus:1986}:
\begin{equation}
\Delta \nu=\frac{\left(\int g(\nu)\,d\nu\right)^2}{\int g(\nu)^2\,d\nu}.
\end{equation}
Note that the maximum bandwidth is attained when $g$ is constant across the band. For a
correlation polarimeter this formula is modified slightly due to the
effects of a frequency-dependent relative phase shift $\phi(\nu)$ between the
arms of the polarimeter.
If $\phi$ has a nonzero value, then the rms noise fluctuations
at the output remain the same, but the response of the polarimeter to a
polarized signal is reduced by a factor of $\cos\phi$, correspondingly degrading the sensitivity.
This effect can be interpreted
as a reduction in the effective bandwidth of a correlation polarimeter:
\begin{equation}
\Delta \nu=\frac{\left(\int g(\nu)\cos\phi(\nu)\,d\nu\right)^2}{\int g(\nu)^2\,d\nu}.
\end{equation}

The $g$ parameter is set by filters and amplifier slopes and is not easily
modified; in order to maximize the effective bandwidth, we must minimize $\phi$
across the band. If $\phi=0$, then the bandwidth of all polarimetry channels
will be approximately 4~GHz. The relative phase shift between the arms of the
polarimeter can be adjusted with shims in the RF or
LO lines and by adjusting the phase tuners in the IF lines.

Since $\phi$ is a function of frequency, it is best measured by
injecting narrow-band radiation
into the polarimeter. If this radiation is linearly polarized along the
appropriate axis, the output of the polarimeter is
$ip=g(\nu)\cos\phi(\nu)$. In order to obtain an independent estimate of $g(\nu)$ and
$\phi(\nu)$, we introduce an additional
$90^{\circ}$ of phase shift between the two arms, such that the response of
the polarimeter is $op=g(\nu)\cos(\phi(\nu)-90^{\circ})=g(\nu)\sin\phi(\nu)$.
With both $ip$ and $op$, we can compute $g=\sqrt{ip^2+op^2}$,
$\tan\phi=op/ip$, and hence the effective bandwidth. Analogous but
more complex methods allow $\phi$ to be estimated
even if the additional phase shift is not exactly $90^{\circ}$.

A number of methods for introducing a $90^{\circ}$ phase difference between the
two arms of the polarimeter have been explored during the development of
PIQUE and CAPMAP. In PIQUE, a phase trimmer existed in one of
the LO lines. This trimmer was calibrated so that it could be set to two
positions differing by exactly $90^{\circ}$. This phase factor was then
carried through the mixers into the IF section of the polarimeter.
For CAPMAP, the LO sections were contained entirely in the dewar, and so
adjustable phase trimmers were impractical. For testing purposes, the
required phase shift was introduced into the IF lines using
$90^{\circ}$ hybrid couplers. However, this invasive procedure could not be
implemented when the polarimeters were fully assembled and installed in
the dewars. Therefore, the final solution was to introduce the extra
phase shift using a linear-to-circular polarization converter in the
source of the narrow-band radiation.
This converter consists of a length of circular
waveguide with a thin piece of dielectric inserted oriented at an angle
of $45^{\circ}$ with respect to the polarization axis of the incident
radiation. The length of this plastic is such that it retards one
component of the electric field by a quarter wavelength relative to
the orthogonal component, introducing the requisite $90^{\circ}$ phase shift between the components of the radiation coupled into the two polarimeter arms.

To measure the phase shift and bandwidth of the polarimeters using
the polarization converter, we first couple a narrow-band, linearly
polarized signal into the polarimeter through a feed horn suspended above
the lens. The input signal sweeps over a range of frequencies, and
the response at each frequency is recorded. The polarization converter is then inserted into the injector
system to present circularly polarized radiation to the polarimeter,
and the response is again recorded. The two data sets are then aligned 
and combined to determine $\phi(\nu)$ and $g(\nu)$.

\begin{figure}
\plotone{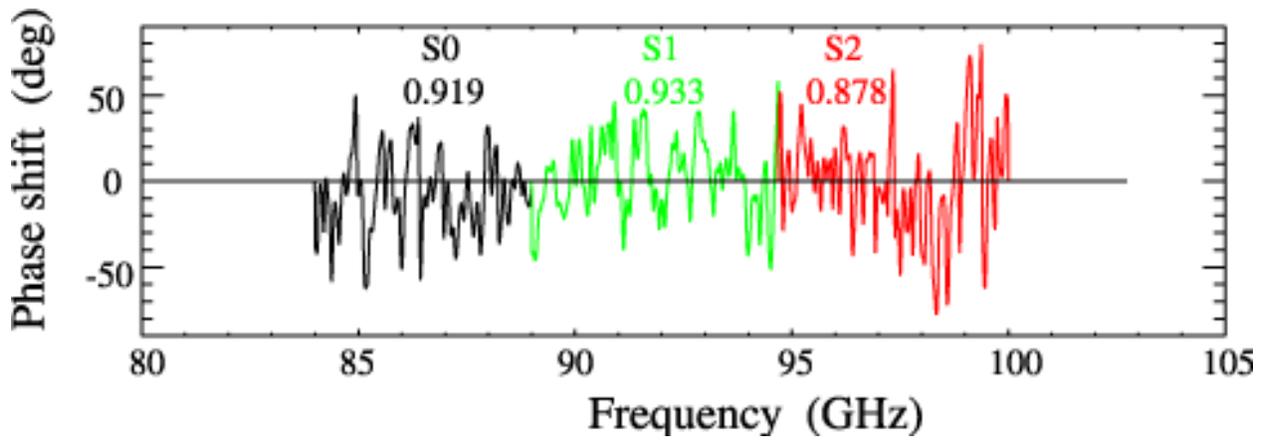}
\caption{\label{exphase}Example of phase-matched phase data from a W-band polarimeter (PIQUE 2001).
The three sub-bands are shown in different shades.
The numbers printed on the plot are the parameter $\left\langle \cos\phi \right\rangle$ for each band. Some of the oscillatory features 
are due to standing waves between the injector and the polarimeter, and do not 
correspond to real phase differences.}
\end{figure}

By adjusting the phase trimmers in the IF section and inserting shims in the
relevant parts of the polarimeter, the phase shift can be
adjusted to be near zero across the entire band of interest.
Figure~\ref{exphase} shows
typical phase data from a phase-matched polarimeter. The phase averages around zero, with variations of $\pm 30^{\circ}$. These variations in $\phi(\nu)$ degrade the effective sensitivity by a factor $\int{g(\nu)\cos\phi(\nu)\,d\nu}/\int{g(\nu)\,d\nu}$, i.e. the
gain-weighted average of $\cos\phi$. In practice, we use the unweighted band average,
$\left\langle \cos\phi\right\rangle=\int\cos\phi(\nu)\,d\nu/\int d\nu$, as an
estimate of how $\phi$ affects the sensitivity of the polarimeter. This parameter is within a few percent of the 
weighted average and is simpler to measure because it is insenstive to 
any frequency-dependent variations in the source intensity.

Typically, the average phase shift is measured multiple times with different
combinations of shims in the LO lines and with the phase tuners in
different positions until the parameter $\left\langle\cos\phi\right\rangle$ is maximized. The final values of this parameter are shown in Table~\ref{phasetab} and range from 0.73 to 0.94 with a median value of 0.87. The
sensitivity of the polarimeters is degraded by only
about 15\% due to these residual phase shifts (see \S~\ref{polsens}).

\subsubsection{Polarized Gain Calibration}
\label{radcal}

The responsivity of the polarimeters is also calibrated prior to their installation on the
telescope. The total power
channels are calibrated during the Y-factor tests described above, but
the polarimetry channels can only be calibrated using specialized
sources. These sources must produce broadband thermal radiation with a
small polarized component ($U < 1$~K). The size of the
polarized component must also be calculable from first principles.

In PIQUE, the primary laboratory calibration source consisted of
two thermally regulated rectangular waveguide loads. Each load produced thermal radiation
polarized along one of two orthogonal axes, which was coupled into the polarimeter
through an OMT rotated by $45^{\circ}$ with respect to the polarimeter's natural coordinate system.
The input polarized signal was simply one-half the temperature difference between the loads,
and the polarimeter could be calibrated by setting the temperatures of these
loads to a variety of values.
While this calibration technique was successful for both PIQUE and CAPMAP
polarimeters, it required replacing the
horn and lens with a system of waveguide, and it could take up to a
day to calibrate a single radiometer due to the load's thermal time constant.

For CAPMAP, a novel calibration method using beam-filling emissive metal plates
was developed, which can be performed quickly on fully
assembled polarimeters with feed systems installed.
Both the thermal emission and the reflected radiation from a metal plate are weakly polarized due to the finite 
conductivity
of the material (see for example \citet{staggs:2002, cortiglioni:1994,Strozzi:2000}). The orientation of the
polarized signal is parallel to the plane of incidence, and the received signal is given by
\begin{equation}
P=\sqrt{4\pi\epsilon_o\nu\rho}\left(\sec\theta-\cos\theta\right)\left(T_{plate}-T_{load}\right),
\label{eq:miniplate}
\end{equation}
where $\rho$ is the resistivity of the metal (in SI units), $\nu$ is the
band center frequency, $\theta$ is the angle of
incidence, $T_{plate}$ is the physical temperature of the plate, and
$T_{load}$ is the temperature of the radiation reflecting off the plate.
For an instrument viewing a liquid nitrogen temperature load through a room temperature
aluminum plate tilted at $45^{\circ}$,
the polarized signal is roughly $100$~mK. Note that the total thermal load on the polarimeter is
comparable to the load when the polarimeter is viewing the sky.

\begin{figure}
\subfigure[]{\label{miniplate}\includegraphics[width=.4\textwidth]{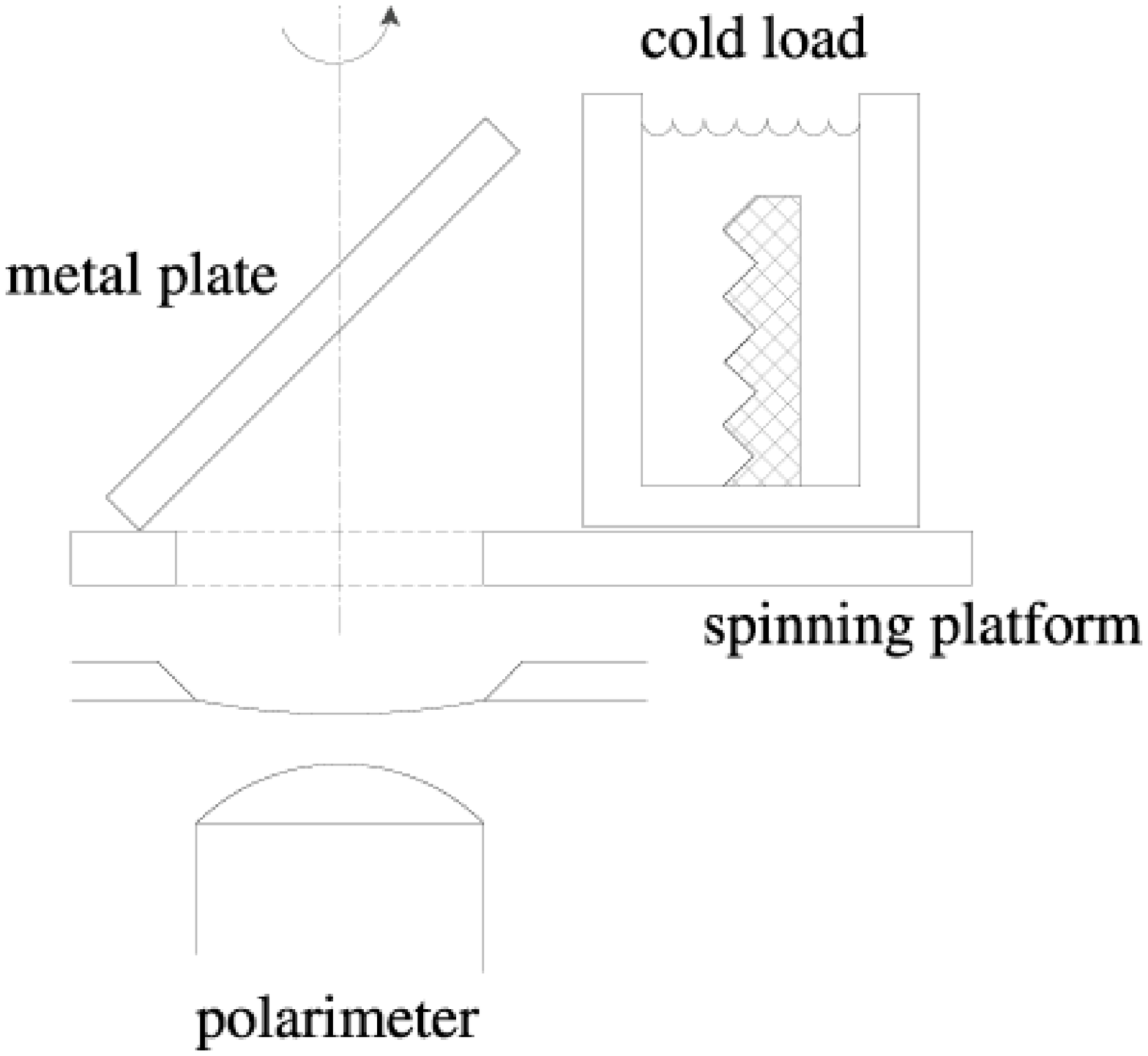}}
\subfigure[]{\label{minidat}\includegraphics*[width=.6\textwidth]{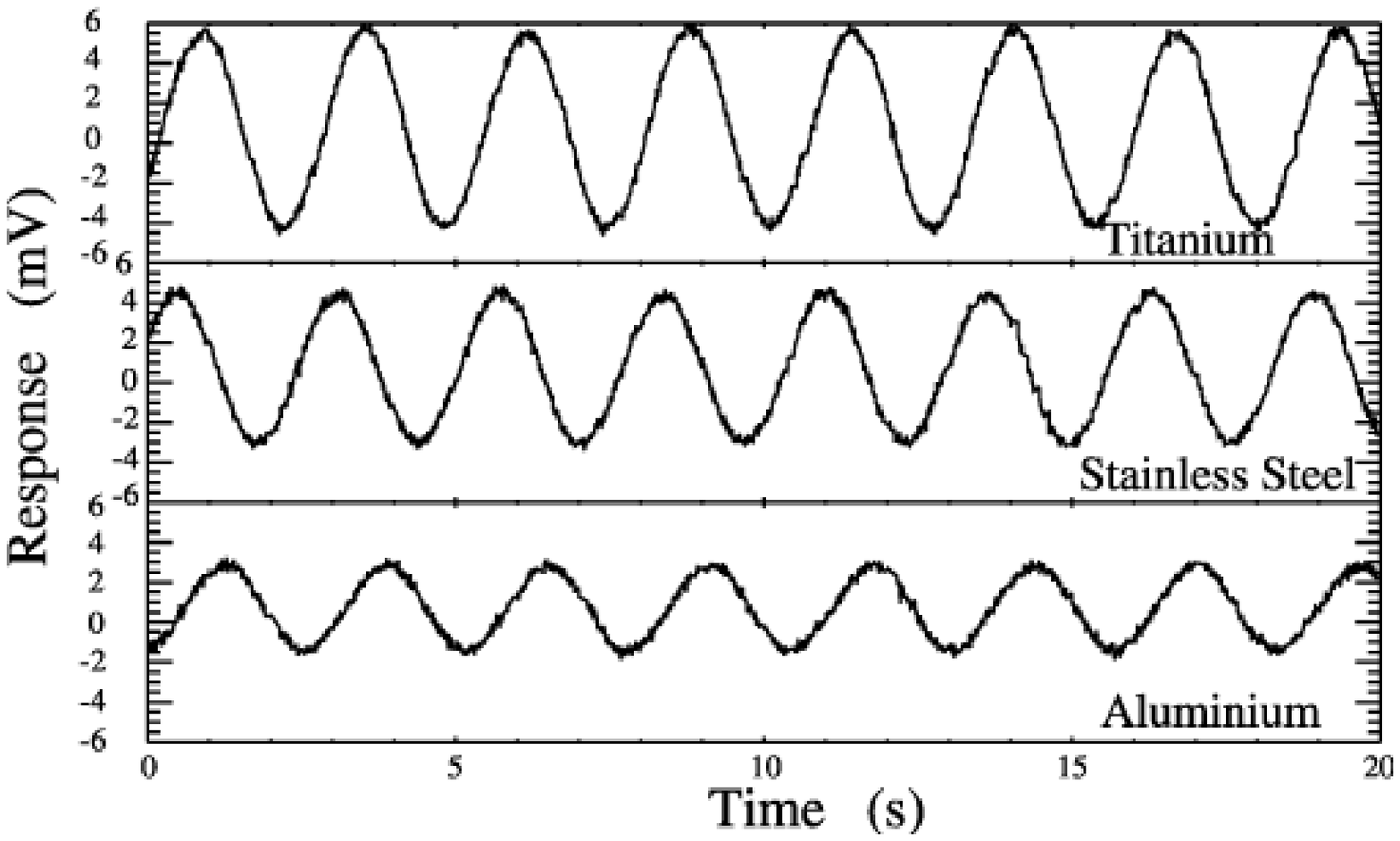}}
\caption{The ``miniplate'' calibration system used for CAPMAP. \textbf{(a)} Radiation
from a cryogenic load reflects off a metal plate into the
polarimeter. The figure is a cut through the plane of incidence. Upon
reflection and emission from the plate, the radiation acquires a calculable
polarized component. Its orientation is parallel to the plane of incidence and its magnitude depends on the composition of the
plate and the temperature difference between the plate and the load. By
rotating the plate about a vertical axis, the orientation of the
polarized signal is modulated. A box of microwave absorber surrounding the
system (not shown) prevents stray radiation from
contaminating the measurement.
\textbf{(b)} Example of data from the miniplate calibration system.
The three panels show time series from a single polarimetry channel
with three different metal plates: grade 5 titanium, stainless steel
305, and aluminum 6061. Note that metals with higher resistivity
produce a larger signal. See \S~\ref{radcal} for a more in-depth
description.}
\end{figure}

This kind of calibration method was used in both CAPMAP and PIQUE to provide
an absolute calibration while the instrument was deployed at the
telescope, using a large metal plate that nutated over a small
range of angles (see \S~\ref{polresp}), with the atmospheric emission
observed at constant elevation providing the incoming signal. After the
first season of CAPMAP, a variation on this technique was developed which
could be implemented in the lab. This ``miniplate'' system
consists of a small metal plate that
reflects radiation from a liquid nitrogen temperature load into the
polarimeter (see Figure~\ref{miniplate}).
The plate and load sit on a mount that rotates about the vertical
axis of the polarimeter. The rotation sinusoidally modulates the polarized
signal, isolating it from any offsets.
Different plates with different resistivities provide
consistency checks on the calibration.

Figure~\ref{minidat} shows data taken using three different plates
with different resistivities (grade 5 titanium, stainless steel 305, and
aluminum 6061, with resistivities of 176, 72, and 4 $\mu\Omega\cdot$cm respectively).
These data show that the signal size increases with increasing resistivity as expected.
Care must be taken in the design of the system to minimize spurious
polarized signals from the room or from the edges of the mount and plate; having data from multiple plates allows any residual contamination from
these sources to be identified and removed. The resulting responsivity 
estimates are consistent with data from the operational telescope (see
\S~\ref{polresp}).

\paragraph{Limited Dynamic Range of Multipliers}
The responsivity measurements not only calibrate the
polarimeter, but also insure that it is
operating within the limited dynamic range of the multipliers. The double
balanced mixers used as multipliers in PIQUE and CAPMAP, like all analog
multipliers, are composed of diodes \citep{Pozar:1998} that operate linearly only over a limited range of input powers.
If the input power is too high, the multiplier output is compressed,
and if it is too low, the Johnson noise of the multiplier contributes
significantly to the output rms fluctuations.

\begin{figure}
\subfigure[]{\label{gainsatA}\includegraphics[width=0.5\textwidth]{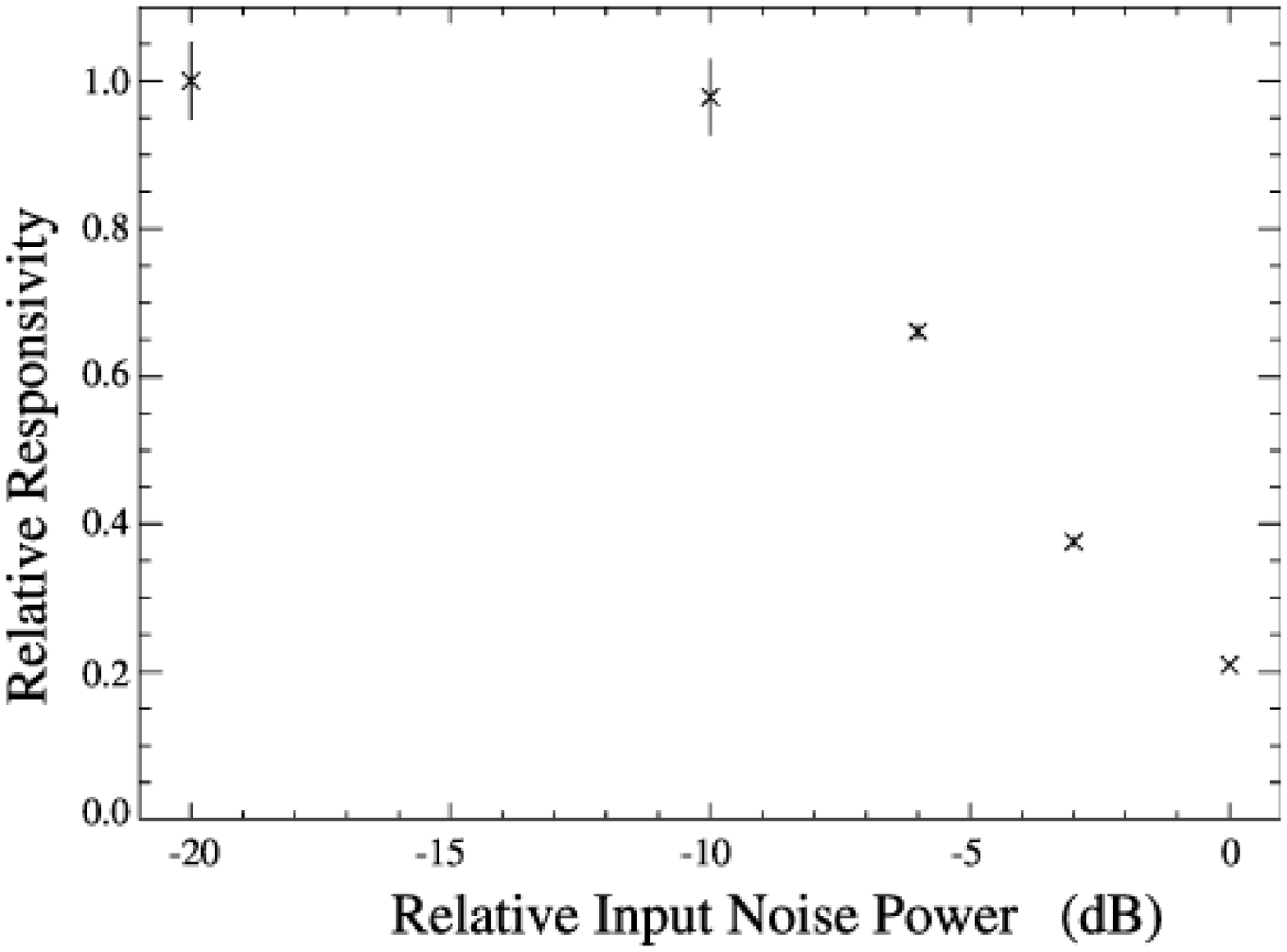}}
\subfigure[]{\label{gainsatB}\includegraphics[width=0.5\textwidth]{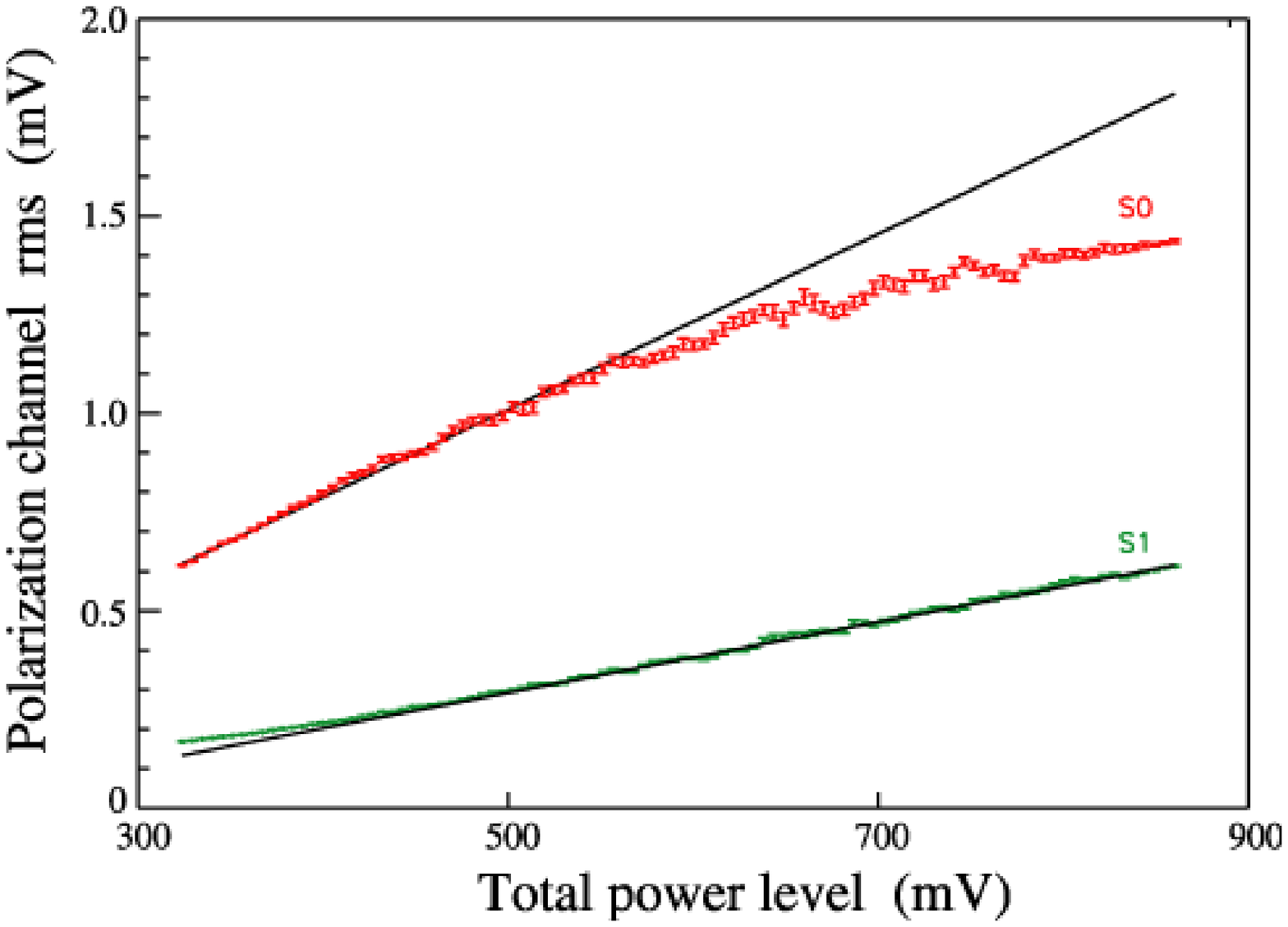}}
\caption{Gain compression of the multipliers with increased total power
loading. \textbf{(a)} Data showing that the relative responsivity (the multiplier's response coefficient)
decreases monotonically for large total power input. The errors for the large input powers are smaller than
the size of the point. Data come from PIQUE lab tests. \textbf{(b)} The rms of the multiplier output versus a
variable total power input. The straight line is fit
to the lower half of the data for the top curve (S0), and to the upper
half of the data for the bottom curve (S1). S0 shows
saturation while S1 shows the onset of Johnson noise domination. Data come from CAPMAP lab
tests. Refer to \S~\ref{radcal} for more details.}
\end{figure}

Compression at high input power levels occurs due to the diodes' nonlinear
I-V curve. The nonlinear behavior of diodes allows them to operate as
multipliers to first order in the input fields. However, if the input
power is too high, additional nonlinear terms couple the responsivity of
the multiplier to the total input power on the polarimeter (see
Figure~\ref{gainsatA}). In this regime, the responsivity of the
instrument to polarized signals decreases monotonically with increasing
system temperature. Only if the total power reaching the multiplier is
kept below roughly 10 $\mu$W will the polarimetry channel responsivity be
independent of loading.

By contrast, if the total power reaching the multiplier is too low, the
intrinsic Johnson noise of the multiplier will contribute
significantly to the rms fluctuations at the output and decrease
the sensitivity of the polarimeter. The Johnson noise of the multiplier
does not significantly affect the sensitivity as long as the total power at
the multiplier is greater than about $1~\mu$W.
The range of gains where the Johnson noise is not significant and the
responsivity of the multiplier is not dependent on the total power is
therefore fairly
narrow (see
Figure~\ref{gainsatB}), and the total power reaching the multipliers must be carefully
adjusted with drop-in attenuators prior to fielding the instrument.

\section{OPTICS}
\label{optics}

Both PIQUE and CAPMAP use off-axis telescopes, which have no central
blockage and little scattered radiation from the support structures.
However, off-axis optical elements also produce a
nonzero polarized signal oriented along the plane of symmetry. The polarimeters in both
experiments are thus oriented so that they are largely insensitive to this
sort of signal---they measure $U$ instead of $Q$ in the telescope's
coordinate system.

The optics of the two experiments are very different. PIQUE used a 1-meter mirror
that was easily installed in a temporary rooftop observatory and was
small enough to be contained in a full-sized set of ground screens. CAPMAP uses
a 7-meter telescope, which is too large to be fully enclosed in
ground screens, but which can provide higher angular resolution and a larger
focal plane. PIQUE and CAPMAP are discussed separately below.

\subsection{PIQUE Optics}

\begin{figure}
\subfigure[]{\includegraphics[width=.45\textwidth]{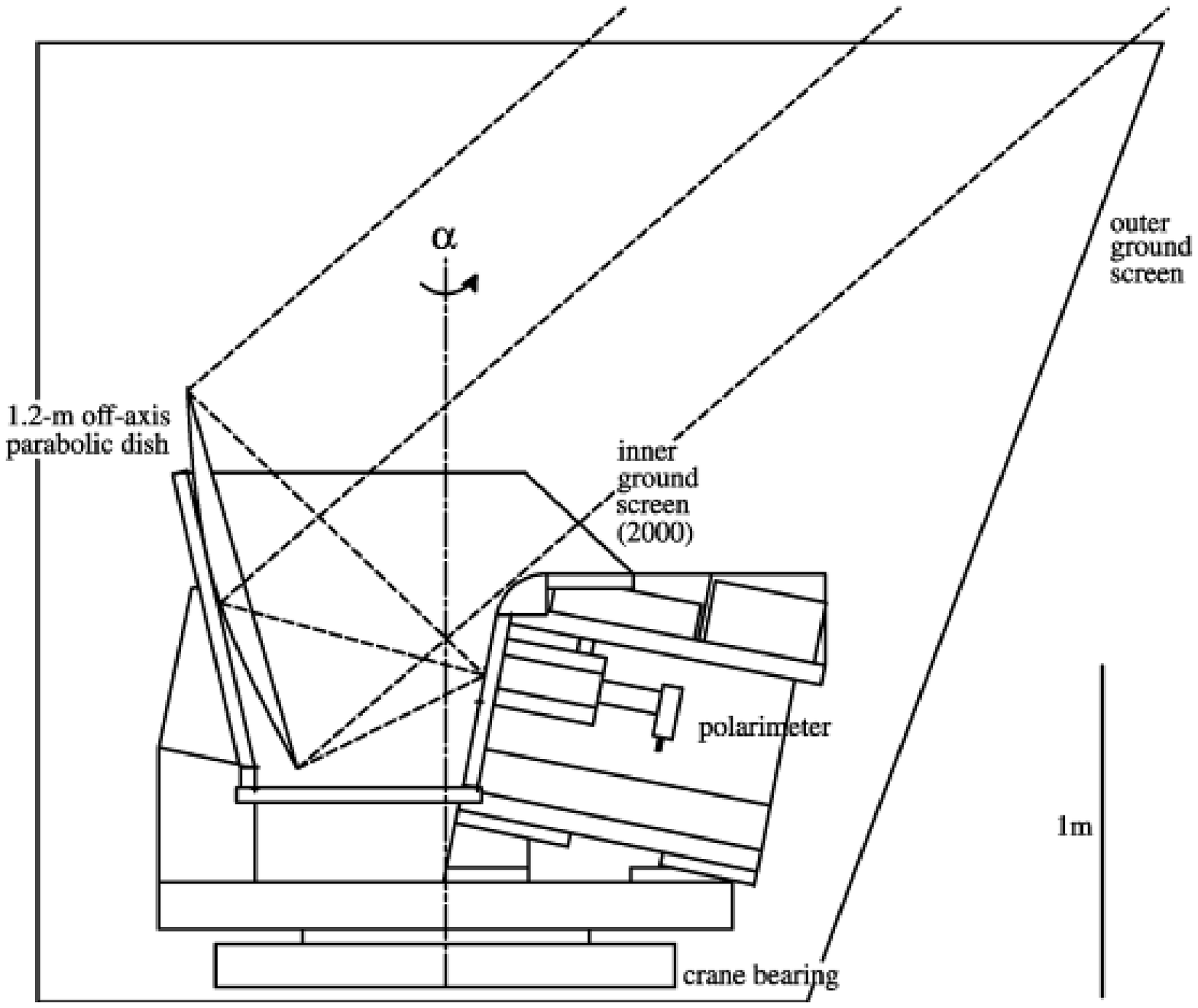}}
\subfigure[]{\includegraphics[width=.48\textwidth]{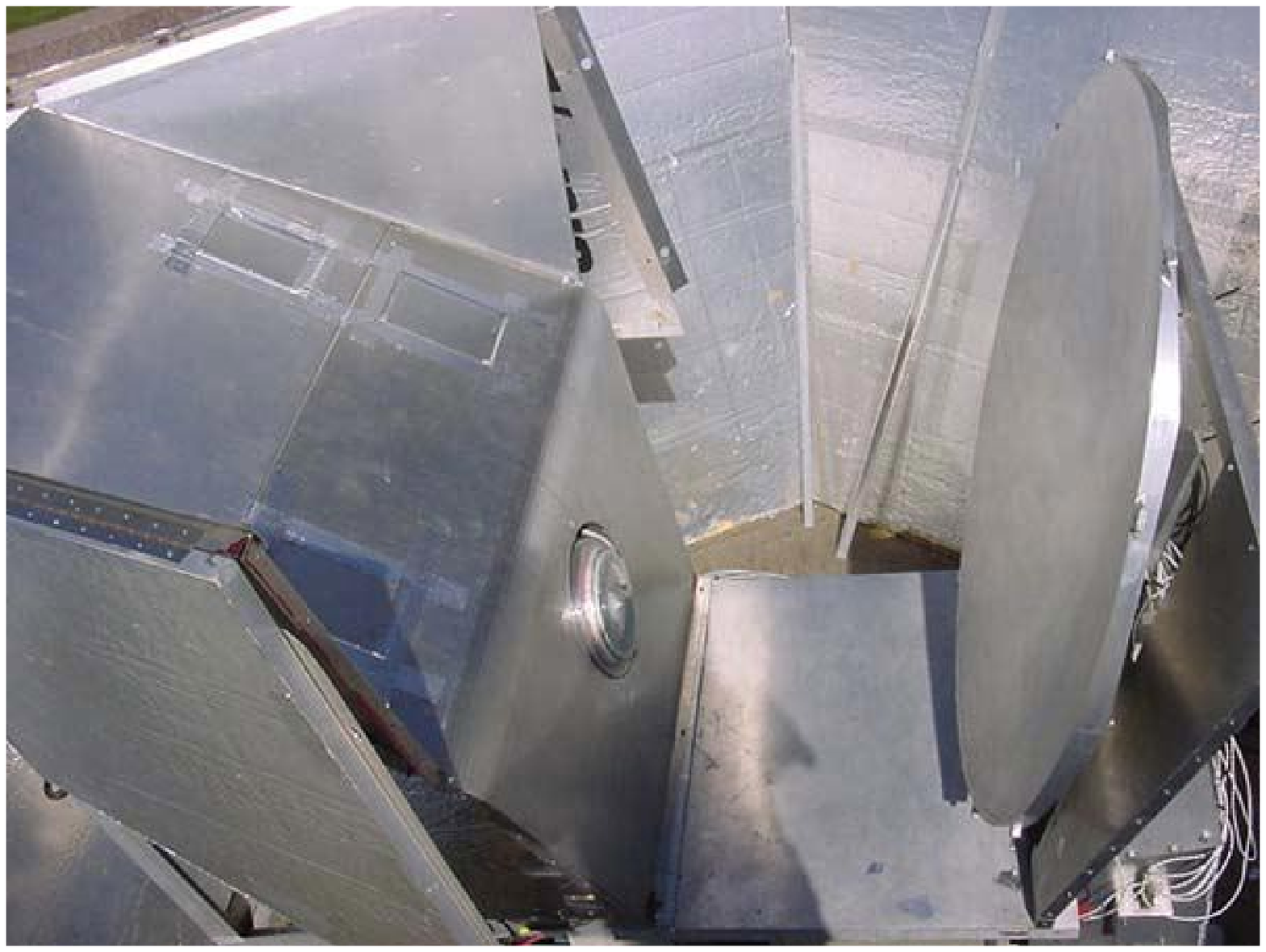}}
\caption{\label{piqueopt}PIQUE optics. \textbf{(a)} A corrugated feed
horn on the polarimeter views a 1.2 m off-axis parabolic mirror. A crane
bearing in the base of the telescope allows it to slew in azimuth, $\alpha$, at a
fixed elevation. Two nested ground screens block radiation from local sources.
The outer ground screen is fixed to the ground while the inner ground
screen rotates with the base. The inner ground screen here was used during
the first observing season. During the second observing season this ground
screen was made twice as tall. \textbf{(b)} Photo of the PIQUE
telescope taken from the top of the outer ground screen. The side
panels of the inner ground screen have been removed for the photo.}
\end{figure}

The PIQUE polarimeter viewed the sky through a corrugated feed
horn and an off-axis parabolic mirror. The entire telescope was
mounted on a crane bearing so the antenna could scan in azimuth at a
fixed elevation of about $41^{\circ}$ (see Figure~\ref{piqueopt}).

\subsubsection{Optical Components}

The parabolic mirror was originally built for the Saskatoon
experiment and is described in detail in \citet{Wollack:1997}. It has a focal length
of 75~cm and an effective diameter of 122~cm, and the optical axis is offset
78~cm from the center of the mirror.
The rms roughness of the surface is better than 50 $\micron$.

\begin{figure}
\plotone{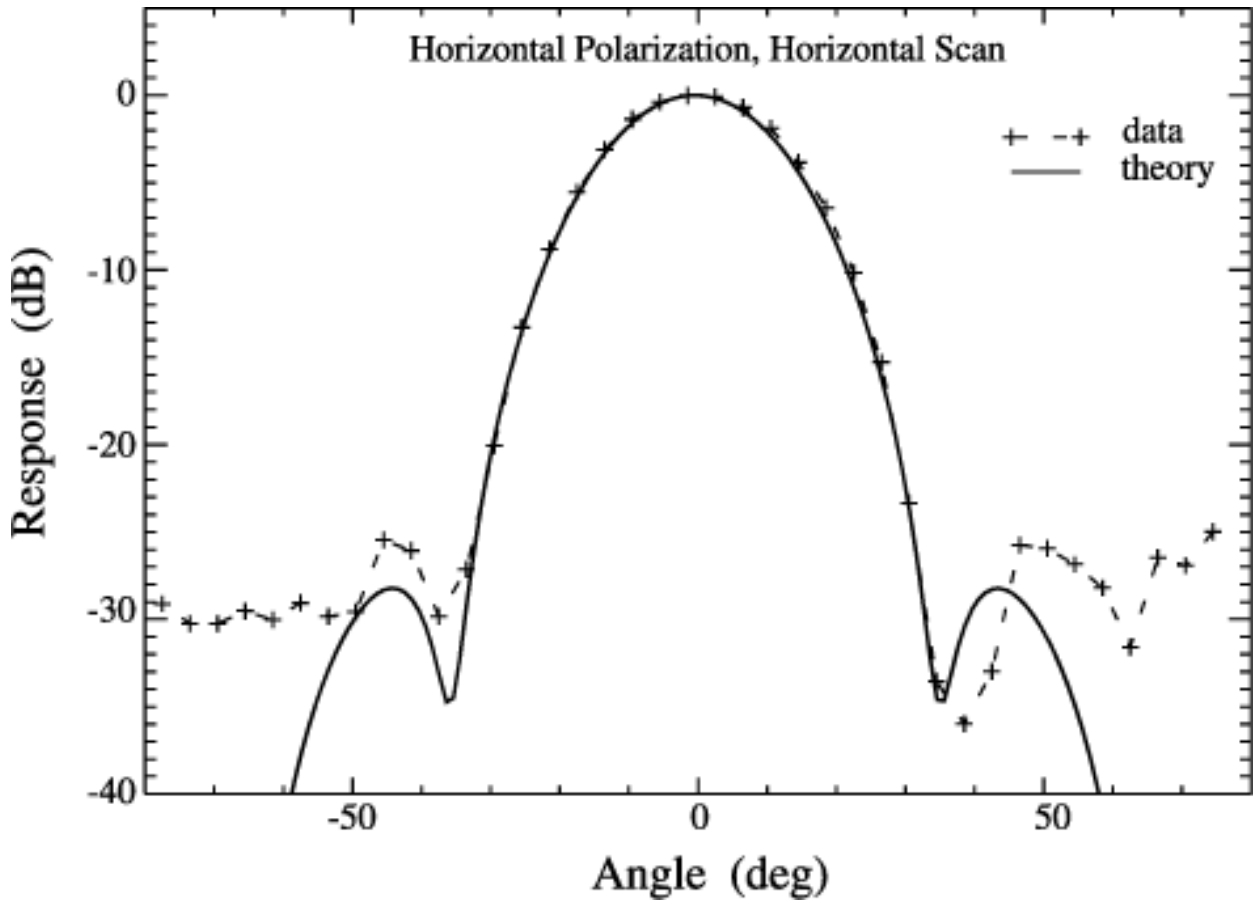}
\caption{The predicted (solid) and measured (dashed) beam patterns of the
PIQUE W-band horn. The $-30$~dB level is the noise floor of the
experimental apparatus used here.}
\label{fig:beampiq}
\end{figure}

The W-band horn was specially made for PIQUE.
This horn was designed using the physical optics program
CCORHRN \citep{James:81} to illuminate the mirror with an edge taper of $-25$~dB.
This criterion determined the length and aperture diameter of the horn,
given in Table~\ref{tbl:optical_param}. To achieve acceptable
polarization fidelity, the horn was corrugated with grooves whose depth
and width were the same as for the WMAP feeds \citep{Barnes:2002}, and the
throat section had progressively deeper and narrower grooves following
the prescription in \citet{Zhang:1993}. The horn was made from electroformed copper
plated in gold. Its measured beam size was $25^{\circ}$ with side-lobe
levels below $-25$~dB, closely matching the theoretical predictions (see
Figure~\ref{fig:beampiq}).

The expected telescope FWHM was $0.25^{\circ}$ at 90 GHz
and $0.50^{\circ}$ at 40 GHz. The exact placement
of the polarimeters was
determined using observations of a near-field source of narrow-band
radiation. The response of the telescope as a function of elevation and
azimuth uniquely determines the location of the polarimeter relative to the
mirror, and by comparing the observed data to the predictions from the
physical optics program DADRA \citep{Imbriale:1991, Imbriale:2003},
it was possible to locate the horn to within 3~mm of the focal point. Observations of Jupiter,
discussed in \S~\ref{tpjup}, confirm that
the telescope was aligned properly.

\subsubsection{Ground Screens}

The PIQUE telescope was located on the roof of Jadwin Hall in
Princeton, New Jersey\footnote{latitude: $40^{\circ} 20\arcmin 45\arcsec$~N, longitude:
$74^{\circ} 39\arcmin 00\arcsec$~W, elevation: 100~m}. Two nested ground screens were
installed around the telescope to prevent stray radiation generated by surrounding buildings from
reaching the telescope (see Figure~\ref{piqueopt}). The outer
ground screen was fixed to the roof and
geometrically blocked all radiation from external sources. It
had the shape of an inverted, truncated hexagonal pyramid with a 3.3~m
footprint, a height of 3.3~m, a vertical southern wall, and
five other walls sloping outward at $20^{\circ}$. The interior walls of this
ground screen were lined with aluminum-covered styrofoam\footnote{Homasote (Ultra/R): \texttt{http://www.homasote.com}},
and each panel was hinged at its mid-height so the
entire structure could be covered in case of bad weather.

The inner ground screen, which was made of aluminum, was attached to the telescope base and
rotated with the telescope. In its initial configuration, it
blocked all radiation emanating from the edge of the outer ground screen and above;
radiation from the outer ground screen itself could reach the polarimeter, producing an
azimuth-dependent polarized offset of a few hundred $\mu$K (see
\S~\ref{offsets}). To reduce these offsets, the inner ground
screen was enlarged for the second observing season to block the
outer ground screen as well; the results of this improvement are discussed in \S~\ref{offsets}.

\subsection{CAPMAP Optics}
\label{capopt}

\begin{figure}
\subfigure[]{\includegraphics[width=.48\textwidth]{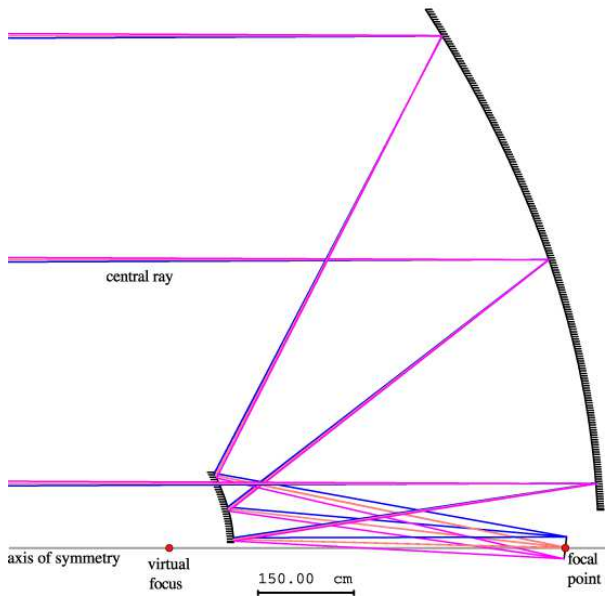}}
\subfigure[]{\includegraphics[width=.45\textwidth]{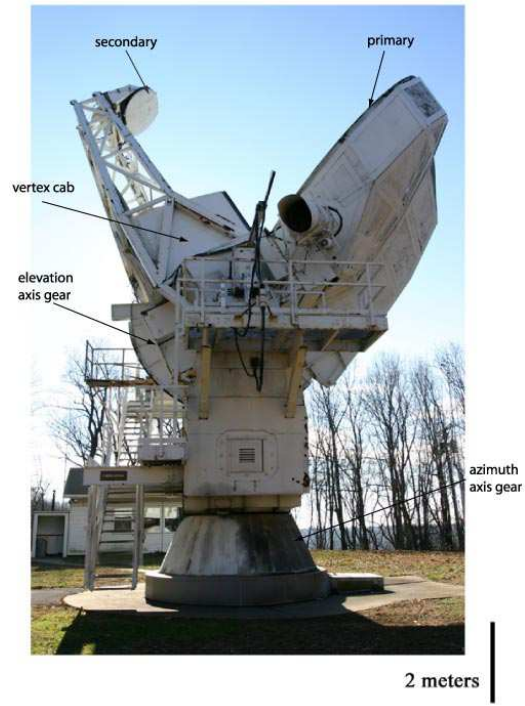}}
\caption{\label{fig:tele_picture}CAPMAP optics. \textbf{(a)} Side view scale drawing of the 7-meter
 telescope pointing at an elevation of $0^{\circ}$. Three bundles of rays are drawn from the Cassegrain focus to the
 secondary with an illumination half angle of $5^{\circ}$. \textbf{(b)} Photograph of the off-axis
 Cassegrain 7-meter telescope. The radiation is focused onto the feed horns in the focal plane. The receivers are enclosed in the vertex cabin around the focal point.}
\end{figure}

CAPMAP uses the Crawford Hill 7-meter
antenna located in Holmdel, NJ\footnote{latitude: $40^{\circ} 23\arcmin 31\arcsec$~N,
longitude: $74^{\circ} 11\arcmin 10\arcsec$~W, elevation: 119~m}. A comprehensive description of the
antenna is given in \citet{Chu:1978}, and
its optical parameters are summarized in Table~\ref{tbl:optical_param}. It
is an off-axis Cassegrain telescope with a 7-meter primary and a 1.2~m $\times$ 1.8~m oval secondary, oversized in the horizontal direction (see Figure~\ref{fig:tele_picture}). The primary reflector is made of 27
aluminum panels arranged in 4 concentric rings. The surface accuracy
of each panel is $\sim 50$~$\micron$, and the surface error of the overall
primary reflector is 100 $\micron$ rms. The telescope is mounted
on a standard alt-az platform with two DC motors on each axis, allowing
a maximum slew speed of 2~deg$\cdot$s$^{-1}$ in azimuth and 1~deg$\cdot$s$^{-1}$ in elevation.
The telescope features a remarkably large and high-quality focal plane with Strehl ratios above 0.97 up to 0.5 m from
the focal point, making it possible to field an array of receivers
while maintaining excellent beam quality.
The telescope also has a cross-polarization response in the main beam below
$-40$~dB \citep{Chu:1978}.

The large 7-meter primary mirror produces a beam narrow enough to resolve the
anisotropy of the CMB polarization, but it also precludes the feasible construction of a PIQUE-style ground screen.
We therefore sacrifice resolution and aggressively
under-illuminate the primary and secondary at approximately $-35$~dB and
$-58$~dB to reduce spillover and produce a $0.065^{\circ}$ (4') FWHM beam at W-band
and a $0.10^{\circ}$ (6') beam at Q-band. This requires a narrow,
$2.3^{\circ}$ beam from the feed system with minimal
sensitivity outside the edge of the secondary ($5^{\circ}$ half angle).
The feed system must also be compact for practical cryogenic cooling.

\subsubsection{Feed System}
\label{sec:feedsystem}
\begin{figure}
\plottwo{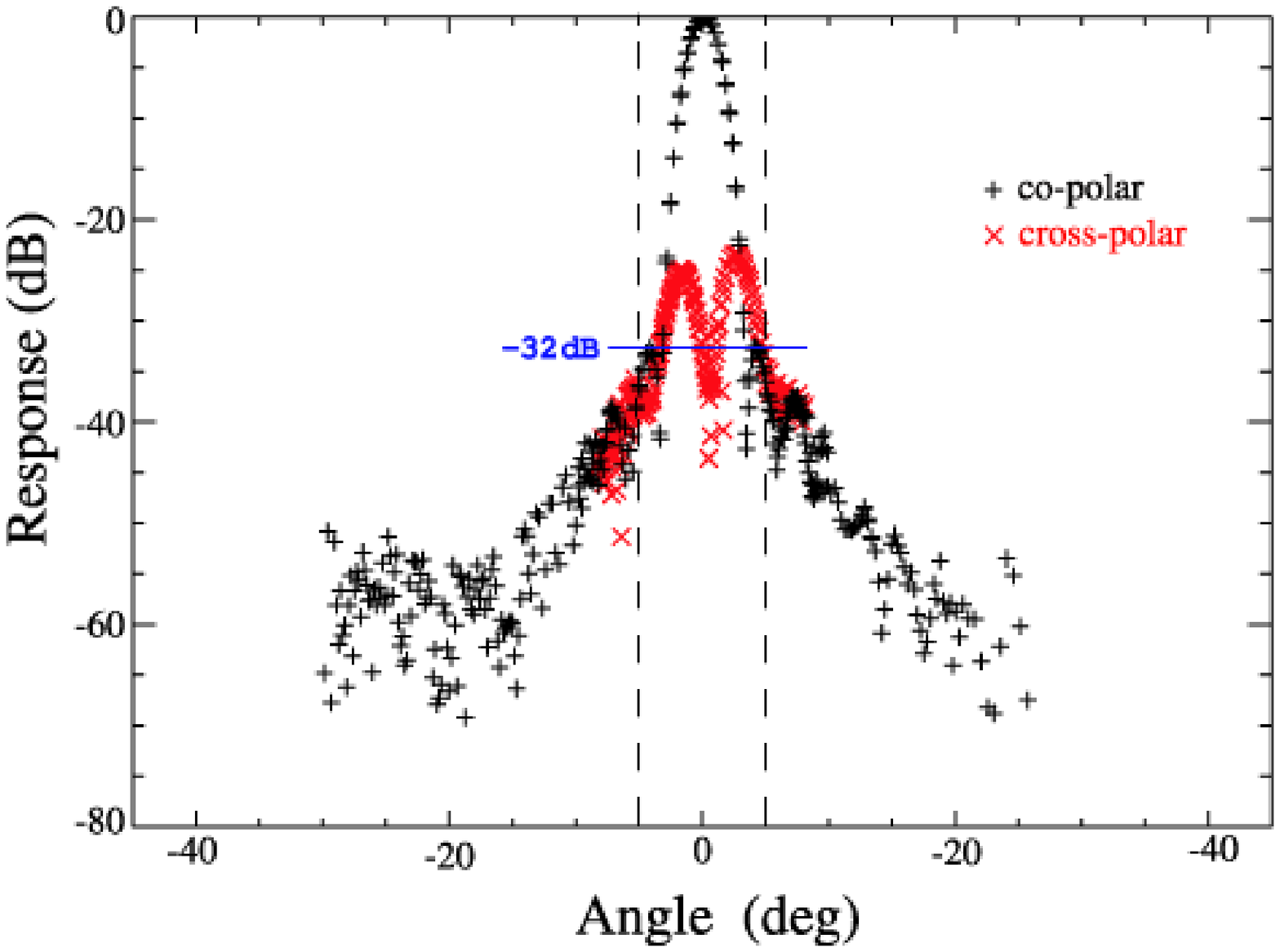}{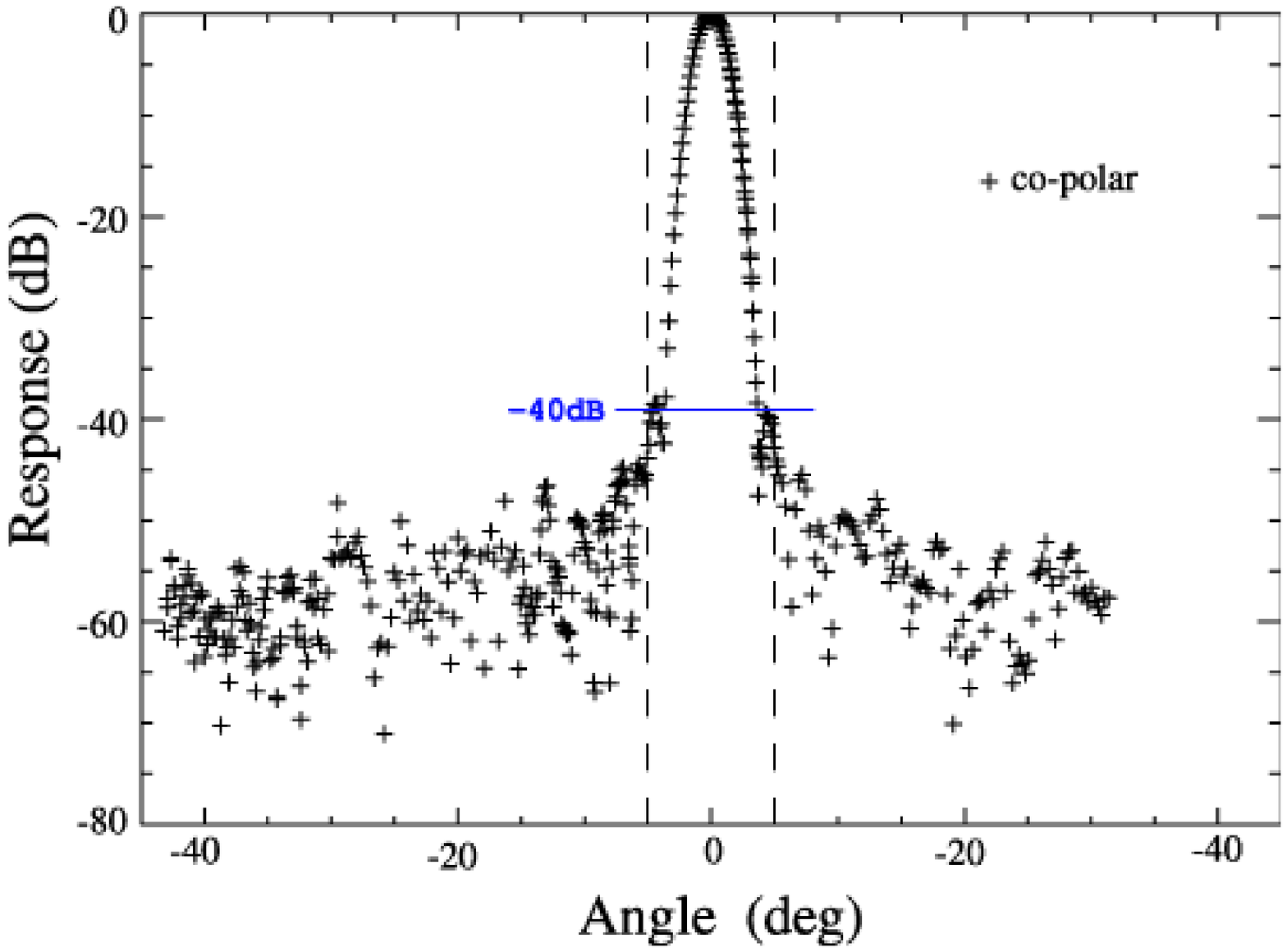}
\caption{\label{fig:optics_beam_maps} Co- and cross-polar beam maps of CAPMAP feed systems during the 2003 (\textbf{left}) and 2004 (\textbf{right}) observing seasons. The measurements were made from the testing range on the roof of Jadwin Hall, where beams can be mapped down to the $\sim -50$~dB noise level. Co-polar maps are shown with plus symbols and cross-polar maps with crosses. The vertical dashed lines at $\pm 5^{\circ}$ show the
illumination at the edge of the secondary mirror. During the first season, the edge taper on the primary was $-32$~dB with a cross-polar maximum at $-25$~dB. The lenses were redesigned for the second season to achieve an edge-taper of $-40$~dB and a cross-polar level $<-40$~dB. (The cross-polar level was too small to be determined reliably from this measurement; it has been inferred from the quadrupole leakage as explained in \S~\ref{loccon}.)}
\end{figure}

To satisfy these criteria, the CAPMAP feed system consists of a compact
profiled horn coupled to a lens.
The design of the CAPMAP horn is similar to that used in PIQUE,
except that the profile is optimized
to reduce the side-lobes from the horn and to improve the Gaussianity of the beam. The CAPMAP
optical system is detailed in \citet{McMahon:2005}. The horn produces a $14.5^{\circ}$ beam that deviates from a Gaussian at the $-35$~dB level.
The beam propagates through free space to a high-density polyethylene (HDPE) lens which
re-converges the beam. The focal length and diameter of the
lens were chosen to minimize the amplitude of the side-lobes
while maintaining a $2.3^{\circ}$ beam. This process produces a first
side-lobe at $-40$~dB below the peak of the main beam, which is Gaussian down to that level (see 
Figure~\ref{fig:optics_beam_maps}).

For the first observing season, the lenses had spherical rear and elliptical
front surfaces. In such a lens, refraction only takes place at
the front surface, where the angle of incidence
is as high as $60^{\circ}$. For normal incidence along $\hat{z}$, the transmission coefficients of the two
orthogonal field components $E_x$ and $E_y$ are the same, but the two coefficients deviate as the angle of incidence increases, since boundary conditions differ for field components perpendicular and parallel to the interface. The difference between the two transmission coefficients was increased in CAPMAP 2003 by the lens anti-reflection (AR) coating, which consisted of concentric near-quarter-wave grooves on the lens surfaces.  Since these grooves are not locally symmetric under $90^{\circ}$ rotations, they have different refractive indices for the two orthogonal polarizations. These effects produce cross-polarization, in which polarized sources viewed off the beam center couple to the orthogonal polarization inside
the horn throat. This causes unpolarized point sources viewed off the beam center to produce a spurious polarized response with a quadrupolar symmetry; the implications for the performance of the experiment are discussed in \S~\ref{loccon}.

For the 2003 season, the level of cross-polarization was $-24$~dB. For the 2004 season, the lenses were re-designed to share the
refraction evenly between the front and back surfaces, reducing the maximum angle
of incidence to $30^{\circ}$ on both surfaces. In
addition, the AR coating was changed to a square array of holes, which---being locally invariant under $90^{\circ}$ rotations---has equal transmission coefficients for each linear polarization.
These modifications significantly improved the cross-polarization of the lenses, which was reduced below $-40$~dB (see
\S~\ref{loccon}).
\subsubsection{Focal Plane}

\begin{figure}
\subfigure[]{\label{fig:focal_plane}\includegraphics[width=.45\textwidth]{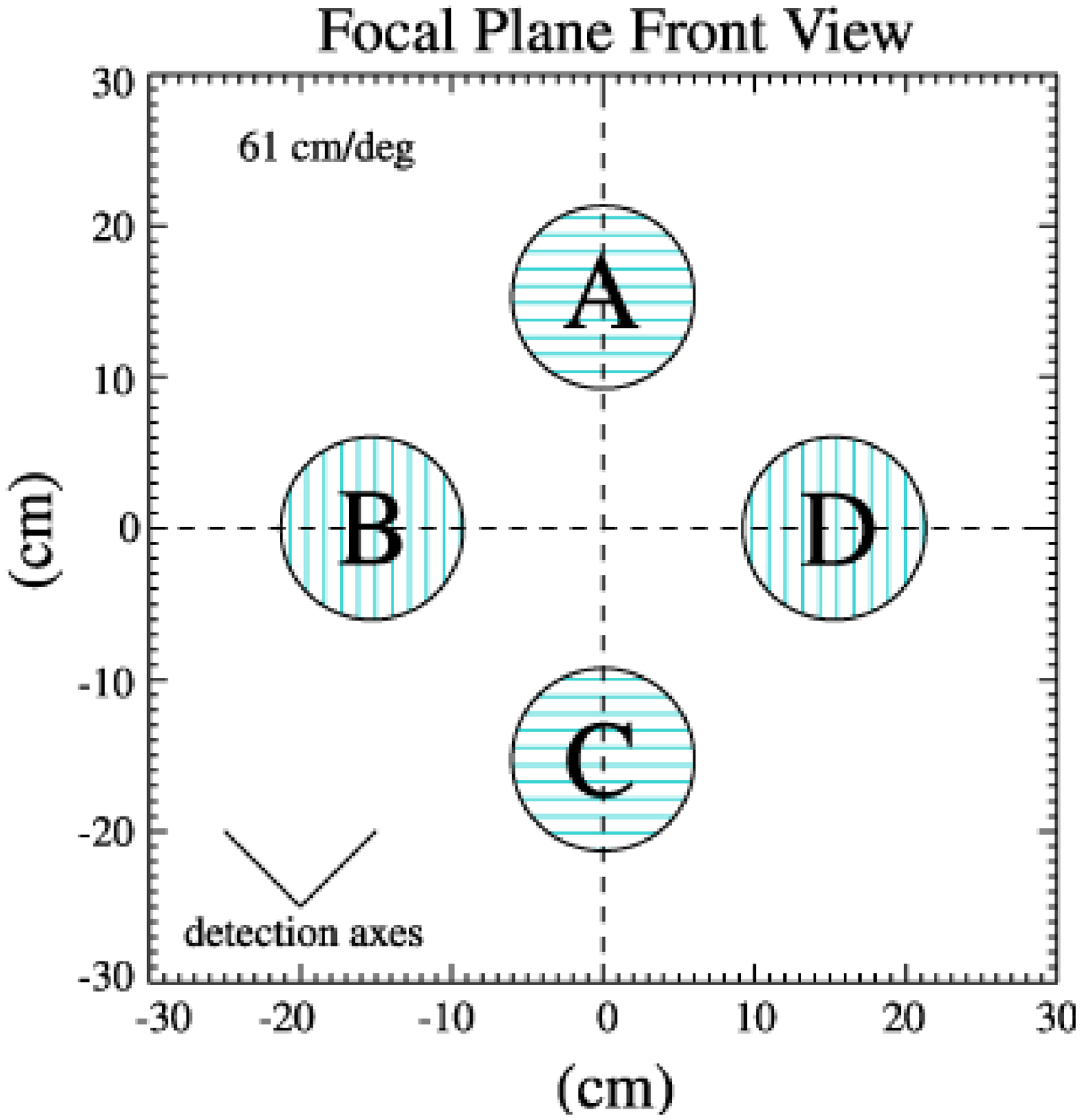}}
\subfigure[]{\label{scan_1}\includegraphics[width=.45\textwidth]{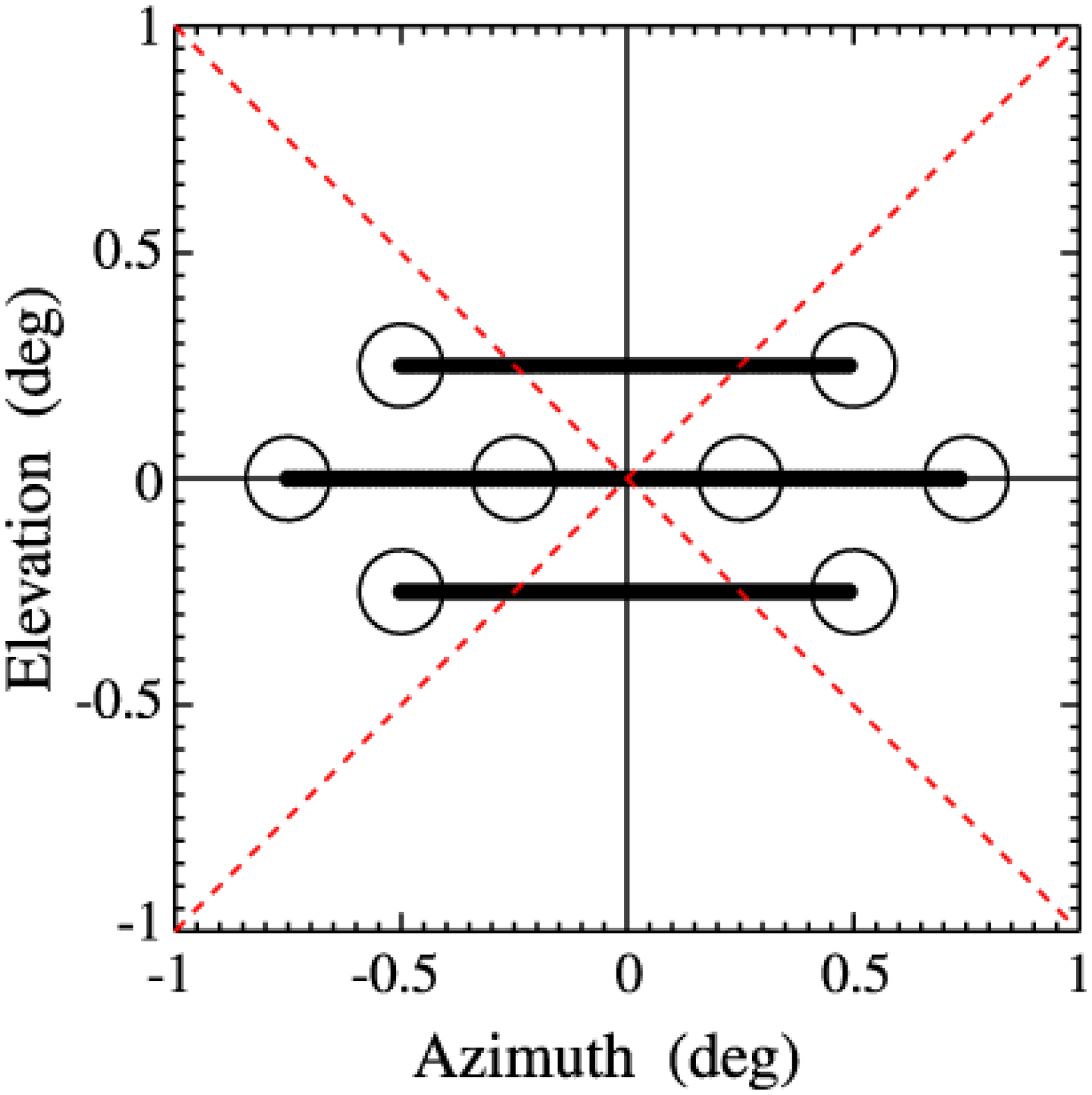}}
\caption{\textbf{(a)} Focal plane layout in the CAPMAP 2003
dewar as seen from the secondary. The cross-hatch indicates the
direction of the E-plane polarization of the main arm of the
OMT. The resulting detection axes are at $45^{\circ}$ from the OMT
axes. Each circle represents the position of the lens. The receivers
are tilted $1.7^{\circ}$ inwards in the focal plane to point towards the
center of the secondary. The position of the beams on the sky with
respect to the center of the array is reversed in both the up-down
and left-right orientation. As seen from the focal point, the
arrangement of the beams on the sky has A on the bottom, B on the
left, C on the top, and D on the right. \textbf{(b)} CAPMAP 2003
azimuth scan pattern on the sky. Each horn is designated by a
circle. The NCP is at the origin. A horn on the thin solid lines
measures U alone, and a horn on the dashed lines measures Q
alone; otherwise each horn measures a combination of Q and U.}
\end{figure}

\begin{figure}
\includegraphics[width=0.8\textwidth]{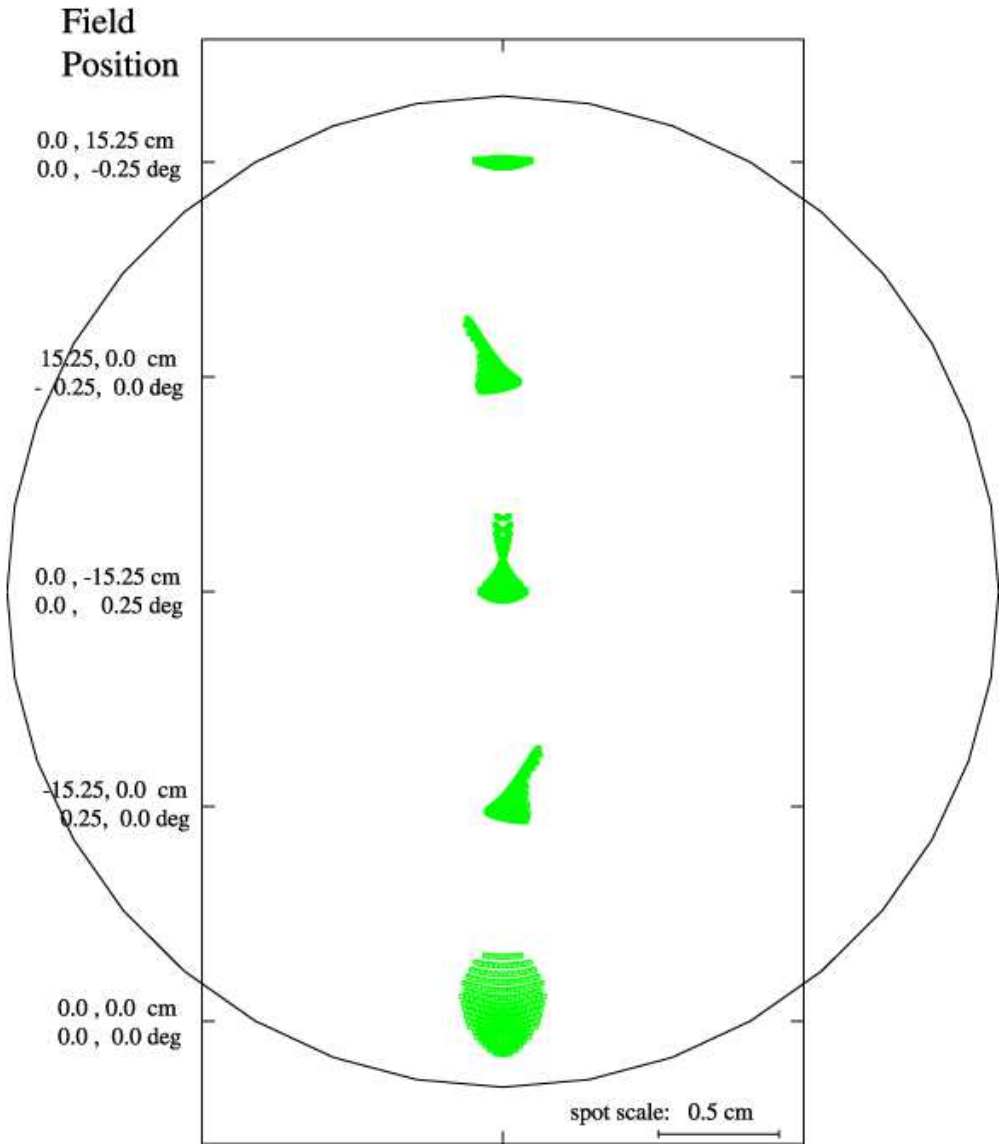}
\caption{Spot diagrams of the Crawford Hill antenna. The spot
diagrams are calculated using the CODE V software. The five spots
displayed are the dispersion of $\sim 300$ rays distributed over
the whole primary and focused to the focal point and the four CAPMAP
2003 horn positions. The large circle is the diffraction beam size
of the fully illuminated telescope for the focal point spot (scaled by the plate scale). 
These data agree with the fact that the beam distortions are very small away
from the focal point. The Strehl ratios for the four beam positions
are better than 0.99.} 
\label{fig:spot_diagram}
\end{figure}

\begin{figure}
\plotone{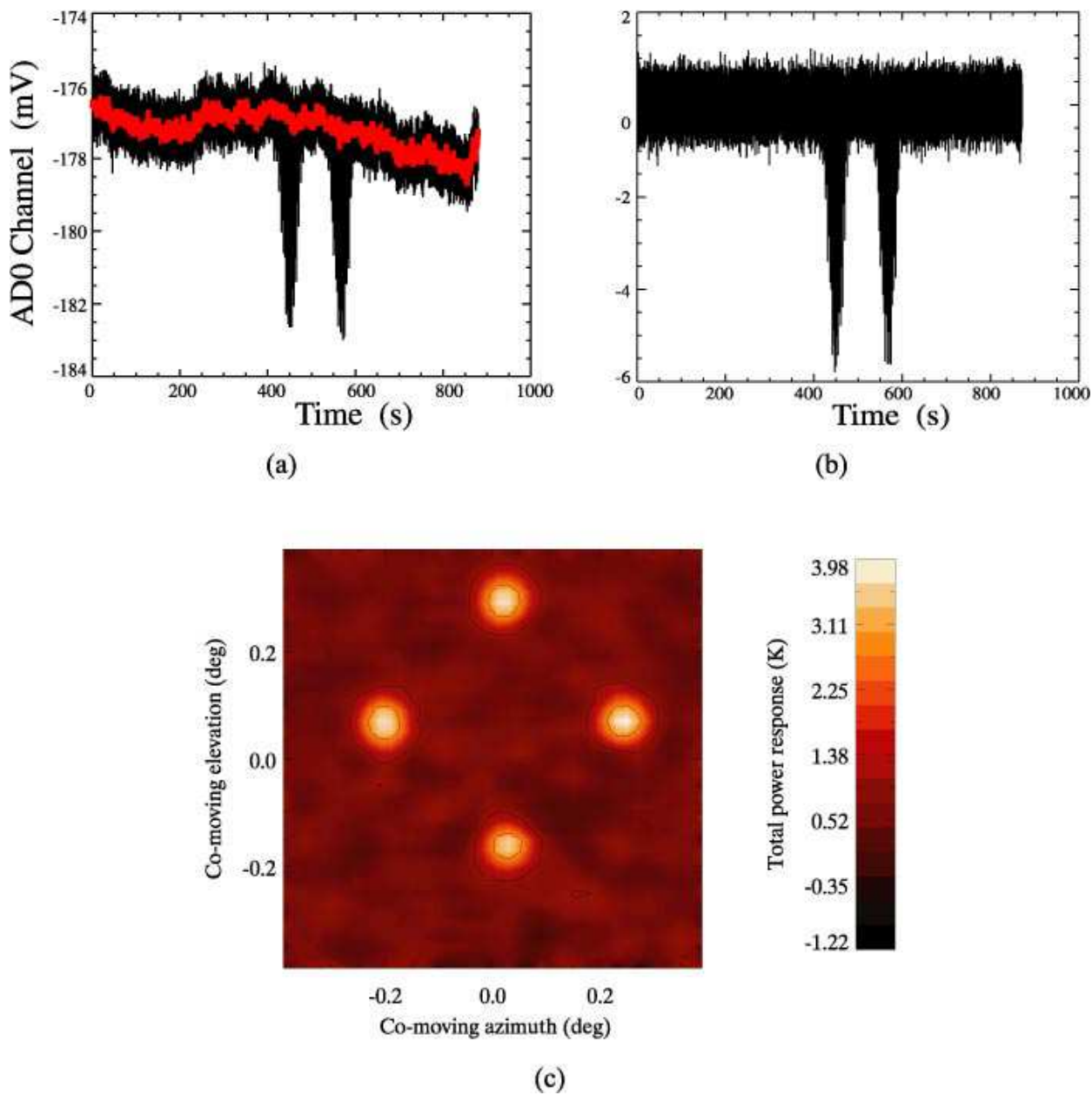}
\caption{\label{fig:baseline}Illustration of the Jupiter and radio sources data processing.
\textbf{(a)} Time series of one total power channel during an observation of Jupiter (black),
with the baseline fit (red) overlayed. Note that each of the
two bumps represents approximately 20 source rasters at
different elevations; the $x$ axis is too large to resolve each
individual Gaussian. \textbf{(b)} The same time stream
with the baseline drift removed. The baseline fit removal is
equivalent to a pre-whitening filter and is only applied to
the total power channels. 
\textbf{(c)} Mosaic map of the CAPMAP 2003 focal plane
from a single 15-minute Jupiter observation taken on 2003
February 12, 
using the time series processing described above. The angular
diameter of Jupiter was 45$\arcsec$ on this date. The black
contours correspond to power levels of 10\% and 50\% of the
peak value.} 
\end{figure}

The CAPMAP 2003 focal plane layout consists of four receivers
located 15.25~cm on either side of the focal point arranged in a
diamond pattern as in Figure~\ref{fig:focal_plane}. Given the 68.9~cm$\cdot$deg$^{-1}$ plate scale, the four horns are spaced $0.25^{\circ}$ equidistant
from the central ray on the sky. Although the receivers are packed
relatively tightly to keep the scan region compact, they are not
packed as tightly as possible to the focal point because the 7-meter
antenna provides a large focal plane with undistorted beams out
to 40~cm away from the focal point. This fact was first simulated with
ray tracing (Figure~\ref{fig:spot_diagram}) and physical optics
software; it was confirmed by beam maps of the receivers on Jupiter,
shown in Figure~\ref{fig:baseline}. In order to minimize the induced
polarization from the telescope mirrors, each receiver is tilted
$1.7^{\circ}$ to point towards the center of the secondary. This geometric
arrangement of the horns also satisfies the azimuth scan
strategy suggested by \citet{Wollack:1997}, which was used for the first observing season.

The azimuth scan swept the telescope in azimuth across the NCP at
constant elevation with an 8 second period. The constant elevation
avoids the $\sec\theta$ total power modulation from the
atmosphere. Figure~\ref{scan_1} shows the path traced out by the four
horns in a single scan across the NCP. The scan amplitude of
$\sim 0.5^{\circ}$ was chosen to optimize the signal-to-noise ratio per pixel given the 
sensitivity of the array. 

Since the Crawford Hill telescope has a large $f/D$ ratio (it is a slow optical system), the
tolerance on the position of the receivers in the focal plane is quite
large. Mechanical alignment methods were therefore
sufficient to position and orient the polarimeters on the telescope for
the first observing season. For the subsequent observing season, the goal for accuracy in the
relative alignment of the receivers was $0.4\arcmin$, or a tenth of a
W-band beam size. This goal corresponds to the requirement that the
dewars be positioned with an accuracy of 4~mm. Observations of
Jupiter were used to verify this alignment.

\section{SITE ATMOSPHERIC CONDITIONS}
\label{atmos}

Both PIQUE and CAPMAP are based in central New Jersey.
Although New Jersey is not considered a prime location for
millimeter wave observation, we have evidence that the atmospheric conditions
are sufficiently good for E-mode polarization measurements at 90 GHz.
Primarily, three factors affect the quality of observations: cloud coverage,
atmospheric temperature, and atmospheric water content.
Table~\ref{tbl:winter_season} lists observed characteristics
for the 2003--2004 winter.

\begin{figure}
\subfigure[]{\includegraphics[width=.495\textwidth]{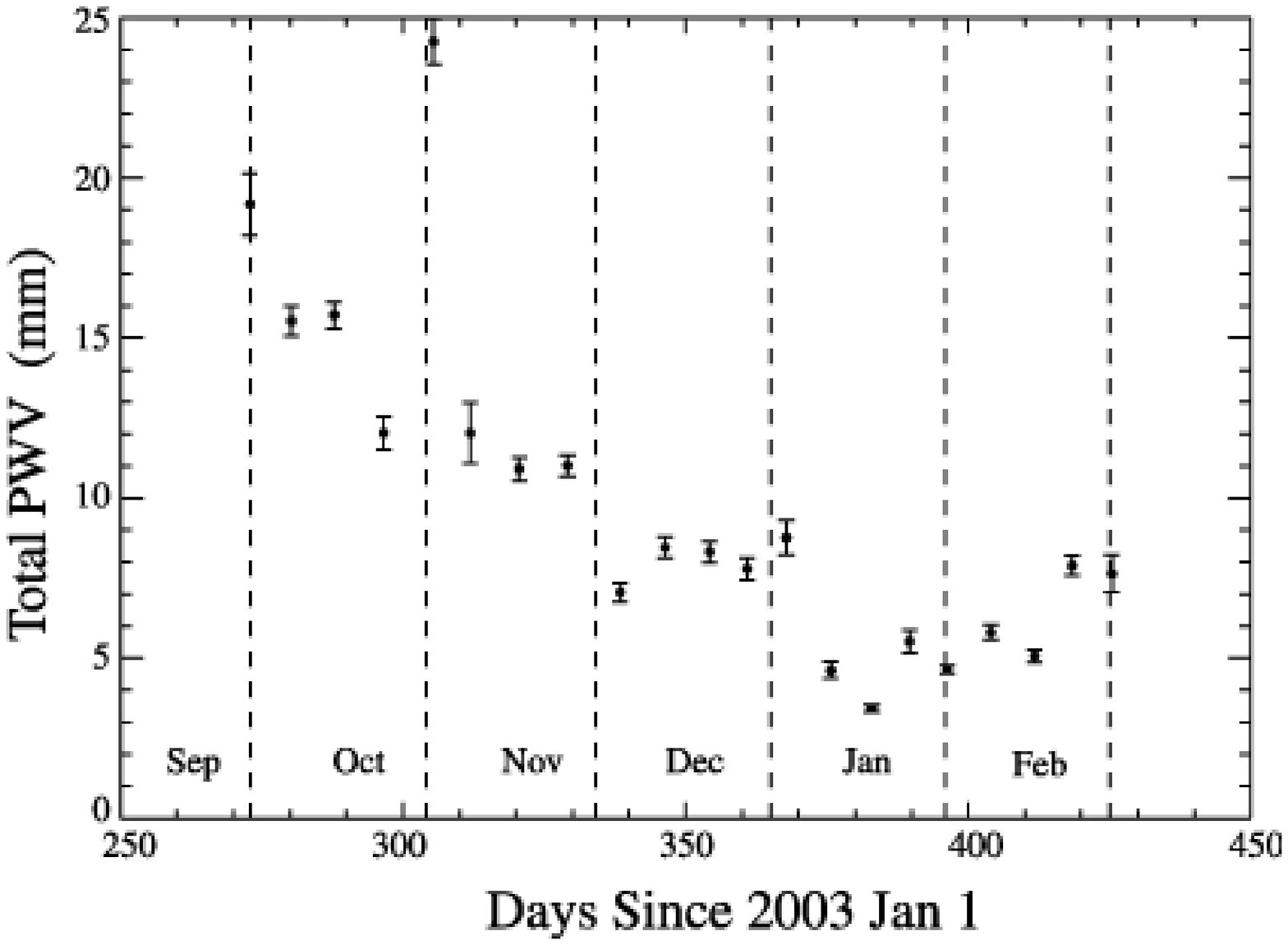}}
\subfigure[]{\includegraphics[width=0.501\textwidth]{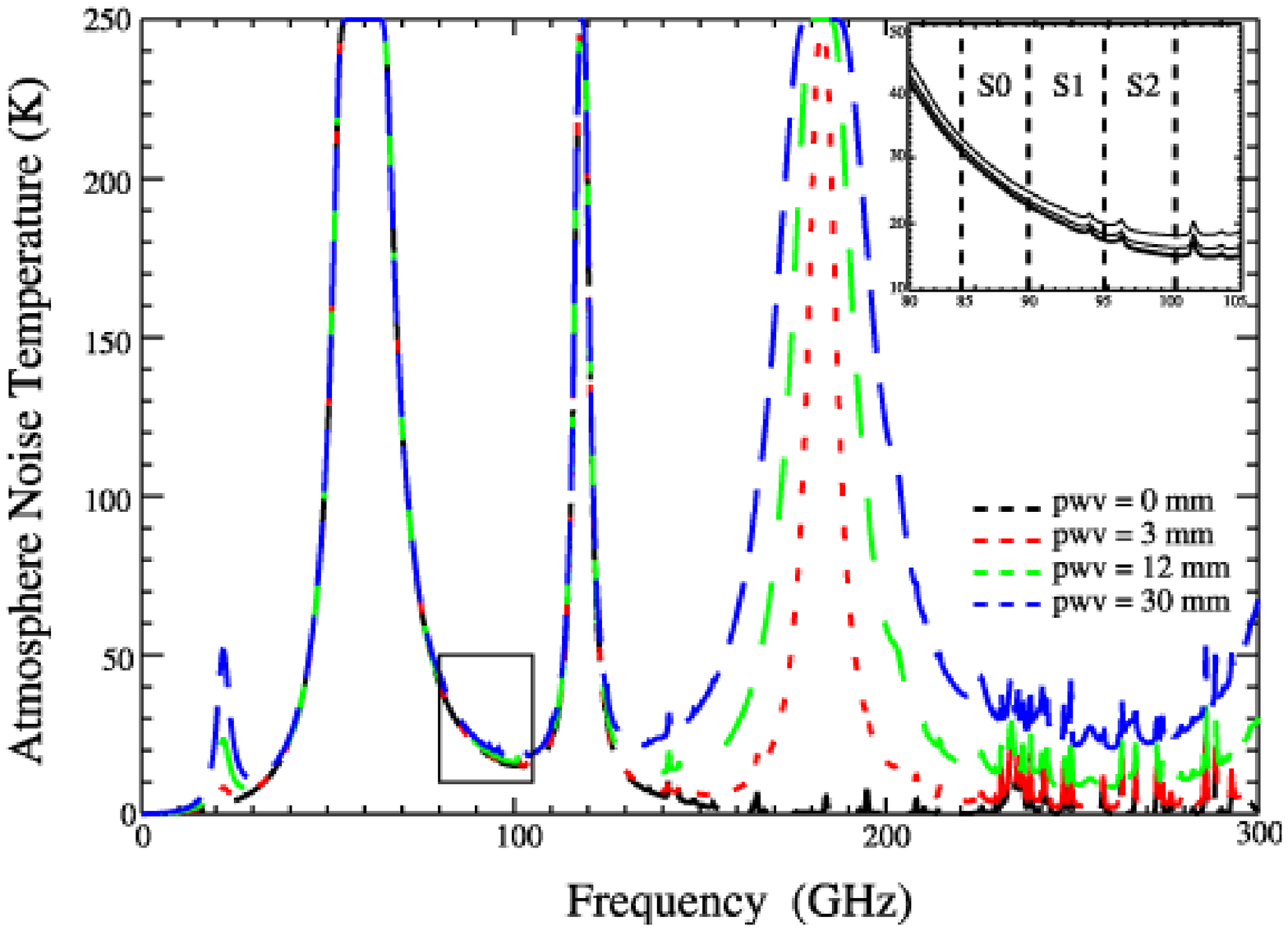}}
\caption{\label{fig:lineshape}Crawford Hill atmospheric environment. \textbf{(a)} Time series of the total PWV during the winter 2004 observing
season, rebinned in 10-day intervals. These data are derived
from the hourly GOES satellite archives. The precipitable water
vapor (PWV) is the average from a 100 km square grid centered on Crawford Hill.
The 50\% quartiles for the months of October through February are 12.8, 12.8, 7.0, 4.0, and 4.8~mm respectively.
\textbf{(b)} Atmospheric zenith emission temperature versus frequency
for different values of PWV. Inset is a zoom on the three CAPMAP
frequency bands.
The three values of PWV (3 mm, 12 mm, and 30 mm) are representative of the best, average, and
worst observing conditions respectively. The line shapes were generated using
E. Grossman AT software~(1989), with the following parameters: full
Lorentzian line profile, elevation~=~100 m, latitude~=~$40^{\circ}$, zenith
angle~=~$0^{\circ}$, physical temperature of the atmosphere~=~250 K. The AT software
produces the atmospheric transmission, $t$, which was converted into a
zenith sky temperature as $T_{atm} = 250 (1-t)$.}
\end{figure}

At 90 GHz, apart from very high and thin cirrus clouds, any cloud
coverage prevents all observations. Table~\ref{tbl:winter_season}
shows that December, January, and February are the clearest months,
with a perfectly clear sky nearly 40\% of the time.
The atmosphere at 90 GHz has a finite opacity,
which is dominated by the wings of the emission lines from molecular
oxygen and water vapor (see Figure~\ref{fig:lineshape}). The
emission from oxygen sets a lower limit on the atmospheric noise
temperature at the site, but the more variable water vapor content
provides a better measure of the data quality.

The precipitable water vapor (PWV)---the total height of liquid
water condensible from a column of atmosphere---is therefore often used as a
standard figure of merit to compare different millimeter wave
observing sites. Figure~\ref{fig:lineshape} shows the time series of the PWV
in Crawford Hill during the 2003--2004 winter.
The three months of December, January, and February consistently have the lowest PWV and
the best data quality. PWV values as
low as 1.6~mm were recorded and lasted for periods of 10 to 24 hours.

\section{ON-SITE PERFORMANCE}
\label{operperf}

The PIQUE and CAPMAP instruments together have taken data during four of the last five winters. The PIQUE telescope first came online
on the roof of Jadwin Hall on 2000 January 1 and operated until 2000 May 8, at which time the
polarimeter was brought to the lab for the summer. The instrument was
installed the next winter on 2000 December 15 and operated until 2001 February 28, at which point the W-band polarimeter was removed, and the Q-band
system took data until 2001 May 10.
The CAPMAP 2003 instrument, consisting of four W-band polarimeters, was
brought to the 7-meter telescope on 2003 January 16. The CMB
observing season lasted from February 18 to April 6. The
instrument remained on the telescope until 2003 June 18 for various
post-season investigations. In the 2003--2004 winter, 9 W-band and
3 Q-band polarimeters operated intermittently at the Crawford Hill telescope, and the full 16-element array was installed for 2004--2005.

Thus far, only data from the PIQUE W-band system and the
first CAPMAP observing season have been fully processed and analyzed. A
breakdown of the time spent on various activities for the various
observing seasons is given in Table~\ref{observing}. Most of the
available time is dedicated to observations of the CMB near the
NCP, which are described more thoroughly in \citet{Hedman:2001, Hedman:2002ck} and \citet{Barkats:2004he}. Additional data
from astronomical and local sources needed to calibrate the operational telescopes are described in the next two sections.

\subsection{Total Power Channel Performance}
\label{tpperf}

The total power channels quantify the receiver noise temperatures and the opacity of
the atmosphere during the CMB observing seasons, so we
calibrate the total power channels using observations of Jupiter and
elevation scans through the atmosphere. These data also provide
estimates of the beam size, pointing solution, and noise temperature of each telescope.

\subsubsection{Jupiter Observations}
\label{tpjup}

Jupiter is a bright, compact, unpolarized source at millimeter
wavelengths \citep{Imke:1999}, so observations of this planet are used to calibrate the
total power channels and to measure the beam size and pointing of the
operational telescope. These observations also provide important
information about the polarization-specific systematic effects discussed
in \S~\ref{polfid}.

PIQUE and CAPMAP have observed Jupiter in slightly different ways. The PIQUE
telescope cannot scan in elevation, so Jupiter observations
consisted of repeated scans in azimuth lasting for a period of about 10
minutes, during which time Jupiter passed through the observed elevation. By
contrast, the
CAPMAP telescope scans back and forth in azimuth over a specified range of
about $1^{\circ}$ and then steps in elevation by about $0.02^{\circ}$.
Repeating this pattern of movements covers a two-dimensional
area centered on the nominal position of Jupiter.

For any of these observations, the data from each total power
channel is processed to produce a two-dimensional beam
map. Figure~\ref{fig:baseline} illustrates the analysis
process. First, any drifts in the time series are removed using a
baseline fit generated
from the data taken when the telescope was pointed at blank sky to the left or right of Jupiter.
After subtraction of this baseline fit (which effectively pre-whitens
the data), the data are binned in coordinates centered on the nominal
position of Jupiter, producing a two-dimensional map of the telescope
response.

Not all observations of Jupiter produced useful maps. Clouds or other
atmospheric phenomena passing through the scan region corrupted
some observations so that it was not possible to extract a clean map of
the signal from Jupiter. These observations are easily identified and
removed, leaving 12 successful observations for each
season of PIQUE and 15 observations for the first season of CAPMAP.

For each successful observation, the map for each channel is fit to a
two-dimensional Gaussian:
\begin{equation}
f(x,y)=A\exp\left\{-\frac{(x-x_0)^2}{2\sigma_x^2}-\frac{(y-y_0)^2}{2\sigma_y^2}\right\},
\label{gaussjup}
\end{equation}
where $x$ and $y$ are the cross- and co-elevation of the
telescope relative to Jupiter, and $A$, $x_0$, $y_0$, $\sigma_x$, and $\sigma_y$ are
fit parameters.

The Gaussian beam widths $\sigma_x$ and $\sigma_y$ for all the
observations with either telescope have a scatter consistent with their
errors determined by the fitting algorithm. There is no evidence that
different total power channels or different polarimeters of the same
band have significantly different beam sizes, so all of these data
are averaged together to obtain the final estimates of the telescope
beam shape (see Table~\ref{optper}). The uncertainties in these estimates
are derived from the scatter of the individual estimates and do not
significantly affect the analysis of the CMB data \citep{Barkats:2004he}. Furthermore,
the measured
beam widths are consistent with the predictions based on physical optics
programs, and there is no significant ellipticity; the evidence indicates that both telescopes were well focused.

The offset parameters $x_0$ and $y_0$ establish the telescope
pointing. For PIQUE the pointing solution was not well constrained, because
Jupiter could only be observed in two positions as it rose or set through
the elevation range visible to the telescope. The measured pointing
offsets at these two locations were not consistent with each other and
indicate some azimuth-dependent offset term that must be extrapolated to the NCP.
This extrapolation increases the uncertainty in the pointing error to approximately $0.1^{\circ}$, which did not
significantly affect our limit on the CMB polarization \citep{Hedman:2001}.
For CAPMAP, Jupiter and other radio sources---Cas A, Tau A, Orion (OMC-1), and Cyg A---were observed 
at a high signal-to-noise ratio at a wide variety of positions.
These data strongly constrain a pointing model and reduce the rms uncertainty in the
telescope pointing near the NCP to less than $30\arcsec$.

Lastly, the parameter $A$ serves to calibrate the total power
channels, given an estimate of the peak
signal from Jupiter 
\begin{equation}
t_{jup}=\tau_{atm}T_{jup}\Omega_{jup}/\Omega_{tel}.
\end{equation}
The effective temperature of Jupiter at 90 GHz $T_{jup}=171$~K is based on WMAP data \citep{Page:2003}.
The solid angle subtended by Jupiter's disk $\Omega_{jup}$ is
computed from ephemeris data. The beam solid angle
$\Omega_{tel}=2\pi\sigma_x\sigma_y$ is calculated from the
beam widths measured above. The atmospheric transmission
coefficient $\tau_{atm}$ is derived directly
from the total power level for the blank sky during the observation.

The responsivity estimates derived from Jupiter observations are
consistent with estimates derived from the Y-factor tests described in
\S~\ref{radopt}. For PIQUE, the two sets of estimates agree to
within 15\%, while for CAPMAP they agree to better than 10\%. The
residual discrepancies between the two methods are statistically
significant---the Jupiter estimates are consistently lower than the
Y-factor estimates---which implies the optical systems have a finite loss
of roughly 6\%. This estimate is consistent with the data from elevation
scans in CAPMAP (see \S~\ref{tpsd}). Even if the discrepancy is due to an unknown
systematic effect in one of the two methods, a 10\%--20\% calibration error
in the total power channels does not significantly affect the analysis
of the CMB data.

\subsubsection{Atmospheric Elevation Scans}
\label{tpsd}

Atmospheric elevation scans are performed by scanning the telescope in
elevation between the horizon and the zenith at a fixed azimuth. These
observations of the atmosphere, which can only be done with the CAPMAP telescope, provide an estimate of the noise
temperature of the entire telescope (including the optics), unlike the
Y-factor tests described previously, which only measure the noise temperature
of the polarimeter itself.

The receiver noise temperature on the telescope ${\widetilde T_{rec}}$ remains constant as the
telescope scans, while the atmospheric noise
temperature varies as $T_z\csc(el)$, where $el$ is the telescope
elevation and $T_z$ is the atmospheric noise temperature at zenith.
The total system temperature, neglecting the CMB contribution, as a function of elevation
is therefore $T_{sys}(el)={\widetilde T_{rec}}+T_z\csc(el)$. The total power channels,
with gains calibrated using the Jupiter observations, provide estimates of $T_{sys}(el)$, which are then fit
to the appropriate functional form to obtain estimates of ${\widetilde T_{rec}}$ and
$T_z$.

Ideally, the estimates of ${\widetilde T_{rec}}$ and $T_z$ from different scans
should be uncorrelated. However, during the first observing season there
was a significant correlation between the estimated receiver noise
and $T_z$. This correlation occurred because side-lobes
behind the primary mirror (discussed in
\S~\ref{offsets}) alter the shape of the $T_{sys}(el)$
curve. The available data are insufficient to model this
effect precisely, so we estimate ${\widetilde T_{rec}}$ simply by extrapolating to the value with $T_z=0$.

The final estimates of the telescope noise temperatures are given in
Table~\ref{noisetemp}, which shows that the telescope noise
temperatures for polarimeters $A$ and $C$ are much
higher than the noise temperatures of the other polarimeters. (This excess noise
was also observed with Y-factor tests at the telescope.) This excess noise
is inconsistent with the polarimetry channel sensitivities (see \S~\ref{polsens}), and is an artifact caused by the IF amplifiers
used in
these polarimeters. While these amplifiers performed well individually,
when they were all installed together in a single box, they produced
significant noise at frequencies below 2 GHz, which can be detected by
the unfiltered total power channels but is outside the polarimetry bands.
Polarimeters $B$ and $D$ used amplifiers with smaller gain and better power
regulation and did not have this problem. After the first observing
season the amplifiers in polarimeters $A$ and $C$ were replaced, and
excess noise in the total power channels was no longer observed. This
out-of-band noise also did not affect the quality of the polarimetry data
from the first observing season in any detectable way.

For polarimeters $B$ and $D$ in 2003, Table~\ref{noisetemp} shows that
${\widetilde{T}_{rec}} - T_{rec} \approx 15$~K, implying that the full
optical path at the telescope has a loss of roughly 6\%. This loss can arise from both absorption through
the windows and diffuse scattering from the mirrors in the optical path; it is consistent with the material properties of these objects.

\subsection{Polarimetry Channel Performance}
\label{polperf}

Astronomical and local sources of polarized radiation
serve to calibrate the polarimetry channels, while observations of the
atmosphere quantify the sensitivity of the polarimeter. These
observations are also used to derive the polarization-specific systematic effects
described in \S~\ref{polfid}.

\subsubsection{Calibration}
\label{polresp}

We have already discussed in \S~\ref{radcal} several methods of
calibrating the responsivity of the polarimetry channels in the
laboratory. Two additional techniques are used to calibrate
the polarimetry data during the observing season: (1) a system based on an
emissive chopper plate, and (2) observations of the polarized source Tau A
(for CAPMAP only).

The chopper plate calibration system works on the same
principle as the miniplate system described in
\S~\ref{radcal}: the reflection and emission of thermal
radiation from a metal surface produce a controlled, calculable
broad-band polarized signal.
In this case the surface is an aluminum
flat 1 meter across by 2 meters tall that nutates sinusoidally about a
vertical axis with a period of $\sim 2$~seconds to modulate the polarized signal. This plate reflects thermal
radiation from the sky at constant elevation into the polarimeters. For PIQUE, the plate was
sufficiently large that it could be viewed by the primary mirror. For
CAPMAP, the plate is installed in front of the secondary so that
radiation from the sky is reflected directly into the polarimeters.

The geometry of the chopper plate is slightly more complicated than that
of the miniplate system because the plate does not rotate about the axis of the feed, 
but the polarized signal is still given by Equation~\ref{eq:miniplate}. In this case the atmosphere
is the load of the system, and we use the total power channels
to estimate $T_{load}$. Statistical uncertainties derived from this method
are small, and the systematic uncertainty in the calibration is about
10\%, dominated by uncertainty in the resistivity of the aluminum plate.
The emissivity of the metal is the same for all polarimeters, so the
relative values of the responsivities for different polarimeters
are better constrained. Based on the repeatability of this test, the
uncertainties in the relative gains of different channels are only a few
percent.

Additional calibration data for CAPMAP 2003 are provided by 10
observations of Tau A, a strong source of polarized radiation at 90 GHz.
Tau A is sufficiently compact that it can be treated as a point
source, and its main polarized component has a well measured position
angle of $155^{\circ}$ \citep{Mayer:68}. Unfortunately, WMAP has not yet provided
an accurate estimate of the polarization fraction of Tau A, and the published data have a relatively large scatter of 7.5\%$\pm$1.0\% \citep{Farese:2003fd}. Thus observations of Tau A
cannot provide a precise calibration, but they serve as a
useful check on the responsivity estimates from the chopper plate.

The Tau A observations are performed and processed like the
Jupiter data to obtain maps of the polarimetry response fit to two-dimensional Gaussians. If the
parallactic angle of the polarimeter during the observation is more than
$23^{\circ}$ from $155^{\circ}$, then the point polarized
source approximation of Tau A breaks down and the observation does not
provide useful calibration data.
The beam widths derived from the successful fits are
significantly different for the different frequency bands, as expected
for diffraction-limited optics (see Table~\ref{optper}). These variations
in beam size are accounted for in the responsivity calculations in the data analysis.

The responsivity estimates from Tau A are not systematically different
from the estimates derived from chopper plate data, and the differences
between the two estimates are in general less than 20\%.

\paragraph{Gain Variations}
The chopper plate and Tau A calibration data also quantify
variations in the responsivity of the polarimetry channels due
to changes in the temperature of various components. For PIQUE, the dewar
was intentionally heated and cooled during one long chopper plate run
near the end of each observing season. These data show that the
responsivities of the polarimetry channels change by a fraction of a percent
for every Kelvin change in the dewar temperature. These variations were
sufficiently minor that they could be ignored in the data analysis.

For the first season of CAPMAP, the temperatures of the IF box and the LO
varied by roughly 10~K over the course of the season. Tau A
data taken with the IF box and LO at different temperatures
quantify the effects of these temperature shifts on the responsivity of
the polarimetry channels. The polarimetry channels are insensitive to
the temperature of the LO, but a 10~K shift in the temperature of the IF box
changes the polarimetry channel gain by roughly 10\%--20\%. This change is
consistent with the temperature coefficient of the IF amplifiers
and must be accounted for in the processing of the CMB data \citep{Barkats:2004he}.

\subsubsection{Sensitivity}
\label{polsens}

Section~\ref{radperformance} describes the procedures used to optimize the
polarimeter performance. Here we compare the sensitivities achieved
with those extrapolated from the characterization. Using typical
values for $T_{sys}$ from Table~\ref{noisetemp}, and values of $\Delta\nu$ which take into account 
the phase factors given in Table~\ref{phasetab}, Equation~\ref{Usys} yields $S \approx 115 \mbox{ K}/ \sqrt{3 \mbox{ GHz}}
\approx 2 \mbox{ mK}\sqrt{\mbox{s}}$ in the Rayleigh-Jeans limit. However,
while the system temperatures and the calibration source power levels are sufficiently high
that the Rayleigh-Jeans limit applies at 90 GHz, the 3~K CMB signal is
not, so the
responsivity and the sensitivity must be increased by a factor of 1.2 at
90~GHz \citep{Bennett:2003ca}. Therefore the sensitivity of a single polarimetry
channel to variations in the CMB should be roughly 2.4 mK$\sqrt{\mbox{s}}$, and the total
sensitivity of a W-band polarimeter with three polarimetry channels
should be 1.4 mK$\sqrt{\mbox{s}}$.

The sensitivities of the polarimetry channels can be measured directly
from the fluctuations in the time stream when the polarimeter views
the sky (in an area free of known sources) and are tabulated in
Table~\ref{noisesens}. The measured sensitivities are consistent with the calculation from
Equation~\ref{Usys}, indicating that the instruments suffer no excess
noise under observational conditions. Note that
the sensitivities of CAPMAP polarimeters $A$ and $C$ are comparable to
the others, demonstrating that the high telescope noise temperatures
given in Table~\ref{noisetemp} are artifacts.

\section{POLARIZATION-SPECIFIC SYSTEMATIC EFFECTS}
\label{polfid}

To detect the small polarized component of the CMB, a polarimeter must
not only have sufficient sensitivity to polarized signals, but it must also
be able to extract the polarized signal from largely
unpolarized radiation. Work is just now beginning on methods to quantify the systematic
effects on the quality of CMB polarization data, which will be extremely
important as increasingly sensitive arrays of instruments push forward
towards the goal of detecting the B-mode component of the 
polarization \citep{gaier:03,Keating:2003,Church:03}.

\citet[hereafter HHZ]{HHZ:2003} have developed a nomenclature that
quantifies polarization-specific systematic effects in terms of
leakages among Stokes parameters. The dominant
instrumental systematic effect in determining E-modes arises from
$I \rightarrow Q$ and $I \rightarrow U$ couplings. Mixing between Q and U can confuse
E- and B-modes by rotating the detection axes of the polarimeter, but such 
effects can be neglected in determining the E-modes because observations of 
polarized sources---such as the moon and Tau A---show that the detection axes are 
within $3^{\circ}$ of the expected orientation. Spurious responses of the polarimeter to
unpolarized incident radiation may be broadly classified into three
types: (1) monopole leakage (often called instrumental polarization,
or ``polarization transfer'' in HHZ)
arising from an axially symmetric unpolarized source; (2) leakage
from off-axis sources (also called dipole/quadrupole leakage or ``local contamination'' in HHZ), arising from
local curvature of the optics and typically dominated by response to
unpolarized dipole and quadrupole source distributions; and (3) polarized far side-lobes.

\subsection{Monopole Leakage}
\label{poltrans}

We follow the convention in HHZ and quantify the monopole leakage
with the parameter $\gamma$, which is the apparent change in the
polarized signal per unit change in the unpolarized intensity of the
axially symmetric incident radiation. This parameter is proportional to the 
Mueller matrix element $m_{\rm UI}$ \citep{Heiles:2001}. For PIQUE and CAPMAP, 
this leakage term arises primarily from the limited isolation of the output
ports of the OMT, as we describe below.

\begin{figure}
\plotone{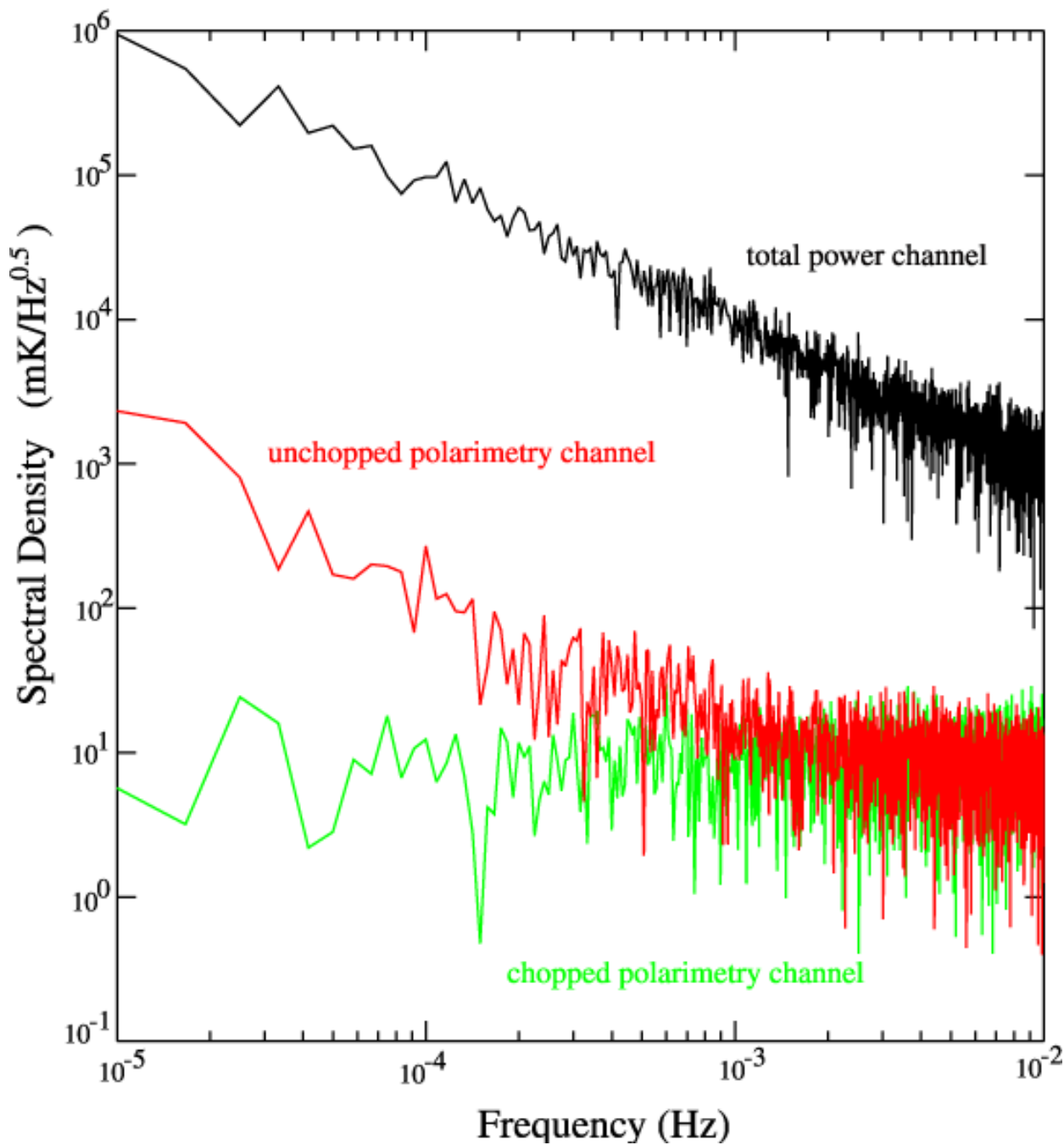}
\caption{\label{powspec}Power spectrum of various PIQUE data channels from a
clear day in February 2000 in NJ. The black curve
shows the power spectrum of a total power channel, which does not have 4
kHz modulation. The spectrum has a pronounced slope characteristic of
$1/f$ noise due to drifts in the temperature of the atmosphere.
 (The contribution from drifts in the polarimeter is sub-dominant on
all time scales displayed here.) The red curve is a polarimetry
channel after demodulation at 4~kHz. The $1/f$ noise
is much reduced, and becomes sub-dominant to the white thermal noise on
time scales less than several minutes.
The PIQUE and CAPMAP systems scan over the sky with periods of roughly
8--12 seconds. The signal from the sky is modulated on this time scale
and has a purely thermal noise spectrum (green line) for
time scales less than a day.}
\end{figure}

The value of $\gamma$ can in principle be estimated by measuring
the response of the polarimetry channels to changes in the temperature of
an unpolarized load. However, for small values of $\gamma$, it is
difficult to create a load sufficiently unpolarized for a reliable
measurement. Observations of the atmosphere at different elevations
do appear promising, but the effects of orientation-dependent signals
from side-lobes are still under investigation. A more
practical method of estimating $\gamma$ is to observe a
quiet patch of sky (see Figure~\ref{powspec}).
The atmosphere is essentially unpolarized \citep{Keating:98,Hanany:2003ms},
but the temperature of the atmosphere drifts such that the power
spectrum of the fluctuations in the total power channels (black line in
Figure~\ref{powspec}) has a strong $1/f$ component. If $\gamma=0$ and if the atmosphere were completely unpolarized, the power spectrum of the
polarimetry channels would have no $1/f$ noise. However, the actual
polarimeters have a nonzero $\gamma$, and their power
spectra have a $1/f$ component that dominates on time scales longer than
1000 seconds (the red curve in Figure~\ref{powspec}). The ratio of the levels of the
$1/f$ components in the total power and polarimetry channels provides a
good estimate of $\gamma$, which is roughly $-23$~dB for both the PIQUE and
CAPMAP polarimeters.

For correlation polarimeters, a nonzero $\gamma$
occurs because some radiation in one arm of the polarimeter leaks into
the other arm before the phase switch and generates a nonzero output voltage from the multiplier. Because
this signal is modulated by the phase switch like a real polarized
signal, it is not eliminated by the demodulation process.
In PIQUE and CAPMAP, the measured value of $\gamma$ is consistent
with leakage through the OMT.
A real OMT couples a small fraction of each polarized component into the
wrong arm of the polarimeter. This ``direct'' leakage between the arms does
not produce a significant $\gamma$ term because the unitarity of a
lossless OMT scattering matrix guarantees that these offsets cancel each other out. Such unitarity
constraints do not apply, however, to signals propagating backwards up
through the OMT. The reflection coefficient of HEMT LNA inputs is
typically around $-5$~dB in power, so a significant fraction of the
power
coupled into one arm reflects back through the OMT. Most of this
radiation escapes through the feed horn, but some leaks through the
OMT into the other arm of the radiometer and produces a spurious output
from the multiplier. The isolation through the OMTs used in
PIQUE and CAPMAP is $-40$~dB in power, so the leaked signal is
suppressed by $-45$~dB in power and the response of the multiplier
is suppressed by $-23$~dB---the geometric mean of the leaked and
original signal power levels---as observed.

A $\gamma$ of $-23$~dB is low enough for E-mode polarization measurements.
Not only is the contamination from CMB temperature anisotropies
more than 10~dB below the expected polarized signal, but the stability of the polarimetry channels is also acceptable
given a practical scan strategy. This value of
$\gamma$ is sufficiently small that the $1/f$ knee of the polarimetry channels occurs at 0.001 Hz (i.e.
time scales of many minutes). This is well below the typical scan
frequency of the telescope (0.1 Hz), so the detector output modulated by the scan is
completely stable for time scales of a day (see the green line in
Figure~\ref{powspec}).

\subsection{Dipole/Quadrupole Leakage}
\label{loccon}
\begin{figure}
\plotone{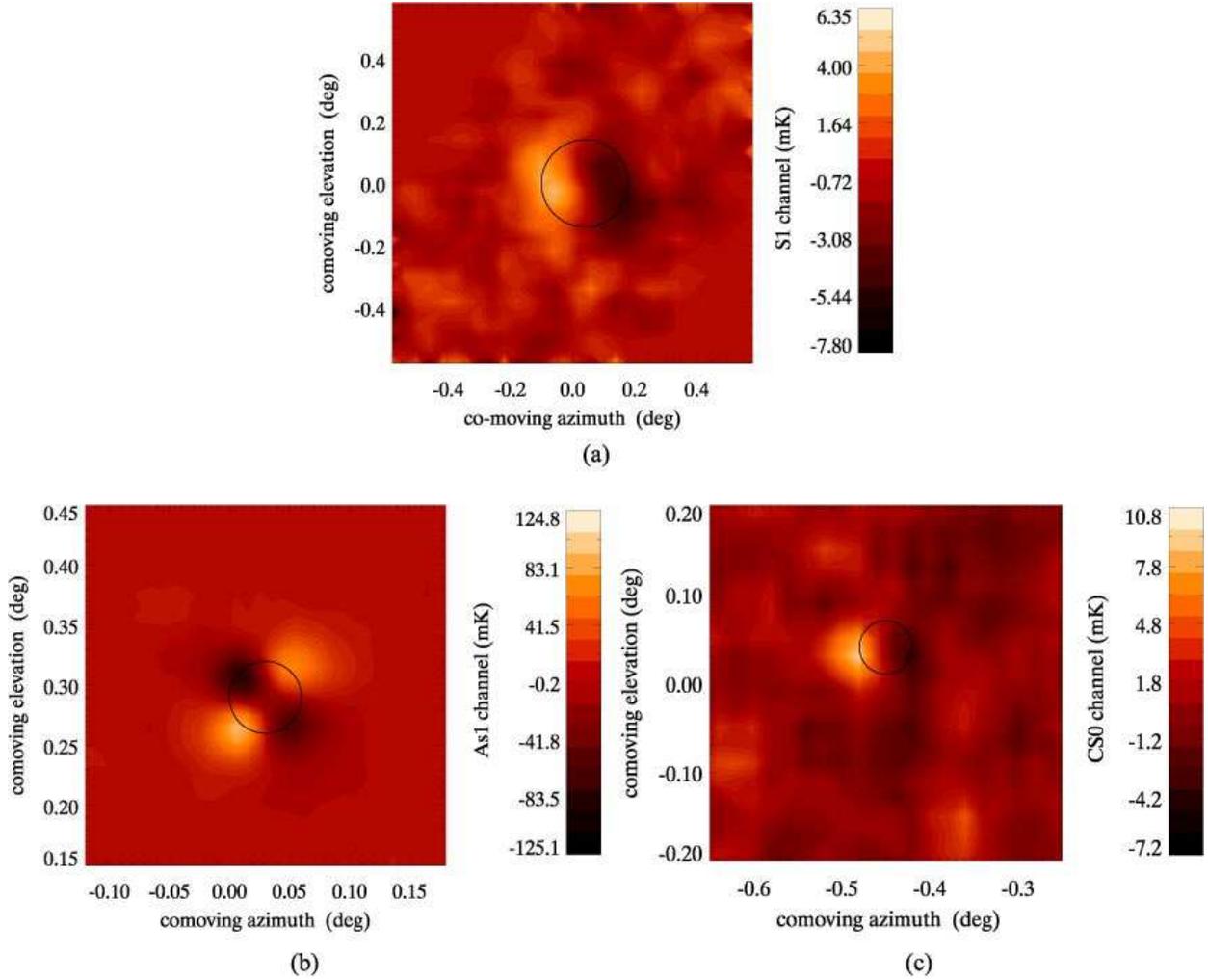}
\caption{Polarimetry channel response to Jupiter measured with the PIQUE
telescope \textbf{(a)} and the CAPMAP telescope \textbf{(b--c)}. The data processing to form these maps is described in \S~\ref{tpjup} and Figure~\ref{fig:baseline}. As detailed in \S~\ref{loccon}, the dipolar and
quadrupolar patterns are good examples of spurious polarized signals
generated by off-axis and on-axis optical elements respectively.
The black circle is the FWHM of the total power beam.
Individual panels are on different scales. The peak of the co-polar
Jupiter maps is 280~mK for PIQUE and $\sim 3$~K for CAPMAP. The rms
noise level is 3~mK for the PIQUE map and 7~mK for both CAPMAP 2003
and CAPMAP 2004 maps. Note that the PIQUE receiver was positioned at
the focal point of the telescope, but that the CAPMAP 2003 and CAPMAP
2004 receivers are as far as 15~cm and 48~cm respectively away from
the focal point. 
The magnitude of these spurious responses is summarized in Table~\ref{polparam}.}
 \label{polmap}
 \end{figure}

 Figure~\ref{polmap} shows the response of
representative PIQUE and CAPMAP polarimetry channels to Jupiter, an
unpolarized point source (the maps are generated using the same procedures
described in \S~\ref{tpjup}). These maps show patterns dominated by either
two or four lobes within the approximate angular scale of the beam.
Each map can be fit by a superposition of dipole and quadrupole
patterns using the following function:
\begin{equation}
\left[ \gamma + d\,\sigma_y \frac{\partial}{\partial y} + q\,\sigma_x\sigma_y
\frac{\partial}{\partial x}\frac{\partial}{\partial y}\right] f(x,y), 
\end{equation}
where $f(x,y)$ is the Gaussian function given in Equation~\ref{gaussjup}, 
$\sigma_x$ and $\sigma_y$ are the Gaussian beam widths---which in practice 
are allowed to float to different values in each term above---and $d$ and $q$ are 
parameters which quantify the size of the dipole and quadrupole 
terms, following the conventions used in HHZ. In this formalism, the 
peak values of the dipole and quadrupole patterns are given by $\pm d/\sqrt{e}$ and $\pm q/e$, where $e$ is the base of natural logarithms. 

Unlike $\gamma$, which is sensitive to the characteristics of the
polarimeters, $d$ and $q$ are determined predominantly by the optical
components of the telescope. The boundary
conditions at optically active surfaces are not the same for the
two orthogonally polarized components of the incident radiation, which are thus coupled
into the polarimeter with slightly different efficiencies depending on
the orientation of the incident wave vector \citep{Gans:1976,Carretti:2004}.
This effect causes an unpolarized source to
generate a spurious polarized signal depending on its position in the sky.
For a differencing polarimeter, this spurious signal arises from
asymmetries in the beam shapes for the two polarizations (HHZ),
while for correlation polarimeters like PIQUE and CAPMAP,
these signals are due to non-zero cross-polar beams which couple one
polarized component of the incident radiation into the other component at
the input to the polarimeter. The electrodynamics behind these two
effects are identical and yield the same values of $d$ and $q$ given
the same optical elements.

The PIQUE and CAPMAP $d$ and $q$ parameters are summarized in
Table~\ref{polparam} and illustrated in Figure~\ref{polmap}.
PIQUE and CAPMAP demonstrate how the geometry of the optics affects the form of the local contamination,
specifically the different patterns generated by on-axis and off-axis
optical elements. Theoretically, on-axis systems do not generate a $d$ term because
the optical elements do not define a preferred direction relative to the
polarization axes. Off-axis systems, by contrast, can generate both $d$
and $q$ terms.

For PIQUE, the dipole term $d \sim -14$~dB clearly dominates. This is
to be expected since the corrugations in feed horns are designed explicitly
to minimize the leakage terms \citep{Goldsmith:1998}, and the short focal length and
strong curvature of the mirror should generate comparatively
high amounts of local contamination.

By contrast, in the CAPMAP 2003 system the quadrupolar term $q \sim -8$~dB is roughly 5~dB higher than $d$.
This suggests that the contamination generated by 
the on-axis optical elements (horn and lens) dominates over the
contamination produced by the mirrors. Since the correlator takes the 
product of two electric fields, the power coupled between polarization 
states should have a peak value of $q^2/e^2 \sim -24$~dB, which is 
consistent 
with the measured cross-polar response of the feed system (see 
Figure~\ref{fig:optics_beam_maps}). 
Also, the $d$ and $q$ terms of polarimeters with AR grooves on both
surfaces of the lenses are twice as big as the terms for the polarimeters
with AR grooves on only one surface (see
Table~\ref{polparam}), and as discussed in \S~\ref{capopt}, we expect
grooves to increase the birefringence
of the lens surfaces and to compromise the
polarization fidelity of the optics \citep{McMahon:2005}.
After discovering the high levels of off-axis leakage generated by these
lenses, we replaced them before the subsequent observing season
with an improved design with very low cross-polarization that produced $d$ and $q$ parameters that were
smaller by a factor of 10 (see Table~\ref{polparam}).

The dipole and quadrupole leakage terms alias sharp features
in the intensity distribution---from the CMB temperature anisotropy or
from point sources---into spurious polarization
fluctuations. Although this aliasing will be a significant issue for
future large arrays (HHZ), it does not pose a significant problem for
PIQUE or CAPMAP. PIQUE's local contamination was sufficiently low that
CMB temperature anisotropies, which are roughly 10 times
greater than the polarization anisotropies, produced	 spurious polarized
signals more than $6$~dB below the expected E-mode signal. The first season of
CAPMAP had higher contamination levels, but the narrow beam width of this
experiment only couples spurious CMB temperature signals from higher
angular scales, where the temperature anisotropy is strongly damped (HHZ).

\subsection{Far Side-Lobe Contamination}
\label{offsets}

Sources far from the optical axis can also produce spurious polarization
signals via the far side-lobes of the instrument. The far side-lobes of a
telescope are features produced by spillover past the
mirrors or scattered radiation from various support structures. These scan-dependent
signals have a significant impact on the performance of
the experiment because they are synchronous with respect to the scan pattern and are not
modulated away like the $1/f$ noise discussed in \S~\ref{poltrans}. The effect of these
spurious signals on the data quality is detailed in \citet{Barkats:2004he}.
The need to remove the scan-synchronous signals effectively reduced the sensitivity of CAPMAP 2003 by a factor of 2.
(Note, however, that with the higher sensitivity of the CAPMAP data obtained in later seasons, the impact is smaller.)

In temperature anisotropy experiments, position-dependent signals
generated by side-lobes are controlled using ground screens which block
radiation from local sources. PIQUE demonstrated that fixed ground
screens are not sufficient to control spurious polarized signals. Such
fixed ground screens are scanned by sidelobes during the telescope
motion and produce scan-synchronous polarized emission for the same
reason our miniplate system does. A
similar phenomenon was observed in the COMPASS experiment \citep{Farese:2003iz}.

During the first PIQUE observing season, the polarized signal measured
at two positions $1^{\circ}$ apart in azimuth differed by
hundreds of $\mu$K. This azimuth-dependent offset occurred because
the angle of
incidence of the radiation reflected off of the fixed ground screens
into the polarimeter was different in the different telescope positions,
changing the polarized component of the scattered radiation (see
\S~\ref{radcal}).
For PIQUE's second observing season the
inner ground screen was enlarged so that any radiation reflected into the
polarimeter would come from this screen and have the same incidence angle
regardless of the
telescope orientation. With this improved system, the azimuth-dependent
signal was reduced by roughly a factor of 2. The remaining signal was
due to a small gap between the bottom of the dish and the floor of the
inner ground screen enclosure. This gap was closed towards the end of
the 2001 season, and the azimuth dependent signals were reduced to
$\sim 30~\mu$K. These residual signals were constant at the 10\% level
required for PIQUE, and so no further improvements were needed. 

A ground screen was not used in CAPMAP, and during its first observing
season, the polarimetry channels registered a scan-synchronous signal (SSS). Note that the level of SSS in CAPMAP was comparable to that in PIQUE,
despite the absence of a groundscreen in CAPMAP; this is due to
the more aggressive under-illumination in CAPMAP, which effectively
uses portions of the mirrors as a co-moving groundscreen.
After a second-order polynomial in azimuth was fit to the data in
20 beam-sized azimuth bins and removed, the polarization data from the
whole season collapsed into the 20 bins revealed channel-specific SSS
with rms levels at 10~$\mu$K--26~$\mu$K. These variations were generated as various side-lobes
moved over sources of emission on or near the ground, including the large atmospheric emission near the horizon.

Maps of the side-lobes derived from observations of an RF source have
recently been performed, indicating side-lobe features sufficiently strong to
cause the observed SSS. More measurements are underway to confirm 
this interpretation.

\begin{figure}
\centering
\includegraphics[width=.7\textwidth]{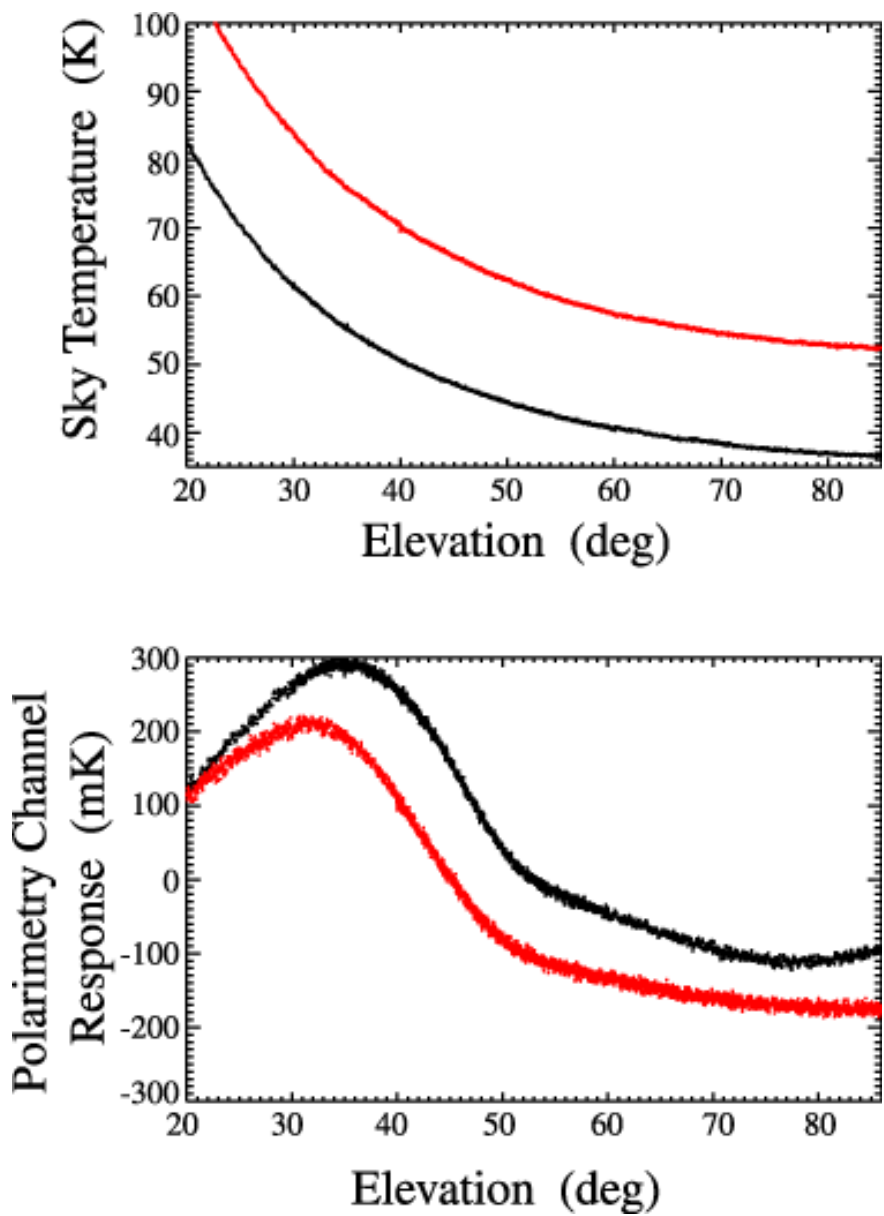}
\caption{\label{sidelobe}Evidence for polarized side lobes. \textbf{(top)} Change in
sky temperature during two elevation scans in 2003 February with a high
(gray/red, $T_z \sim 50$~K) and a low (black, $T_z \sim 35$~K) zenith noise temperature. The sky temperatures are 
derived from the total power channels by removing the receiver temperature. \textbf{(bottom)} Corresponding 
polarimetry channels for the same sky dips. The level shift around 45$^{\circ}$ is
consistent with side-lobes due to spillover past the primary moving
from the sky to the ground (see \S~\ref{offsets} for details). A similar level shift is not observed in the total 
power channels, which indicates that this is not simply monopole leakage of unpolarized signals from the horizon.}
\end{figure}

An indirect measurement of these side-lobes is found in
the elevation scans performed during the first observing season. Since
the atmosphere is unpolarized, polarimetry channels should not
change as the elevation of the telescope changes (aside from the trend
generated by the nonzero $\gamma$). In fact, the level in all the
polarimetry channels shifts to a more negative value when the telescope
elevation exceeds $\sim 45^{\circ}$ (see Figure~\ref{sidelobe}). The
shape of the transition
depends on the temperature of the atmosphere and is consistent with
side-lobes due to spillover past the primary moving from the sky through the
horizon to the ground.

A number of improvements to the optics were made
prior to the second season to reduce the size of these side-lobes, including a
screen around the secondary and new lenses to sharpen the
beam. The results of these modifications will
be presented in a later paper.

\section{CONCLUSION}

The performance of the PIQUE and CAPMAP experiments in the field
shows that the concept of the phase-switched
correlation polarimeter is well suited for precise CMB polarimetry. We have presented various techniques to
improve and evaluate the calibration and systematic effects associated
with this type of polarimeter. As the CMB polarization field advances towards higher sensitivities, these
techniques will become even more critical.
CAPMAP continues to observe the polarized CMB emission with its full set of receivers,
and we expect by the summer of 2005 to have increased our observation
time with respect to the CAPMAP 2003 data set by an order of magnitude. 

\acknowledgements
We acknowledge Lucent Technologies for use of the Crawford Hill 7-meter antenna,
and Bob Wilson, Greg Wright, and Tod Sizer for their unrelenting
assistance with the all aspects of the telescope. We
thank Norm Jarosik, Lyman Page, Steve Meyer, Paul Waltz, Mike Niemack,
and Toby Marriage for helpful discussions. We thank Michelle Yeh, Ashish Gupta, Jamie Hinderks, and Marc Schreiber for their help with telescope control and data acquisition
software. We are grateful to Mary Wells for assembling the LNAs and to
Northrop Grumman Space Technology (Richard Lai and Ronald
Grundbacher). We gratefully acknowledge design and construction
help from Ted Griffiths, Laszlo Vargas, Bill Dix, Mike Peloso, Glenn
Atkinson, Harold Sanders, and Fukun Tang. We also thank Peter
Hamlington, Maire Daly, Hannah Barker, Trina Ruhland, Al Dietrich, Chris Herzog, Ariane Billing,
Richard Cendejasm, Phil Marfuta, George Costow, Jennifer Hou, Dan
Crosta, Alex Dahlen, Jamie Gainer, Matt Goss, Jae-Young Lee, Michael
Matejek, Celia Muldoon, Serena Rezny, Chris Rogan, Michael Rosen,
Nicole Rowsey, Sameer Shariff, Yunior Savon, and Phyo Thant for their
help with building and characterizing various parts of the instruments. This work was supported by NSF grants PHY-9984440, PHY-0099493, PHY-0355328, AST-0206241, and PHY-0114422, and the Kavli Foundation. Portions of this work were carried out at the Jet Propulsion Laboratory,
California Institute of Technology, operating under a contract from the National Aeronautics and Space Administration.
\pagebreak
\bibliographystyle{ms}
\bibliography{ms}

\begin{thebibliography}{57}
\expandafter\ifx\csname natexlab\endcsname\relax\def\natexlab#1{#1}\fi

\bibitem[{{Barkats}(2004)}]{Barkats:04T}
{Barkats}, D. 2004, PhD thesis, Princeton University

\bibitem[{Barkats {et~al.}(2004)}]{Barkats:2004he}
Barkats, D., {et~al.} 2004, \apj, 619, L127

\bibitem[{{Barnes} {et~al.}(2002)}]{Barnes:2002}
{Barnes}, C., {et~al.} 2002, \apjs, 143, 567

\bibitem[{{Bennett} {et~al.}(2003a)}]{Bennet:2003a}
{Bennett}, C.~L., {et~al.} 2003a, \apjs, 148, 1

\bibitem[{Bennett {et~al.}(2003b)}]{Bennett:2003ca}
Bennett, C.~L., {et~al.} 2003b, \apjs, 148, 97

\bibitem[{{Carretti} {et~al.}(2004){Carretti}, {Cortiglioni}, {Sbarra}, \&
  {Tascone}}]{Carretti:2004}
{Carretti}, E., {Cortiglioni}, S., {Sbarra}, C., \& {Tascone}, R. 2004, \aap,
  420, 437

\bibitem[{{Chu} {et~al.}(1978){Chu}, {Wilson}, {England}, \& {Legg}}]{Chu:1978}
{Chu}, T.~S., {Wilson}, R.~W., {England}, D.~A., \& {Legg}, W.~E. 1978, Bell
  System Technical Journal, 57, 1257

\bibitem[{{Church} {et~al.}(2003)}]{Church:03}
{Church}, S., {et~al.} 2003, \nar, 47, 1083

\bibitem[{{Cortiglioni}(1994)}]{cortiglioni:1994}
{Cortiglioni}, S. 1994, Rev. Sci. Instrum., 65, 2667

\bibitem[{{de Pater}(1999)}]{Imke:1999}
{de Pater}, I. 1999, {Encyclopedia of the Solar System} (Academic Press),
  735--772

\bibitem[{{Farese} {et~al.}(2003)}]{Farese:2003iz}
{Farese}, P.~C., {et~al.} 2003, \nar, 47, 1033

\bibitem[{{Farese} {et~al.}(2004)}]{Farese:2003fd}
---. 2004, \apj, 610, 625

\bibitem[{{Gaier} {et~al.}(2003){Gaier}, {Lawrence}, {Seiffert}, {Wells},
  {Kangaslahti}, \& {Dawson}}]{gaier:03}
{Gaier}, T., {Lawrence}, C.~R., {Seiffert}, M.~D., {Wells}, M.~M.,
  {Kangaslahti}, P., \& {Dawson}, D. 2003, \nar, 47, 1167

\bibitem[{Gans(1976)}]{Gans:1976}
Gans, M.~J. 1976, Bell System Technical Journal, 55(3), 289

\bibitem[{{Goldsmith}(1998)}]{Goldsmith:1998}
{Goldsmith}, P.~F. 1998, Quasioptical Systems: Gaussian Beams Quasioptical
  Propagation and Applications (IEEE Press)

\bibitem[{{Hamaker} {et~al.}(1996){Hamaker}, {Bregman}, \&
  {Sault}}]{Hamaker:1996}
{Hamaker}, J.~P., {Bregman}, J.~D., \& {Sault}, R.~J. 1996, \aaps, 117, 137

\bibitem[{{Hanany} \& {Rosenkranz}(2003)}]{Hanany:2003ms}
{Hanany}, S., \& {Rosenkranz}, P. 2003, \nar, 47, 1159

\bibitem[{{Hedman}(2002)}]{Hedman:02T}
{Hedman}, M.~M. 2002, PhD thesis, Princeton University

\bibitem[{{Hedman} {et~al.}(2002){Hedman}, {Barkats}, {Gundersen}, {McMahon},
  {Staggs}, \& {Winstein}}]{Hedman:2002ck}
{Hedman}, M.~M., {Barkats}, D., {Gundersen}, J.~O., {McMahon}, J.~J., {Staggs},
  S.~T., \& {Winstein}, B. 2002, \apjl, 573, L73

\bibitem[{{Hedman} {et~al.}(2001){Hedman}, {Barkats}, {Gundersen}, {Staggs}, \&
  {Winstein}}]{Hedman:2001}
{Hedman}, M.~M., {Barkats}, D., {Gundersen}, J.~O., {Staggs}, S.~T., \&
  {Winstein}, B. 2001, \apjl, 548, L111

\bibitem[{{Heiles} {et~al.}(2001)}]{Heiles:2001}
{Heiles}, C., {et~al.} 2001, \pasp, 113, 1274

\bibitem[{{Hu} {et~al.}(2003){Hu}, {Hedman}, \& {Zaldarriaga}}]{HHZ:2003}
{Hu}, W., {Hedman}, M.~M., \& {Zaldarriaga}, M. 2003, \prd, 67, 043004

\bibitem[{{Huei} {et~al.}(2001)}]{Wang_huei:2001}
{Huei}, W., {et~al.} 2001, IEEE Trans. Microwave Theory Techniques, 49, 9

\bibitem[{IAU(1973)}]{IAU}
IAU. 1973, Trans. IAU, 15B, 166

\bibitem[{{Imbriale}(2003)}]{Imbriale:2003}
{Imbriale}, W.~A. 2003, {Large Antennas of the Deep Space Network} (Wiley and
  Sons)

\bibitem[{Imbriale \& Hodges(1991)}]{Imbriale:1991}
Imbriale, W.~A., \& Hodges, R.~E. 1991, Applied Computational Electromagnectic
  Society Journal, 6, 74

\bibitem[{{Jackson}(1998)}]{jackson:book}
{Jackson}, J.~D. 1998, {Classical Electrodynamics}, 3rd edn. (Wiley)

\bibitem[{{James}(1981)}]{James:81}
{James}, G.~L. 1981, IEEE Trans. Microwave Theory Techniques, 29, 1059

\bibitem[{{Jarosik} {et~al.}(2003)}]{Jarosik:2003}
{Jarosik}, N., {et~al.} 2003, \apjs, 145, 413

\bibitem[{{Johnson} {et~al.}(2003)}]{Johnson:03}
{Johnson}, B.~R., {et~al.} 2003, \nar, 47, 1067

\bibitem[{{Kamionkowski} \& {Kosowsky}(1998)}]{Kamionkowski:1998av}
{Kamionkowski}, M., \& {Kosowsky}, A. 1998, \prd, 57, 685

\bibitem[{{Keating} {et~al.}(1998){Keating}, {Timbie}, {Polnarev}, \&
  {Steinberger}}]{Keating:98}
{Keating}, B., {Timbie}, P., {Polnarev}, A., \& {Steinberger}, J. 1998, \apj,
  495, 580

\bibitem[{{Keating} {et~al.}(2003)}]{Keating:2003}
{Keating}, B.~G., {et~al.} 2003, in Polarimetry in Astronomy, ed. S.~Fineschi,
  Vol. 4843 (SPIE), 284--295

\bibitem[{{Kovac} {et~al.}(2002){Kovac}, {Leitch}, {Pryke}, {Carlstrom},
  {Halverson}, \& {Holzapfel}}]{Kovac:2002}
{Kovac}, J.~M., {Leitch}, E.~M., {Pryke}, C., {Carlstrom}, J.~E., {Halverson},
  N.~W., \& {Holzapfel}, W.~L. 2002, \nat, 420, 772

\bibitem[{{Kraus}(1986)}]{Kraus:1986}
{Kraus}, J.~D. 1986, Radio Astronomy (Cygnus Quaasar Books)

\bibitem[{{Leitch} {et~al.}(2002)}]{Leitch:2002}
{Leitch}, E.~M., {et~al.} 2002, \nat, 420, 763

\bibitem[{{Leitch} {et~al.}(2004)}]{Leitch:2004gd}
---. 2004, \apj, submitted, arXiv: astro-ph/0409357

\bibitem[{{Mayer} \& {Hollinger}(1968)}]{Mayer:68}
{Mayer}, C.~H., \& {Hollinger}, J.~P. 1968, \apj, 151, 53

\bibitem[{{McMahon}(2005)}]{McMahon:2005}
{McMahon}, J.~J. 2005, in preparation

\bibitem[{{Montroy} {et~al.}(2003)}]{Montroy:03}
{Montroy}, T., {et~al.} 2003, \nar, 47, 1057

\bibitem[{{O'dell}(2002)}]{O'dell:02}
{O'dell}, C. 2002, PhD thesis, University of Wisconsin, arXiv: astro-ph/0201224

\bibitem[{{Page} {et~al.}(2003)}]{Page:2003}
{Page}, L., {et~al.} 2003, \apjs, 148, 39

\bibitem[{{Pozar}(1998)}]{Pozar:1998}
{Pozar}, D.~M. 1998, {Microwave Engineering}, 2nd edn. (Wiley and Sons)

\bibitem[{Readhead {et~al.}(2004)}]{Readhead:2004xg}
Readhead, A. C.~S., {et~al.} 2004, Science, 306, 836

\bibitem[{{Rohlfs} \& {Wilson}(1996)}]{rohlfs:1996}
{Rohlfs}, K., \& {Wilson}, T.~L. 1996, {Tools of Radio Astronomy}, 2nd edn.
  (Springer-Verlag)

\bibitem[{{Seljak} \& {Hirata}(2004)}]{Seljak:2004}
{Seljak}, U., \& {Hirata}, C.~M. 2004, \prd, 69, 043005

\bibitem[{{Seljak} \& {Zaldarriaga}(1997)}]{Seljak:1997gy}
{Seljak}, U., \& {Zaldarriaga}, M. 1997, \prl, 78, 2054

\bibitem[{{Seljak} \& {Zaldarriaga}(1999)}]{Seljak:1998nu}
---. 1999, \prd, 60, 043504

\bibitem[{{Smith} {et~al.}(2004){Smith}, {Hu}, \& {Kaplinghat}}]{Smith:2004}
{Smith}, K.~M., {Hu}, W., \& {Kaplinghat}, M. 2004, \prd, 70, 043002

\bibitem[{{Staggs} {et~al.}(2002){Staggs}, {Barkats}, {Gundersen}, {Hedman},
  {Herzog}, {McMahon}, \& {Winstein}}]{staggs:2002}
{Staggs}, S.~T., {Barkats}, D., {Gundersen}, J.~O., {Hedman}, M.~M., {Herzog},
  C.~P., {McMahon}, J.~J., \& {Winstein}, B. 2002, in AIP Conf. Proc. 609:
  Astrophysical Polarized Backgrounds, 183--186

\bibitem[{{Strozzi} \& {McDonald}(2000)}]{Strozzi:2000}
{Strozzi}, D.~J., \& {McDonald}, K.~T. 2000, arXiv: physics/0005024

\bibitem[{{Tegmark} {et~al.}(2000){Tegmark}, {Eisenstein}, {Hu}, \& {de
  Oliveira-Costa}}]{Tegmark:2000}
{Tegmark}, M., {Eisenstein}, D.~J., {Hu}, W., \& {de Oliveira-Costa}, A. 2000,
  \apj, 530, 133

\bibitem[{{Tinbergen}(1996)}]{Tinbergen:96}
{Tinbergen}, J. 1996, {Astronomical Polarimetry} (Cambridge University Press)

\bibitem[{Weinreb {et~al.}(1999)Weinreb, Lai, Erickson, Gaier, \&
  Wielgus}]{weinreb:1999}
Weinreb, S., Lai, R., Erickson, N., Gaier, T., \& Wielgus, J. 1999, IEEE MTT-S
  International, 1, 101

\bibitem[{{Wollack} {et~al.}(1997){Wollack}, {Devlin}, {Jarosik},
  {Netterfield}, {Page}, \& {Wilkinson}}]{Wollack:1997}
{Wollack}, E.~J., {Devlin}, M.~J., {Jarosik}, N., {Netterfield}, C.~B., {Page},
  L., \& {Wilkinson}, D. 1997, \apj, 476, 440

\bibitem[{{Zaldarriaga} \& {Seljak}(1998)}]{Zelda:1998}
{Zaldarriaga}, M., \& {Seljak}, U. 1998, \prd, 58, 023003

\bibitem[{Zhang(1993)}]{Zhang:1993}
Zhang, X. 1993, IEEE Trans. Microwave Theory Techniques, 41(8), 1263

\end{thebibliography}
\clearpage
\clearpage
\clearpage
\clearpage
\clearpage
\clearpage
\clearpage
\clearpage
\clearpage
\clearpage

\clearpage

\clearpage
\clearpage
\clearpage
 \clearpage
\clearpage
\begin{table}
\begin{tabular}{lr}
\tableline
\tableline
Acronym & Meaning \\
\tableline
ADC & Analog-to-Digital Converter \\
AR & Anti-Reflection \\
CAPMAP & Cosmic Anisotropy Polarization MAPper \\
CBI & Cosmic Background Imager \\
CMB & Cosmic Microwave Background \\
DASI & Degree Angular Scale Interferometer \\
FWHM & Full Width at Half Maximum\\
HEMT & High Electron Mobility Transistor \\
IF & Intermediate Frequency (2--18 GHz at W-band) \\
LNA & Low Noise Amplifier \\
LO & Local Oscillator \\
MMIC & Monolithic Microwave Integrated Circuit \\
OMT & Ortho-Mode Transducer \\
PIQUE & Princeton IQU Experiment \\
PWV & Precipitable Water Vapor\\
RF & Radio Frequency (84--100 GHz at W-band) \\
SSS & Scan-Synchronous Signal\\
WMAP & Wilkinson Microwave Anisotropy Probe \\
\tableline
\end{tabular}
\caption{Commonly used acronyms\label{acronyms}}
\end{table}
\clearpage
\begin{table}
\begin{tabular}{lccccl}
\tableline
\tableline
& \multicolumn{2}{c}{PIQUE} & \multicolumn{2}{c}{CAPMAP} \\
Characteristic & W-band & Q-band & W-band & Q-band \\
\tableline
$T_{sys}$\tablenotemark{a} (K)&140&50&115&50\\
Bandwidth (GHz)&11&8&12&8\\
Sensitivity\tablenotemark{b} (mK$\sqrt{\mbox{s}}$)&1.6&0.6&1.3&0.6\\
No. Detectors & 1 & 1 & 12 & 4 \\
Total Sensitivity (mK$\sqrt{\mbox{s}}$)&1.6&0.6&0.4&0.3\\
Flux Conversion Factor\tablenotemark{c} ($\mu$K/Jy)&230&270&3200&6100\\ 
\tableline
\end{tabular}
\caption{Typical polarimeter performance parameters for PIQUE and
CAPMAP\label{tbl:senfac}}
\tablecomments{For more detailed performance
specifications see Tables~\ref{gainstab}--\ref{phasetab} and~\ref{noisesens}.}
\tablenotetext{a}{$T_{sys}$ includes 50~K (25~K) at W-band (Q-band) from the sum
of the atmospheric contribution and the CMB.}
\tablenotetext{b}{The sensitivity is given here in thermodynamic temperature
units, uncorrected for atmospheric opacity, using
Equation~\ref{Usys}. The absorbtion of the atmosphere has to be
accounted for in the CMB analysis \citep{Barkats:2004he}. It varies
over the season from 10\% to 40\%, amounting in average to a 20\%
degradation of the sensitivity for the selected data.} 
\tablenotetext{c}{The flux conversion factor is calculated as $\Gamma=A_{e}/2k_{B}$, where $A_{e}=\Omega/\lambda^2$ is the effective beam aperture \citep{rohlfs:1996}, and is given to 2 significant figures.}
\end{table}
\clearpage
\begin{table}
\begin{tabular}{ccccc}
\tableline
\tableline
Component & Gain\tablenotemark{a} & Effective Noise & Signal Level & Signal Level \\
& & Temperature   & (Polarization)\tablenotemark{b} & (Total Power) \\ \tableline
Sky + CMB &\nodata & \nodata& 5 $\mu$K & 50 K \\ \tableline
Lens\tablenotemark{c} & $-0.08$~dB\tablenotemark{d} & 3~K & $1.1\times10^{-12}~\mu$W &$1.1\times10^{-5}~\mu$W \\
Horn & $<-0.05$~dB\tablenotemark{d} & $<0.3$~K & $1.1\times10^{-12}~\mu$W &
$1.1\times10^{-5}~\mu$W \\ \tableline
OMT & $-0.05$~dB & 0.3 K & $1.1\times10^{-12}~\mu$W & $1.1\times10^{-5}~\mu$W \\
LNAs & 35 dB & 50 K & $3.5\times10^{-9}~\mu$W & $8.8\times10^{-2}~\mu$W \\
WG Bend & $-0.3$~dB  & $<1$~K & $3.3\times10^{-9}~\mu$W & $8.3\times10^{-2}~\mu$W \\
Stainless Steel WG & $-0.7$~dB & $<1$~K & $2.8\times10^{-9}~\mu$W &
$7.0\times10^{-2}~\mu$W \\
Filter & $-0.6$~dB & $<1$~K & $2.4\times10^{-9}~\mu$W & $6.0\times10^{-2}~\mu$W \\
Mixer &$-10$~dB & $<1$~K & $2.4\times10^{-10}~\mu$W & $6.0\times10^{-3}~\mu$W\\ \tableline
IF Amplifier & 35~dB & 3~K & $7.2\times10^{-7}~\mu$W & $18~\mu$W \\
Splitter & $-3$~dB & \nodata & $3.6\times10^{-7}~\mu$W & $9~\mu$W \\
Detector Diode & 1 mV/$\mu$W & \nodata & \nodata & 3 mV \\
Filter Bank & $-5$~dB & \nodata & $1.2\times10^{-7}~\mu$W & \nodata \\
Multiplier & 1.2 mV/$\mu$W & \nodata & 0.15 nV & \nodata\\
Pre-Amplifier & 100 & \nodata & 15 nV & 300 mV \\ \tableline
\end{tabular}
\caption{Summary of W-band polarimeter responsivities and signal levels\label{gainstab}}
\tablecomments{See Table~\ref{noisetemp} for the final receiver temperatures.}
\tablenotetext{a}{The gains and losses included here are rough, band-averaged estimates, based on a combination
of actual measurements, typical data, and the specifications of the parts in question.}
\tablenotetext{b}{This column estimates the polarized signal from the CMB only,
assumed to be about 5~$\mu$K; the (larger) spurious polarized signals are not included.}
\tablenotetext{c}{{}Lenses are present only in the CAPMAP experiment.}
\tablenotetext{d}{Gain in this case refers to gain reduction due to reflections and losses.}
\end{table}
\clearpage
\begin{table}
\begin{tabular}{lrr}
\tableline
\tableline
Component & W-band Parts & Q-band Parts \\
\tableline
OMT &Vertex RSI 111590 & Vertex RSI 111588 \\
HEMT$^P$& Custom NRAO & Custom NRAO \\
HEMT$^C$& Custom JPL & Custom JPL \\
Filter & MRI FRWS-94 & MRI FLQS-40.0 \\
Mixer &Spacek Labs P94-10-LN& Spacek Labs M-40.5-15 \\ \tableline
LO$^P$ & Custom Millimeter Wave & Millimeter Wave 28CSO-30.5 \\
LO$^C$& Spacek GW-820 & Spacek GW-305 \\
Power Amp$^C$& Custom JPL & \nodata \\
Phase Switch$^P$& Pacific Millimeter W180 & Pacific Millimeter 2640 MC-30.5\\
Phase Switch$^C$ & Pacific Millimeter 75MS-82 & Pacific Millimeter 2640 MC-30.5\\
Phase Tuner$^P$ & Millitech VPS-10-R0000 & Millitech VPS-28-S0000 \\ \tableline
IF Amp$^P$& DBS Microwave DB97-0421 & Miteq AFS44-02001800-25-KCR-S-44\\
IF Amp$^C$& Custom Miteq & Custom Miteq \\
$0^{\circ}$ Power splitter & MAC Technology P248-2& MAC Technology P248-2 \\
$90^{\circ}$ Hybrid Coupler$^P$ & \nodata & Sage 2375-9\\
Detector Diode &Agilent HP 8472-B & Agilent HP 8472-B \\
Filter Bank & ES Microwave 3SM-7/12.7-10PM & ES Microwave 2SM-9.75-10PM \\
Phase Tuners$^P$& M/A-COM s054-6002-00 & Weinschel 917-12 \\
Phase Tuners$^C$ & Weinschel 917-12 & Weinschel 917-12 \\
Multiplier & Miteq DB0218LW2&Miteq DB0218LW2\\ \tableline
\end{tabular}
\caption{Components list for PIQUE and CAPMAP polarimeters\label{compolist}}
\tablecomments{For each component, the name of the manufacturer as well as the part number is given. Superscripts $P$ and $C$ denote
the components used for PIQUE and CAPMAP respectively.}
\end{table}
\clearpage
\begin{table}
\begin{tabular}{lcccc}
\tableline
\tableline
Component & PIQUE 2000 & PIQUE 2001 & CAPMAP 2003 & CAPMAP 2004\\
\tableline
Lens&\nodata&\nodata&160&130\\
Horn&40&20&40&25\\
OMT/LNAs&30&18&40&25\\
Filters/Mixers&80&50&300&300\\
LO&315&315&305&315\\
IF Section&305&305&288&315\\ \tableline
\end{tabular}
\caption{Operating temperatures of polarimeter components in K\label{temps}}
\end{table}
\clearpage
\begin{table}
\begin{tabular}{ccccc}
\tableline
\tableline
Polarimeter & Channel & LNA & Receiver & Telescope \\
\tableline
PIQUE 2000 & D0 & \nodata & 73 & \nodata \\
      & D1 & \nodata & 68 & \nodata\\
PIQUE 2001 & D0 & \nodata & 66 & \nodata \\
      & D1 & \nodata & 90 &\nodata\\ \tableline
CAPMAP 2003 A & D0 & 55 & 47 & 80\tablenotemark{a} \\
       & D1 & 53 & 52 & 115\tablenotemark{a} \\
CAPMAP 2003 B & D0 & 53 & 51 & 66 \\
        & D1 & 57 & 46 & 61 \\
CAPMAP 2003 C & D0 & 52 & 53 & 119\tablenotemark{a} \\
        & D1 & 54 & 65 & 76\tablenotemark{a} \\
CAPMAP 2003 D & D0 & 52 & 47 & 72 \\
       & D1 & 55 & 45 & 67 \\
\tableline
\end{tabular}
\caption{Radiometer noise temperatures in Kelvin\label{noisetemp}}
\tablecomments{As described in \S~\ref{radopt}, the first column contains the noise temperature of the LNA alone. The second column is $T_{rec}$, the receiver noise temperature tested in lab. We have found that the noise temperatures of the LNAs measured in the test chamber at JPL are consistently 5--10~K higher than those measured in the assembled receivers. This is due to a known systematic error for which a correction was never determined, because the JPL tests were designed to find the optimal bias settings rather than to make a precise noise temperature measurement. The third column is ${\widetilde T_{rec}}$, the noise temperature of the receivers on the telescope, which includes all the optical elements.}
\tablenotetext{a}{These numbers are elevated due to an offset generated by the IF
amplifiers. These high values do not affect the sensitivity of the polarimetry
channels (see \S~\ref{tpperf} and Table~\ref{noisesens}).}
\end{table}
\clearpage
\begin{table}
\begin{tabular}{lccc}
\tableline
\tableline
Polarimeter & \multicolumn{3}{c}{$\left\langle\cos\phi\right\rangle$} \\
 & S0 & S1 & S2 \\
\tableline
PIQUE 2000 & 0.90 & 0.91 & 0.94 \\
PIQUE 2001 & 0.92 & 0.93 & 0.88 \\
\tableline
CAPMAP 2003 A & 0.73 & 0.90 & 0.82 \\
CAPMAP 2003 B & 0.87 & 0.90 & 0.77 \\
CAPMAP 2003 C & 0.92 & 0.90 & 0.74 \\
CAPMAP 2003 D & 0.94 & 0.82 & 0.77 \\
\tableline
\end{tabular}
\caption{Radiometer phase statistic $\left\langle\cos\phi\right\rangle$\label{phasetab}}
\tablecomments{As described in \S~\ref{radopt}, the signal-to-noise ratio of a polarimeter is degraded by the gain-weighted band average of the cosine of the phase difference between its arms.}
\end{table}
\clearpage
\begin{table}
\begin{tabular}{lrr}
\tableline
\tableline
Parameter & Crawford Hill 7-m Telescope & PIQUE Telescope\\
\tableline
Primary Diameter (cm)&700&$140 \times 122$\\
Primary Focal Length $f_1$ (cm)&656.59&75\\
Secondary Diameter (cm)&$120\times180$&\nodata\\
Secondary Near Focal Length $f_2$ (cm)&86.87&\nodata\\
Secondary Far Focal Length $f_3$ (cm)&522.62&\nodata\\
Effective Focal Length $f$ (cm)&3945.59&75\\
Focal Ratio $f/D$&5.63&0.61\\
Plate Scale (cm$\cdot$deg$^{-1}$)&68.9&1.30\\
\tableline
Horn Aperture Diameter (cm)&2.54&0.980\\
Horn Length (cm)&11.43&7.62\\
Beam Width&14.5$^{\circ}$ & 25$^{\circ}$ \\
Return Loss (dB)&$-50$&$-50$\\
\tableline
Lens Diameter (cm) & 12.27&\nodata\\
Lens Focal Length (cm)&16.97&\nodata \\
Lens Beam Width & 2.3$^{\circ}$ & \nodata \\
Lens Reflection (dB)&$-25$& \nodata \\
\tableline
\end{tabular}
\caption{Optical parameters of the PIQUE and CAPMAP telescope
systems\label{tbl:optical_param}}
\tablecomments{The horn parameters refer to the W-band
horn. See also \citet{Chu:1978} and \citet{Wollack:1997}.}
\end{table}
\clearpage
\begin{table}
\begin{tabular}{ccccc}
\tableline
\tableline
Month & Fraction of the time & Average temperature\tablenotemark{b} & Average zenith & 50\% quartile \\
& the sky is clear\tablenotemark{a} & high/low ($^{\circ}$C) & sky temperature\tablenotemark{c} (K) &PWV\tablenotemark{d} (mm) \\
\tableline
Aug & 27\% & 30/20 & \nodata & \nodata \\
Sep & 32\% & 25/16 & \nodata & \nodata \\
Oct & \nodata & 19/9 & \nodata & 12.8 \\
Nov & 31\% & 13/4 & \nodata & 12.8 \\
Dec & 40\% & 6/$-2$ & \nodata & 7.0 \\
Jan &38\% & $-3$/$-5$&30 & 4.0 \\
Feb &46\% & 5/$-4$ &38 & 4.8\\
Mar &22\% & 10/1 & 48 & 9.6\\
\tableline
\end{tabular}
\caption{Atmospheric properties of the Crawford Hill site\label{tbl:winter_season}}
\tablenotetext{a}{Archives from \texttt{http://www.cleardarksky.com} for the winter 2003-2004.}
\tablenotetext{b}{Winter 2002-2003 archives from \texttt{http://www.weatherunderground.com} recorded at Belmar-Farmingdale, NJ.}
\tablenotetext{c}{Average of the eight total power channels during the nominal CMB observations. The channels' receiver temperatures have been removed, and the results have been multiplied by $\cos 40.3^{\circ}$ to refer to the zenith temperature.}
\tablenotetext{d}{Median PWV from GOES--12 satellite hourly archive at \texttt{ftp://suomi.ssec.wisc.edu/pub/rtascii/tpwtext12} for the
winter of 2003--2004 in a 100~km square grid centered on the observing site.}
\end{table}
\clearpage
\begin{table}
\begin{tabular}{lrrr}
\tableline
\tableline
Status & PIQUE 2000 & PIQUE 2001 & CAPMAP 2003 \\
\tableline
Initial tests and tuning & 240 & 96 & 384 \\
Post-season tests & 840 & 0  & 1700 \\
Active observing period & 2000 & 1700 & 1500 \\
\tableline
Various electromechanical problems & 50 & 80  & 150 \\
Scheduled calibration\tablenotemark{a} & 110 & 55 & 60 \\
Planet calibration scans\tablenotemark{b} & 20 & 10 & 20 \\
Bad weather (snow, fog, rain) & 910 & 890 & 730 \\
CMB observations & 810 & 660 & 540 \\
CMB data after cuts & 310 & 190 & 430 \\
\tableline
\end{tabular}
\caption{Time summary (in hours) of the observing seasons\label{observing}}
\tablenotetext{a}{Chopper plate tests, Y-factor tests.}
\tablenotetext{b}{Observations of Jupiter, the moon, and other celestial objects.}
\end{table}
\clearpage
\begin{table}
\begin{tabular}{lcc}
\tableline
\tableline
Instrument/Channel & $\theta_x$ & $\theta_y$ \\
\tableline
PIQUE W-band\tablenotemark{a} & $14.27\arcmin \pm 0.28\arcmin$ & $13.84\arcmin \pm 0.28\arcmin$ \\
CAPMAP 2003\tablenotemark{a} & $3.94\arcmin \pm 0.05\arcmin$ & $3.81\arcmin \pm 0.04\arcmin$ \\
CAPMAP 2003 S0\tablenotemark{b} & \multicolumn{2}{c}{$4.15\arcmin \pm 0.24\arcmin$} \\
CAPMAP 2003 S1\tablenotemark{b} & \multicolumn{2}{c}{$3.81\arcmin \pm 0.16\arcmin$} \\
CAPMAP 2003 S2\tablenotemark{b} & \multicolumn{2}{c}{$3.58\arcmin \pm 0.22\arcmin$} \\
\tableline
\end{tabular}
\caption{FWHM beam sizes\label{optper}}
\tablecomments{Measured beam FWHMs based on Jupiter and Tau A
observations. The FWHMs quoted in the table are calculated from the beam
sizes derived from the Gaussian fits using
$\theta_{\rm FWHM}=\sqrt{8\ln{2}}\,\sigma $. Also listed is the estimated 1$\sigma$ dispersion of the beam sizes
for each channel. See \S~\ref{tpjup} for more details.}
\tablenotetext{a}{Each beam size is the average from both total power channels (D0 and
D1).}
\tablenotetext{b}{Each beam size is the average from each polarimetry sub-band.} 
\end{table}
\clearpage
\begin{table}
\begin{tabular}{lcccc}
\tableline
\tableline
Polarimeter & S0 & S1 & S2 & Combined \\
\tableline
PIQUE 2000 & 2.3 & 2.3 & 3.4 & 1.5 \\
\tableline
CAPMAP 2003 A & 3.2 & 2.1 & 2.7 & 1.5 \\
CAPMAP 2003 B & 3.4 & 2.1 & 2.0 & 1.3 \\
CAPMAP 2003 C & 2.9 & 2.1 & 3.5 & 1.5 \\
CAPMAP 2003 D & 3.1 & 2.2 & 2.5 & 1.5 \\
\tableline
\end{tabular}
\caption{Radiometer sensitivities in thermodynamic units (mK$\sqrt{\mbox{s}}$)\label{noisesens}}
\tablecomments{The sensitivities are measured from a short
 data sample with the polarimeters viewing a fixed spot on the sky,
 when the atmospheric noise temperature was $\sim 50$~K. The
 sensitivities are calculated from the rms of the calibrated
 polarimetry channels on time scales where the noise is white, and neglecting
atmospheric opacity affects. Refer to \S~\ref{polsens} for more details.}
\end{table}
\clearpage
\begin{table}
\begin{tabular}{lcccc}
\tableline
\tableline
Parameter & PIQUE & \multicolumn{2}{c}{CAPMAP 2003} & CAPMAP 2004 \\
 & no lens & double AR\tablenotemark{a} &single AR\tablenotemark{a} & optimized\tablenotemark{b} \\
\tableline
$\gamma$ & $-23$~dB & \multicolumn{2}{c}{ $-23$~dB} & $-23$~dB \\
$d$ & $-14$~dB & $-11$~dB & $-15$~dB & $\leq -21$~dB\\
$q$ & \nodata & $-7$~dB &$-10$~dB & $\leq -19$~dB\\
\tableline
\end{tabular}
\caption{Typical leakage terms for the PIQUE and CAPMAP experiments\label{polparam}}
\tablecomments{The monopole leakage term $\gamma$ is
dominated by the OMT performance (see \S~\ref{poltrans}); the dipole term $d$ for PIQUE arises
primarily from the off-axis mirrors, and the quadrupole term $q$ for CAPMAP stems
from the lenses. The high $d$ term during 2003 is attributed to small alignment errors in the otherwise on-axis horn and lens feed system and is thus correlated with the $q$ term. These leakage terms are consistent with the measured cross-polar beams, which were $-25$ dB peak for the CAPMAP 2003 lenses, and $\leq -40$ dB peak for the re-designed 2004 lenses. During the 2003 season the cross-polarization was dominated by the lens. The errors on the parameters are of order 1~dB.
The CAPMAP 2004 $d$ and $q$ parameters are the values for the worst
channels. See \S~\ref{polfid} and \S~\ref{loccon} for further details on the definitions of these parameters and the methods used to determine them.}
\tablenotetext{a}{For the CAPMAP 2003 season, half of the lenses had AR coating
on only one surface and the other half had AR coating on both.}
\tablenotetext{b}{The lenses and AR coatings were optimized for the CAPMAP 2004 season as explained in \S~\ref{sec:feedsystem}.}
\end{table}
\end{document}